\documentclass{ietbook}

\usepackage{color}
\usepackage{placeins}
\usepackage{amsthm}
\usepackage{mathtools}
\usepackage{amssymb,amsmath,amsthm,bm,amsfonts}
\usepackage{eucal}

\usepackage[linesnumberedhidden,ruled,vlined]{algorithm2e}
\usepackage{algpseudocode}
\usepackage{graphicx}
\usepackage{textcomp}
\usepackage{multirow}
\usepackage[table]{xcolor}
\usepackage{threeparttable}
\usepackage{array}
\newcolumntype{?}{!{\vrule width 1pt}}
\usepackage{verbatim}
\usepackage{cite}
\usepackage{url}
\usepackage{cases}
\usepackage{capt-of}
\usepackage{upgreek}
\usepackage[normalem]{ulem}
\usepackage{nicefrac}
\usepackage{balance}

\begin{document}

\rhbooktitle{New Methodologies for Understanding Radar Data}
\markboth{New Methodologies for Understanding Radar Data}{Modern GPR Target Recognition Methods}
\cauthor{Fabio Giovanneschi\thanks{Fraunhofer FHR, Germany}, 
Kumar Vijay Mishra\thanks{United States Army Research Laboratory, Adelphi, MD 20783, USA} and \\
Maria Antonia Gonzalez-Huici\thanks{Fraunhofer FHR, Germany}
}

\setcounter{chapter}{6}
\chapter{Modern GPR Target Recognition Methods}

In some humanitarian, commercial, and military applications, information on imaging, detection, and localization of shallow buried targets is highly desirable \cite{daniels2004ground}. Often there is a necessity to sense and retrieve this information reliably and safely in, for instance, landmine recognition, archaeological excavations, planetary expeditions, and construction engineering. While a wide variety of technologies are available for subsurface exploration namely, radiometric, seismic, and electromagnetic (EM), only a ground penetration radar (GPR, hereafter) provides non-invasive, safe, efficient, and high-resolution sensing \cite{daniels2004ground}. This advantage has led to significant advances in research on GPR acquisition and information retrieval for the past two decades.

In an \textit{impulse GPR} system \cite{persico2014introduction}, the transmit antenna emits an ultra-wideband EM pulse at the surface. As the signal propagates through the subsurface layers, their magnetic and electrical properties such as permittivity, permeability and conductivity induce changes in the phase and amplitude of the scattered wave. While the amplitude of the reflected wave is directly proportional to the complex reflectivity of the buried target, the phase provides information on the target's location relative to the radar mounted on the surface. The GPR receiver collects and processes the reflected echoes to extract the reflectivities and ranges of several targets. 

In GPR applications, a target is located within the close range of the radar. As a result, the received signal is composed of multiple reflections from different parts of the same object. This \textit{extended} target model implies retrieval of multiple ranges and amplitudes of portions corresponding to the same larger object. By exploiting the spatial correlation among the point values of the  backscattered signal, GPR generates a \textit{target signature} that is usually specific to the object being explored and, hence, useful to identify it \cite{jol2008ground}. In this chapter, we focus on this \textit{target recognition} aspect of GPR processing.

Often terms such as target recognition, classification, and identification are interchangeably used for techniques to classify targets from measured radar data. However, for the purposes of this chapter, \textit{target classification} implies isolating the target to a general class, while \textit{target identification} is used to distinguish the target more precisely. For example, labeling a target as a landmine or soil clutter is a classification step. Identifying a landmine target as PMN/PMA is its identification. The target recognition encompasses both classification and identification \cite{tait2005introduction}.


 In practice, the recorded GPR signals generally suffer from several unwanted contributions arising from system effects such as antenna coupling, multiple surface reflections, system instability, timing jitter, limited spatial resolution, and amplitude variations \cite{gonzalez2014comparative}. Further, the rough ground surface in which the targets of interest are buried generates the undesired \textit{clutter} signals which are comparable in strength to the target signatures. As a result, GPR target recognition is a difficult task.

Traditional GPR target recognition methods (see \cite{jol2008ground} for a review) include preprocessing the data by removal of noisy signatures, dewowing (high-pass filtering to remove low-frequency noise), filtering, deconvolution, migration (correction of the effect of survey geometry), and can rely on the simulation of GPR responses. These techniques usually suffer from the loss of information, inability to adapt from prior results, and inefficient performance in the presence of strong clutter and noise. 

To address these challenges, several advanced processing methods have been developed over the past decade to enhance GPR target recognition. In this chapter, we provide an overview of these modern GPR processing techniques. In particular, we focus on the following methods:

\begin{itemize}
\item  \textit{adaptive receive processing} of range profiles depending on the target environment \cite{giovanneschi2019dictionary}
\item adoption of \textit{learning-based methods} so that the radar utilizes the results from prior measurements \cite{shao2013sparse, giovanneschi2019dictionary}
\item application of methods that exploit the fact that the target scene is \textit{sparse} in some domain or dictionary \cite{mishra2019sub,krueger2014compressive}
\item application of advanced \textit{classification} techniques \cite{giovanneschi2013parametric}
\item \textit{convolutional coding} which provides succinct and representatives features of the targets \cite{tivive2017gpr}
\end{itemize}
We describe each of these techniques or their combinations through a representative application of landmine detection.

The rest of the chapter is organized as follows. In the next section, we provide a brief overview of GPR technology and complex scattering behaviour of EM waves in the soil. We describe the GPR signal model and list both classical and modern processing methods in Section~\ref{sec:GPR_sigpro}. We explain the dataset for our illustrative application of landmine detection in Section~\ref{sec:landmine_meas}. Section~\ref{sec:SR} focuses on various sparse representation techniques which are enabled by dictionary learning algorithms detailed in Section~\ref{sec:DL}. We outline the adaptive statistical evaluation carried out by the radar processor in order to selectively apply one of the algorithms in Section~\ref{sec:parameval}. Finally, we present results of all of these techniques for target recognition in Section~\ref{sec:target_recog} before concluding in Section~\ref{sec:summ}.

Throughout this chapter, we reserve boldface lowercase and uppercase letters for vectors and matrices, respectively. The $i$th element of vector $\textbf{y}$ is $\mathbf{y}_i$ while the $(i,j)$th entry of the matrix $\textbf{Y}$ is $\textbf{Y}_{i,j}$. We denote the transpose by $(\cdot)^T$. We represent the set of real and complex numbers by $\mathbb{R}$ and $\mathbb{C}$, respectively. 
The notation $\left\Vert\cdot\right\Vert_p$ stands for the $p$-norm of its argument; $\left\Vert\cdot\right\Vert_F$ is the Frobenius norm; $\langle\cdot,\cdot\rangle$ indicates the dot product between the two arguments; and $|\cdot|$ is the cardinality of the set argument. A subscript in the parenthesis such as $(\cdot)_{(t)}$ is the value of the argument in the $t$-th iteration. The  convolution product is denoted by $\ast$. The function $\text{diag}(\cdot)$ outputs a diagonal matrix with the input vector along its main diagonal. We use $\text{Pr}\{\cdot\}$ to denote probability, $\text{E}\left\lbrace \cdot \right\rbrace $ is the statistical expectation, and $|\cdot|$ denotes the absolute value. The functions $\text{max}(\cdot)$ and $\text{sup}(\cdot)$ output the maximum and supremum value of their arguments, respectively. The $\textrm{i}=\sqrt{-1}$ is the imaginary unit; $\vec{(\cdot)}$ is a vector field; $\mathbf{\nabla} \cdot$ and $\mathbf{\nabla} \times$ denote the divergence and curl vector operators, respectively.

\section{GPR for Buried Target Recognition}
\label{sec:GPR}
The GPR is a sophisticated technology with decades of development history. We refer the reader to excellent expositions in \cite{daniels2004ground,jol2008ground}. In the following, we summarize key principles of GPR transmission and acquisition that are relevant to this chapter. 
\subsection{Subsurface EM propagation}
\label{subsec:scat_th}
A GPR probes the underground by repeatedly transmitting EM pulses into the subsurface at a wavelength $\lambda$ of near-constant power $P_{t}$. The wave travels through the medium and reflects back to the radar when there is a change in the dielectric properties caused by variation in the medium \cite{daniels2004ground,gonzalez2013accurate}. 

Define the vector $\vec{E}$ (in $V/m$) as the electric field intensity, $\vec{D}$ ($C/m^2$) as the electric flux density, $\vec{B}$ ($T$) as the magnetic flux density, $\vec{J}$ ($A/m^2$) as the electric current density, and $\rho_c$ ($C/m^3$) as the electric charge density. The $\epsilon_0$ ($F/m$) and $\mu_0$ ($H/m$) are the electric and magnetic field constants, respectively; $\epsilon_r$ and $\mu _r$ are the relative dielectric permittivity and the relative magnetic permeability, respectively; $\sigma_r$ ($S/m$) is the electrical conductivity and the parameter $\omega$ ($rad/s$) is the angular frequency. Provided some boundary conditions are defined and assuming the general case of anisotropic (dispersive) media, the electric and magnetic fields vectors of the transmit EM wave at a given point in space characterized by the position vector $\mathbf{r}$ at time instant $t$ is described by the \textit{Maxwell equations}  \cite{balanis2005antenna,gonzalez2013accurate}
\begin{align}
\mathbf{\nabla} \times \vec{E} (\mathbf {r},\omega )&=-\mathrm{i}\omega\vec {B} (\mathbf {r},\omega ),\label{eq:Maxwell1}\\
\mathbf{\nabla} \times \vec {H} (\mathbf {r},\omega )&=\left( \sigma (\mathbf {r},\omega )+\mathrm{i}\omega \epsilon_0 \epsilon_r (\mathbf {r},\omega ) \right) \vec{E}(\mathbf {r},\omega ),\label{eq:Maxwell2}\\
\mathbf{\nabla} \cdot \left( \epsilon_0\epsilon_r (\mathbf {r},\omega)\vec {E} (\mathbf {r},\omega )\right)&=0, \label{eq:Maxwell3}\\
\mathbf{\nabla} \cdot \vec {B} (\mathbf {r},\omega )&=0,\label{eq:Maxwell4}
\end{align}
and the related \textit{constitutive relations} 
\begin{align}
\vec{D}(\mathbf{r},\omega )&=\epsilon _{0} \epsilon _{r} (\mathbf{r},\omega ) \vec{E}(\mathbf{r},\omega ),\label{eq:const1}\\
\vec{B}(\mathbf{r},\omega )&=\mu_{0} \mu_{r} (\vec{r},\omega ) \vec{H}(\mathbf{r},\omega ),\label{eq:const2}\\
\vec{J}(\mathbf{r},\omega )&=\sigma (\mathbf{r},\omega )\vec{E}(\mathbf{r},\omega ).\label{eq:const3}
\end{align}
We choose to provide the frequency-domain representations of these equations because, unless the medium of propagation is non-dispersive (e.g. vacuum), the constitutive equations (\ref{eq:const1})-(\ref{eq:const3}) do not generally hold for time-dependent fields. 

The quantities $\epsilon_r$, $\mu _r$ and $\sigma$ are generally complex and frequency dependent. Assuming the media to be homogeneous, they are spatially independent. However, for a typical GPR scenario with operating frequency from 10 MHz to a few GHz, the magnetic permeability is negligible ($\mu _r=1$), the imaginary part of the electric conductivity is ignored and its real part is considered frequency-independent and equal to the direct current (DC) conductivity \cite{knight2005introduction}. The real part of $\epsilon_r$ (with $\epsilon_r=\epsilon'_r+i\epsilon''_r$) represents the electric permittivity of the soil and the imaginary part incorporates the losses for the conductivity and frequency. 

The speed of the EM wave inside the soil is affected by its dielectric properties. In general, the higher the dielectric permittivity, greater is the reduction in the speed \cite{daniels2004ground}. The \textit{phase velocity} (in $m/s$) of the EM wave in the subsurface is $v_{p} = \frac{1}{\sqrt{\epsilon \mu}} \phantom{1} \approx \phantom{1} \frac{c_0}{\sqrt{\epsilon_r}}$, where $c_0 = \frac{1}{\sqrt[]{\epsilon_0 \mu_0}}$ is the phase velocity in free space with $c_0=3\times10^8$ m/s. For most GPR applications, the imaginary part of $\epsilon_r$ is ignored and only the real part affects the attenuation and phase constant of the transmitted EM wave. Hence, ${v_p}$ is inversely proportional to the square root of the real part of the dielectric permittivity ${v_p} \approx {c_0}/{\sqrt{\displaystyle\epsilon'_{\operatorname{r}}}}$.

In practice, GPR media are not homogeneous and often an EM wave encounters interfaces of different dielectric media. The classical optical geometry equations are not sufficient to describe the EM waves behavior at the interface when the shallow objects are smaller than the GPR wavelength or interfaces in the close proximity of the illuminating source are present. 
The electrical properties of the soil, which affect the GPR signal propagation, depend on various factors such as the volumetric water content, texture of the soil particles, bulk density, and temperature. Among these, the water content is the most predominant factor influencing the electrical permittivity of the soil. 
The spatial variability expressed by the correlation length of the soil, i.e. the measure of the range over which fluctuations in one region of space are correlated with those in another region, also characterizes the EM propagation in GPR. Finally, the roughness of the air-ground interface influences the magnitude of the backscattered energy and depends on the surface characteristics and radar's wavelength.

The operation of GPR is similar to conventional ultra-wideband (UWB) radar systems except that the GPR signal propagates into the subsurface. GPR systems can be employed for frequencies above 1MHZ, where the EM behavior is not inductive and can be described by EM waves. At these frequencies, other factors such as the moisture content of the subsurface or, more generally, the dielectric properties of the soil hamper the penetration depth. In applications such as landmine detection, higher bands ($>1$ GHz) are employed because they provide sufficiently wide bandwidth to achieve the necessary range resolution for discriminating very small targets \cite{daniels2004ground} at the cost of a lower penetration depth. The attenuation coefficient ($\alpha$) of an EM wave travelling in the subsurface is \cite{daniels2004ground}
\begin{align}
\label{eq:GPR_att}
\alpha = \sqrt[]{\frac{\mu}{\epsilon}} \phantom{1} \frac{\sigma}{2} \approx Z_0 \phantom{1} \frac{\sigma}{2 \phantom{1} \sqrt[]{\epsilon_r}},
\end{align}
where $\epsilon=\epsilon_r\epsilon_0$ is the permittivity of the soil, $\mu=\mu_0\mu_r$ is its magnetic permeability, and $Z_0 = \sqrt{\frac{\mu_0} {\epsilon_0}}$ is the free space impedance (in $\Omega$). Note that we employed the approximation $\mu_r \approx 1$ for the operating frequencies here. 

The radar signal is reflected when there is a change in the EM properties of the soil, especially $\epsilon$. The reflected signal is the superposition of the contributions of a specific target at a certain depth and several unwanted effects as listed below.
\begin{description}
 \item[Ground reflection] The antennas of GPR systems are usually mounted just above the surface leading to significantly strong reflections from the air-ground interface.
 \item[Antenna cross-coupling] The cross coupling between the receive and the transmit antennas causes the transmit signal to interfere with the received echoes. This is mitigated only partially by using absorber materials in the radar.
 \item[Antenna ringing] This happens when transmit antenna continues radiating after the exciting source has expired. In the popular GPR bow-tie antenna, internal reflections of the charges that occur at the ends of the bow-tie wings cause ringing. At the excitation points, this corrupts the transmit pulse waveform and reduces range resolution.
 \item[Surface roughness] As per the Rayleigh criterion \cite{lambot2006effect}, the roughness of the surface generates undesired scattering patterns which are not associated with subsurface features. This is especially true for shallow surface exploration (e.g. landmine detection) where the frequencies of operations are \texttildelow$1$-$2$ GHz.
 \item[Moisture content] 
 The signal attenuation caused by the water content depends on the frequency of operation and its strictly related to the dielectric permittivity of the soil \cite{daniels2004ground}. 
 \item[Low permittivity contrast] The intensity of reflections depends on the permittivity contrast between the soil and features of interest \cite{gonzalez2013accurate}. Small non-metallic targets (such as landmines) have very low permittivity contrast thereby making it difficult to identify them. 
\end{description}

\subsection{GPR Systems}
\label{subsec:gpr_syst}
Broadly, GPR systems are classified as time- and frequency-domain systems. In the former, GPR transmits a short pulse and the receiver applies the time-domain processing to the backscattered reflection. In the latter, the radar emits a series of individual frequencies and receives the signal via a frequency conversion receiver. In this chapter, we focus on time-domain pulsed GPR systems.

\begin{figure}[t]
\centering
  \includegraphics[width=0.55\textwidth]{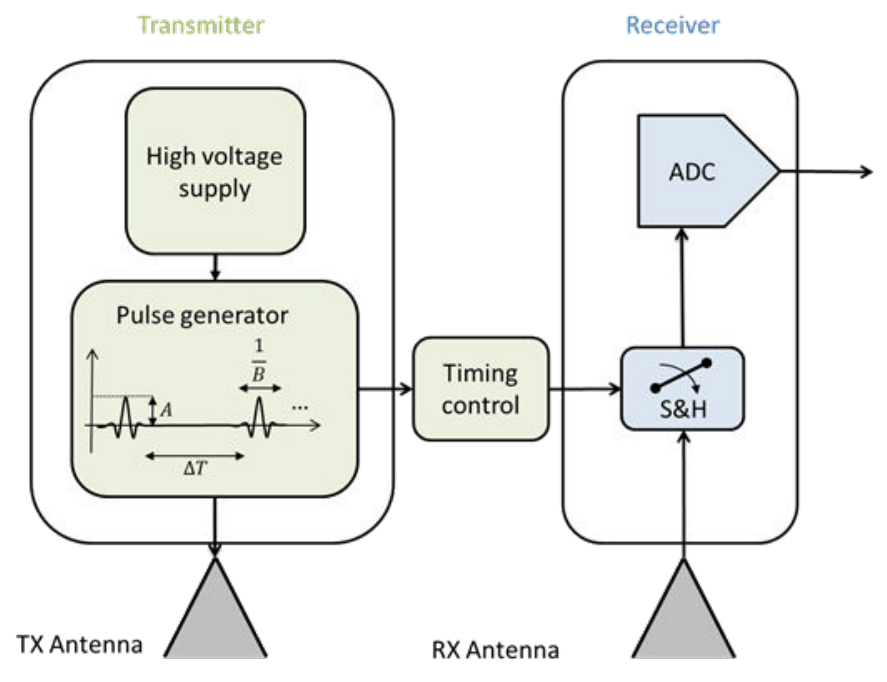}
  \caption{Simplified block diagram of a pulsed UWB GPR. The pulse generator employs a technique of rapid discharge of the stored energy (from the high voltage supply) into a short transmission line for generating the pulse. The sample-and-hold (S\&H) receiver quantizes the signal to obtain digital samples using a flash analog-to-digital converter (ADC). The pulsed GPR antennas are usually dipoles and bow-ties because they are wideband, easy to design, non-dispersive, and linearly polarized \cite{lestari2005adaptive}. They are sealed in a shielding box filled with the absorbing material to prevent coupling.}
\label{fig:tdGPR}
\end{figure}

A pulsed GPR generates a series of short pulses (with pulse-widths ranging from $200$ ps to $50$ ns) with a repetition interval of the order of hundreds of $\mu$s to 1 ms. A UWB GPR pulse (Fig~\ref{fig:tdGPR}) has a central frequency ($f_c$) (few MHz to 1 GHz) with bandwidth equal to $\Delta f$. 
The depth resolution $\Delta_z$ (in $m$) of a GPR along the depth axis ($z$ in Cartesian coordinate system) is $\Delta_z = \frac{v_{p}}{B}$. Since the resolution depends on the phase velocity, it is affected by the dielectric permittivity of the material in which the target is embedded. Hence, a radar with receiver bandwidth of $2$ GHz resolves targets spaced $1$-$10$ cm apart from each other depending on the value of $\epsilon_r$.

\begin{figure}[t]
\centering
  \includegraphics[width=0.75\textwidth]{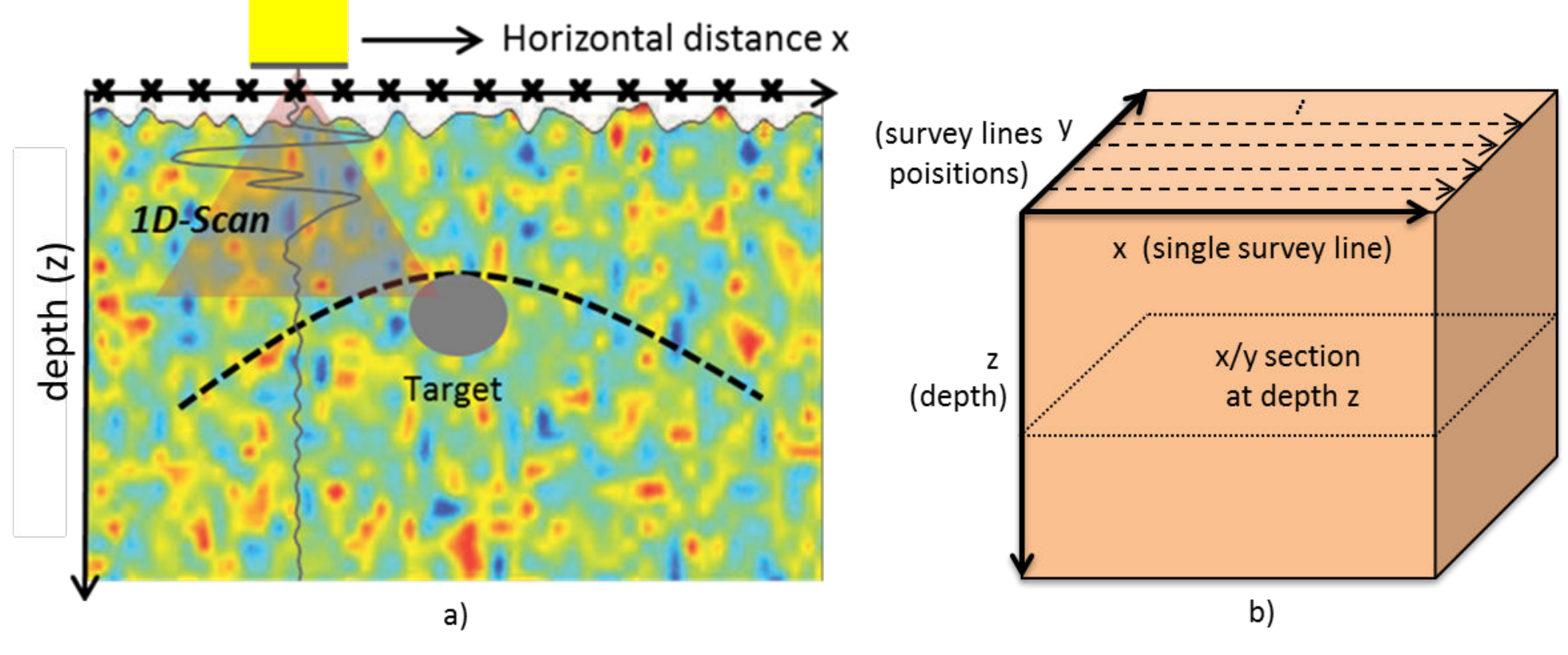}
  \caption{(a) Illustration of a down-looking GPR system operation indicating a single range profile (b) Directions of a single survey line ($x$-axis), survey lines positions ($y$-axis) and depth ($z$-axis) in a Cartesian coordinate system. The 2-D section (C-Scan) is indicated at depth $z$.}
\label{fig:GPRprinciples}
\end{figure}
\begin{figure}[t]
\centering
  \includegraphics[width=0.65\textwidth]{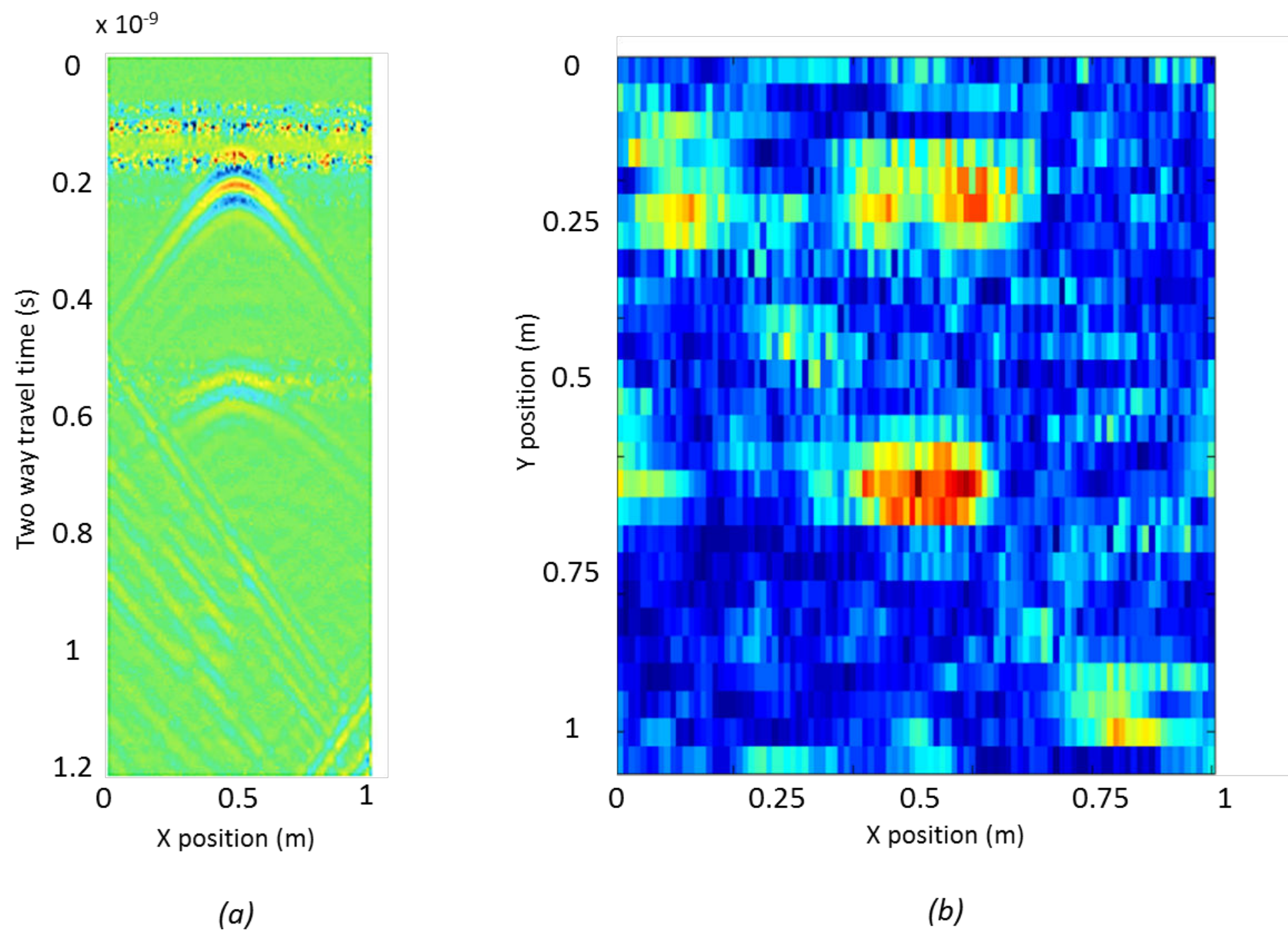}
  \caption{(a) B-scan of a simulated spherical target buried in sand material (b) C-Scan from real data at $15$ cm depth of two landmine simulants buried in highly non-homogeneous soil.}
\label{fig:bscancscan}
\end{figure}

Figure~\ref{fig:GPRprinciples} illustrates the geometry of GPR operation. The radar moves along a survey line. For each horizontal position (indicated with crosses), the radar sends down a pulse (or a series of pulses) and records the backscatter (or integrate a series of received echoes). The GPR could be moved by a single operator or pulled by a vehicle. 
The acquisition is usually triggered as a function of the traveled distance (for instance, by an odometer connected to the wheel of the moving platform). For each horizontal position, we obtain a single range profile known as one-dimensional (1-D) or A-scan (Fig.~\ref{fig:GPRprinciples}) usually generated from the integration of many received pulses in order to increase the signal-to-noise ratio (SNR). If we stack many range profiles together along the survey line, we obtain a 2-D visualization or a B-Scan (Fig.~\ref{fig:bscancscan}a). Due to the wide antenna beamwidth, the GPR receives reflections from a target before it is right above it. This results in the well-known hyperbola-like response in the B-scan where the edge of the hyperbola is associated to the real depth of the target. If the acquisition has been repeated for many survey lines, we could stack all the data together to form a \textit{data-cube}. Figure~\ref{fig:bscancscan}b shows an X/Y section of the aforementioned data-cube for a certain depth. This visualization is called C-scan.

\section{GPR Signal Processing}
\label{sec:GPR_sigpro}
One of the most employed transmit pulse of time-domain GPR systems is the \textit{monocycle}. Given a Gaussian waveform
\begin{align}
\label{eq:Gauss}
s_G(t)= Ae^{-2\pi^2 f_c^2 (t-\tau)^2 },\;t\in[0,+\infty],
\end{align}
where $f_c$ is the central frequency, $A$ is the peak amplitude and $\tau=1/f_c$, the monocycle waveform is its first derivative
\begin{align}
\label{eq:Mono}
s_T(t)= -4 \pi^2 f_c^2 A (t-\tau) e^{-2\pi^2 f_c^2 (t-\tau)^2 },\;t\in[0,+\infty].
\end{align} 
In these UWB systems both the central frequency and the bandwidth are approximately the reciprocal of the pulse length. 
The scattering of UWB radar signals from complex targets that are composed of a finite number of scattering centers, can be described in terms of the channel impulse response (CIR). Here, the CIR is considered as a linear, time invariant, causal system which is a function of the target shape, size, constituent materials, and scan angle. Without taking in account the influence of the soil, the CIR $h(t)$ of a GPR target, with $M$ scatterers, can expressed as a series of time-delayed and weighted Gaussian pulses
\begin{align}
h(t)= \sum\limits_{m=1}^{M}\alpha_m e^{-4\pi\left(\frac{t-t_m}{\Delta T_m}\right)^2},
\end{align}
where each scatterer located at range $r_m$ from the radar is characterized by the reflectivity $\alpha_m$, duration $\Delta T_m$, relative time shift $t_m = 2r_m/v_s$, where $v_s=c/\sqrt{\epsilon_r}$ is the speed of the EM wave in the soil, $c=3\times 10^8$ m/s is the speed of light, and $\epsilon_r$ is the dielectric constant which depends on the soil composition and moisture.

The response of the target to the Gaussian monocycle is then the received signal
\begin{align}
\label{eq:rng_prof}
y(t)= s_T(t)\ast h(t),
\end{align}
also regarded as the \textit{range profile}. For each X/Y position, the system receives a radar echo (range profile) from the transmitted pulse. In order to deal with the exponential signal attenuation during the propagation through the soil medium, the dynamic range of the signal is enhanced via stroboscopic sampling \cite{daniels2004ground,pasculli2015real}. This technique comprises integrating $N$ receiver samples (generated by transmitting a sequence of $N$ pulses) at the ADC receiver sampling rate but with a small time offset $\delta$ for each of them. To achieve the desired stroboscopic sampling rate $T_s$, the time offset must be selected accordingly, i.e., $\delta= T_s/ N$. 

\subsection{Classical Pre-Processing}
\label{subsec:class_gpr}
In order to extract useful information from the raw data contaminated with noise and system instability, classical pre-processing in time-domain pulsed GPR is grouped into following four categories: \textit{basic handling} (editing/gating, dewow, background removal, and resampling) \cite{jol2008ground}, \textit{filtering} (spatial filtering and predictive deconvolution) \cite{kim2007removal}, \textit{imaging} (velocity analysis, elevation correction, backprojection, and Stolt migration) \cite{gonzalez2014comparative}, and \textit{modeling} (2-D forward modeling and simulation of GPR responses) \cite{bitri2008frequency}. We provide details of some of these techniques below.
\begin{description}
\item[Time gating]
In practice, the location of ground reflections and/or starting depths of targets may be known from prior measurements. Then, time gating is used to edit the range/time profiles to limit the prospection depth to a specified interval. However, this may also end up removing shallow target (i.e. landmines) responses. Time gating is also used in post-processing.
\item[Dewow]
This process comprises a running average filter to remove the initial DC and low-frequency component. A common way to perform dewow is to calculate the mean of a few GPR profiles (A-Scans) along a section and then subtract it from every single profile in the section. The length of the section depends on GPR wavelength.
\item[Time gain]
Some GPR architectures incorporate a time-varying gain to compensate from the attenuation of the signal during its propagation through the ground. The gain curve is customized in order to account to a more smooth or edge transition along the depth. This could also be applied in post-processing to improve data visualization.
\item[Background subtraction]
Background subtraction is commonly implemented by subtracting the mean profile (over an entire survey line) from each profile in the line, usually through Principal Component Analysis (PCA) \cite{wall2003singular}. The PCA performs singular value decomposition to extract the first principal features of the data associated with the background and subtract it from profiles under consideration. Note that background removal is different from dewow because it takes an entire survey line instead of a ``moving window".
\item[Migration]
Due to the beam-width of transmit and receive antennas and the differences in round-trip travel time of the pulse caused by the movement of the antenna along the measurement line, the reflections from scatterers will appear as hyperbolic curves in the recorded data. These hyperbolic structures can be migrated (focused) into the real position of the corresponding scatterer via different migration techniques \cite{gonzalez2014comparative} to increase SNR and better localize/detect the targets. To apply these algorithms successfully a correct estimation of the velocity structure of the propagation medium, i.e., the dielectric permittivity of the soil, needs to be done. The phase-shift migration is a Fourier transform based technique which is also referred to as frequency-wavenumber migration (F-K migration). It exploits the wave equation to back-propagate the received signal into the soil back to the scattering source, and obtain an image of the subsurface reflectors. In practice, the 2-D case and a monostatic setting is generally assumed (a valid approximation when the transmitter and receiver are close to each other) but the extension to 3-D is straightforward. Consider a 2-D spatio-temporal Fourier transform of the range profile $y(z_{0}=0, x, t)$ at time t along the x-axis:
\begin{align}
D\left(z_{0}=0, k_{x}, \omega\right)=\iint d\left(z_{0}=0, x, t\right) e^{\mathrm{j} k_{x} x} e^{-\mathrm{j} \omega t} \mathrm{d} x \mathrm{d} t,
\end{align}
where $\omega$ is the angular frequency, $z_{0}$ is the antenna vertical position, and $k_{x}$ is the wavenumber along the x-axis. Here, it is assumed that $y$ satisfies the wave equation. Then, to determine the field at a range of depths, a phase shift is applied, which depends on the propagation constant. This phase shift operation is an extrapolation along x-axis, such that at $z=z_1=z_0+\Delta z$,
\begin{align}
D\left(z_{1}, k_{x}, \omega\right)=D\left(z_{0}=0, k_{x}, \omega\right) e^{\mathrm{j} \sqrt{\frac{4 \omega^{2}}{v_p^2}-k_{x}^{2}} \Delta z}.
\end{align}
By recursively extrapolating the field in steps $\Delta z$ and using the result of each step as input for the next iteration, the F-K distribution of the field is reconstructed. Finally, the migrated data $\tilde{y}$ are obtained via the inverse fast Fourier transform (IFFT) of the wavenumber data over $k_{z}$ and $\omega$ for the imaging condition $t=0$
\begin{align}
\tilde{y}\left(z=z_1, x, t=0\right)=\frac{1}{4 \pi^{2}} \iint D\left(z=z_{1}, k_{x}, \omega\right) e^{j k_{x} x} \mathrm{d} k_{x} \mathrm{d} \omega,
\end{align}
where $z_{1}$ is the depth of the migrated scene. In order for the IFFT to solve the above integral, the data matrix $D$ previously needs to be evenly mapped into the $k$-space via interpolation. When $v$ is assumed constant along depth, it is called \textit{Stolt migration} or \textit{Stolt mapping} \cite{gonzalez2014comparative}.
\end{description}


\subsection{Modern GPR Processing Techniques}
\label{subsec:mod_gpr}
The classical processing techniques involve several approximations and heuristics that lead to loss of information during editing and filtering operations. Often a constant intervention of the operator is required to ensure correct interpretation of the data. When the sampled data is limited, it is very difficult to classify targets. Modeling based on electromagnetic equations leads to inaccuracies in the presence of coupling and ringing effects. To address these drawbacks, we leverage recent advances in signal processing to enhance target recognition. We consider following techniques in subsequent sections.
\paragraph{Role of Sparsity}
In many radar applications, the received signal is sparse in a certain domain or dictionary \cite{mishra2019sub}. This property is not only useful for applying Compressive Sensing based algorithms in signal reconstruction obtained at reduced rates, it is also helpful in classifying the signal with fewer samples \cite{giovanneschi2019dictionary}. In Section~\ref{sec:SR}, we introduce the basics of sparse representation (SR) to construct the dictionary in which the GPR signal occupies fewer basis points. The selection of the dictionary matrix is crucial for obtaining a sparse(r) representation of a signal, i.e., a representation with a minimum number of non-zero coefficients. Depending on the application, one can build a dictionary with an arbitrary basis (such as Fourier, Wavelets, etc.) or by collecting empirical or synthetic target signatures. The SR has shown improvement in augmented resolution, clutter reduction, and target classification. Note that SR and CS share the same framework of techniques except that the former tries to exploit the inherent sparsity of data by representing it on an appropriate basis without necessarily reduce the amount of measurements.
\paragraph{Learning-Based Methods} 
In recent years, learning methods have been widely adopted in radar processing. Since GPR signals are not naturally sparse in the common domains of time, space, and frequency, it is critical to learn the arbitrary sparse basis of such signals. In Section~\ref{sec:landmine_meas}, we present dictionary learning (DL) algorithms aimed at target classification using GPR range profiles (A-Scans). Here, we are particularly interested in \textit{online} DL algorithms because of their fast computation times.  

Another state-of-the-art learning technique for classification is the use of deep neural networks. These methods have the ability to extrapolate new features from a limited set of features contained in a training set and, thus, are valuable tools in the face of non-availability of critical parameters. Architectures such as Convolutional Neural Network (CNN) are very effective in object classification. However, defining these networks involves many design decisions. Further, huge data-sets are required to train the network which may be a major difficulty in downlooking GPR applications \cite{sakaguchi2015recognizing,besaw2015deep,bralich2017improving}. Possible solutions may be to extend the training set using synthetically generated images \cite{lameri2017landmine} or to augment the training dataset with suitable transformations or augmentation strategies \cite{reichman2017some}. Later in the chapter, we compare both CNN and DL-based classifications when the samples of the original range profiles are randomly reduced. 
Comparisons of the classification performance using CNN reveal that sparse decomposition based techniques with DL generally perform better than CNN alone when the input signals are randomly sub-sampled.
\paragraph{Adaptive and Cognitive Processing}
The cognitive processing in radar entails a framework of techniques to realize the sense-learn-adapt cycle \cite{mishra2020toward} that operates at different layers of the radar architecture to achieve pre-determined tasks tuned to the changes in the target environment. The lower level of skill-based performances includes all techniques that permit the adaptation of both optimal waveforms and other design parameters. At the level of rule-based behavior, extraction of adaptive features from the received data along with applications such as target identification and classification are included. This often involves use of learning techniques and artificial intelligence. The final knowledge-based processing layer \cite{gurbuz2019overview} includes the modes of operation, such as specific scans, beam scheduling, and waveform selection, employed to enable cognition \cite{ender2015cognitive,bruggenwirth2016design}.

The aforementioned adaptive architectures have also been investigated and developed for GPR. The localizing GPR system described in \cite{cornick2016localizing} employs online data processing to keep autonomous ground vehicles in a lane by feeding them with a map of subsurface features. This map‐based vehicle localization using GPR works in conjunction with the existing sensors such as a Global Positioning System (GPS), lidar, and camera mounted at the bottom of the vehicle. When these devices fail to provide accurate maps because of low visibility in the presence of snow, dust, gravel or dirt on the road, the GPR cognitively functions to provide complementary data to the vehicle. In \cite{williams2014crevasse}, supervised machine learning with pre-trained models was used in a GPR system to probe glacier surfaces and automatically identify crevasses. It employs hidden Markov models to adaptively prescreen the data and mark locations in the collected data. Then, only marked files are used for crevasse detection.

Fully cognitive GPRs employ a feedback mechanism to dynamically tune radar operational parameters and continuously improve sensing performance. In this context real-time GPR data processing requires significant computing and storage capability, often limiting its applicability. To tackle this problem, edge computing is used to reduce the computing latency by pushing the computation, communication, and storage resources from a remote data center to the edge of network \cite{mishra2018sustainable}. Recently, \cite{ omwenga2019autonomous } describes development of an edge computing and reinforcement learning framework that enables autonomous cognitive GPR. 

The GPR target environment is riddled with various contamination sources. There is, therefore, interest in making the data processing adaptive based on the received data so as to maximize the probability of detection \cite{mishra2020toward}. In this context, time-domain GPRs are preferred over the frequency-domain systems. The latter, such as the ones employing stepped-frequency continuous-wave (SFCW) transmitter, are more efficient than impulse GPRs but offer a poorer resolution. The frequency-modulated continuous-wave (FMCW) GPR has the advantage of a low-rate ADC at the receiver because of the low bandwidth of the beat signal \cite{mishra2019toward}. However, these systems are expensive, inflexible for adaptive processing, and, in case of multiple transmitters, suffer from high level of self-interference.

The impulse radar is more suitable for enabling cognition and scenario-awareness in modern GPR systems, which use digital transmit and receive modules build using field-programmable gate arrays (FPGA) to impart high reconfigurability. The receiver directly samples backscattered echoes at radio-frequency (RF). At present, GPRs generate only 1D-scans and contiguous linear scans must be stacked together after acquiring more measurements. The manual errors that creep in while choosing the start/ending point generate drifts in the acquired data. To scan different depths, transmit-receive pairs from antenna array elements in a multiplexed manner\cite{cornick2016localizing} or separate antennas \cite{GSSI30} are employed.

In \cite{srivastav2020highly}, an illumination system that uses pseudorandom codes with ideal autocorrelation properties is proposed for GPR. This enables receiver processing based on pulse-compression that improves the range resolution and maximum depth \cite{mishra2020rasster}. To reduce scan drifts and improve target localization, this system deploys multi-static antenna with 8 transmitters and 8 receivers.

\begin{figure}[t]
\centering
  \includegraphics[width=0.75\textwidth]{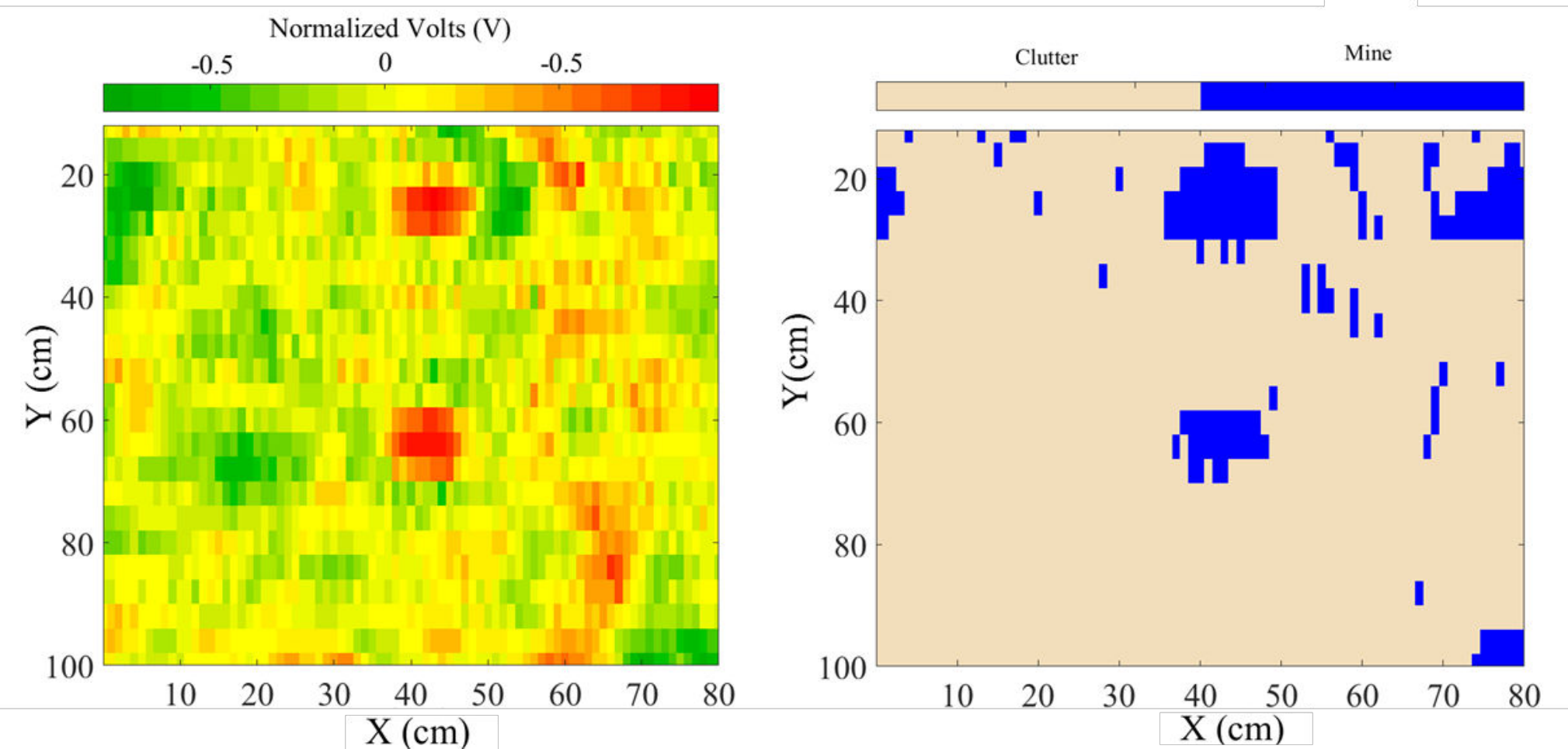}
  \caption{Left: Raw data at $15$ cm depth from an L-band GPR contains two landmine simulants. Right: Corresponding classification map of the survey area.
  }
\label{fig:class_example}
\end{figure}

In the context of DL, we describe an adaptive technique in Section~\ref{sec:parameval}, where we evaluate various DL algorithms based on statistical metrics to select the optimal input parameters for DL. For the specific application of landmine detection, we demonstrate the adaptive assessment using statistical distances \cite{massart1990tight}. These metrics allow greater fine-tuning of parameters respect to the conventional bulk statistics such as root mean square error (RMSE).
\color{black}
\paragraph{Advanced Target Recognition and Classification} 
GPR classification approaches aim to discriminate target/anomalies from the ground clutter, possibly indicating their precise position in the analyzed surveys. Unlike GPR imaging approaches (such as backprojection and migration) which aim at improving the visualization of the targets in the final GPR image, the output of classification methods can be seen as a map of declared classes along the survey area (see, for example, Fig.~\ref{fig:class_example} which shows such a map from real data).
GPR classification approaches work on raw or pre-processed data, in time or frequency domain, and evaluate single A-scans or a collection of received GPR data (such as entire B- or C-scans). A variety of signal processing algorithms have been proposed for detection of low metal-content targets in realistic scenarios; approaches based on feature extraction and classification are found to be the most successful (see e.g. \cite{giannakis2016model}), yet false-alarm rates remain very high. Many classification approaches rely on a database of GPR signatures. This can be synthetically generated (for example, by a modeling software like GprMax \cite{giannopoulos2005modelling} or Comsol) taking in account all the necessary parameters (soil, antenna, transmitted signals, etc) of the scenario under test. However, obtaining a complete and general database of GPR signatures is usually very challenging; in many cases it is more convenient to extract salient features from a \textit{representative} database of the target of interest and use them as an input for the chosen classifier. A representative database is smaller with respect to a general database because it contains only an accurate selection of the classes of interest for the desired application. Some GPR signal representation that were promising for buried targets/structures discrimination purposes include wavelets \cite{shao2013sparse}, target resonances and top/bottom reflection ratios\cite{giovanneschi2013parametric}, edge histogram descriptors \cite{frigui2009detection}, and sparse representation coefficients \cite{giovanneschi2015preliminary}. 

In this chapter we show the use of low-dimensional features for classification instead of entire B-scans or range profiles. Specifically, we employ SR coefficients extracted from the GPR range profiles. These coefficients strongly depend on the selected basis for the representation, which is a key for the success of the classification approach. Figure~\ref{fig:class_example} schematically shows the proposed strategy for  landmine classification. The final detection could be based on hypotheses testing \cite{pambudi2019forward} or data-dependent heuristics \cite{giovanneschi2019dictionary}.
\paragraph{Convolutional Methods} 
 It is possible to further improve the performance of SR, which conventionally assumes that different input data (A-Scans) are independent of each another. In practice, successive GPR scans are spatially correlated and, therefore, processing larger datasets (e.g., using B-Scans) may be more appropriate. Application of SR and DL to massive datasets is computationally expensive because of tight memory constraints. Although patch-based SR (DL) \cite{papyan2015multi} reduce the computations by processing smaller data blocks, they ignore the inter-patch correlation and, consequently, the overall reconstructed signal could be erroneous. 
  
  Recently, convolutional sparse coding (CSC) \cite{rey2020variations} has attracted interest because it allows regularization to be performed directly on the global signal by imposing a banded convolutional structure to the dictionary. The dictionary is arranged as a sequence of convolution filters, each of which represents a single convolution with the given data. The decomposition generates more succinct data features for target recognition and/or clutter suppression. 
  
  To learn dictionaries with the aforementioned structure, convolutional dictionary learning (CDL) strategies have been proposed \cite{garcia2018convolutional}. In the most common formulation for the sparse coding via CSC or the dictionary update via CDL, an $\mathcal{l_1}$-$\mathcal{l_2}$ minimization is solved through different variations of the Alternate Direction Method of Multipliers (ADMM) algorithm \cite{boyd2011distributed}. While CSC has been applied to computer vision and pattern recognition, its application to GPR is more recent. In \cite{tivive2017gpr}, separation of target returns from clutter is formulated as a constrained optimization problem to estimate the low-rank and sparse components of the data. Whereas the low-rank portion contains a strong ground surface reflection and is extracted with a dictionary based on discrete cosine transform, the sparse part is associated with the targets and is estimated by CSC with a dictionary comprising convolutional Ricker wavelet filters.
  \color{black}  


\begin{figure}[t]
\centering
  \includegraphics[width=0.95\textwidth]{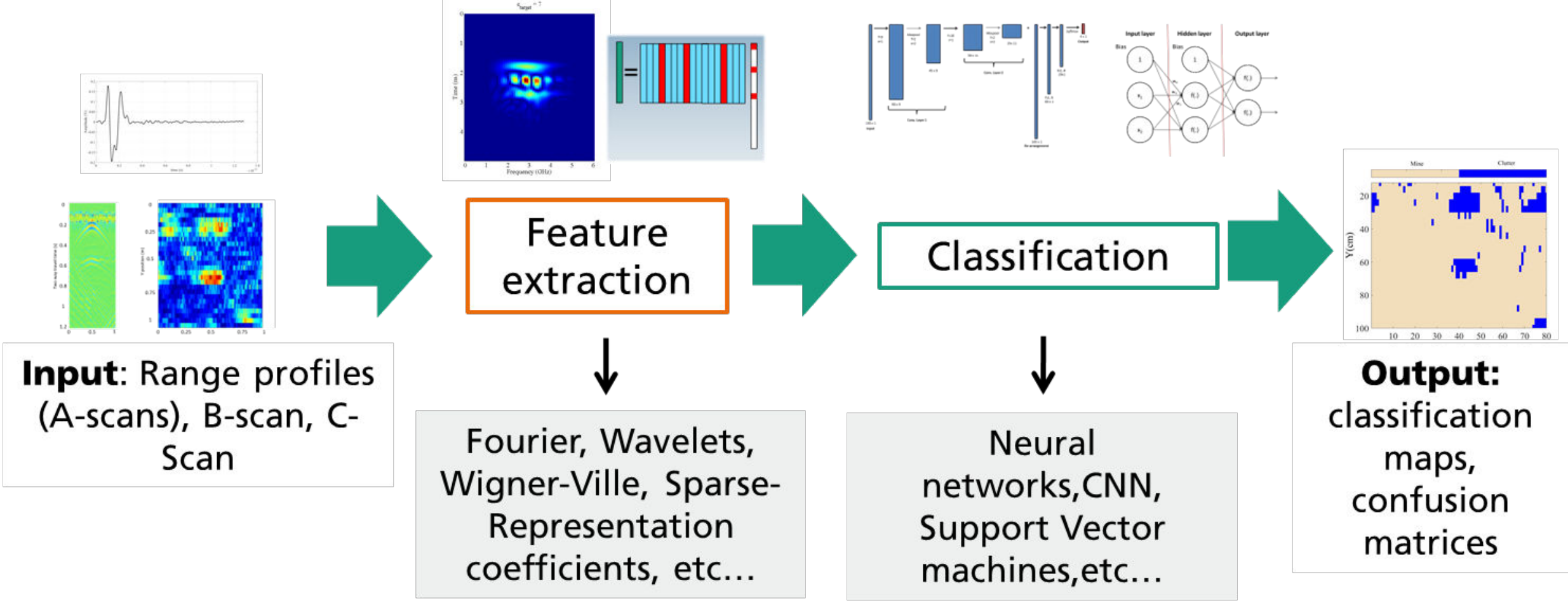}
  \caption{Simplified flow of modern GPR target classification. An appropriate data representation is necessary to extract features from the range profiles which are then fed to an advanced classifier to separate and identify targets from clutter.}
\label{fig:class_over}
\end{figure}

We now illustrate aforementioned techniques for GPR-aided landmine detection. For this purpose, in the next section, we briefly describe the deployed system and field campaign to obtain data.

\section{GPR Measurements for Landmine Recognition}
\label{sec:landmine_meas}
The detection of Anti-personnel landmines (APM) is appropriate not only from our representative application point-of-view but also its general utility for humanitarian purposes. APMs are a serious threat to civilian population and one of the worst kinds of global pollution nowadays. 
Broadly, APMs include blast mines, fragmentation mines or even improvised explosive devices (IED), the latter are not constructed by conventional military designs and are usually associated to insurgent guerrillas and commando forces. 
The UWB GPR is a promising non-invasive technology to tackle APM detection. Mine detection GPR usually operates in L-band ($1$-$2$ GHz) with UWB transmit signals that allow resolving small targets ($5$-$10$ cm diameter) at shallow depths (\texttildelow$15$-$30$ cm) \cite{yarovoy2002ultra, giovanneschi2013parametric}. 
The constituting material of many models of landmines is largely plastic and has a very weak response to radar signals because of its low dielectric contrast with respect to soil. 
In the following, we describe the specific GPR system and its field measurement campaign to obtain mine data for our processing application.

\subsection{Deployed System}
\label{subsec:sprs}
\begin{figure}[t]
	\centering
	\includegraphics[width=0.35\textwidth]{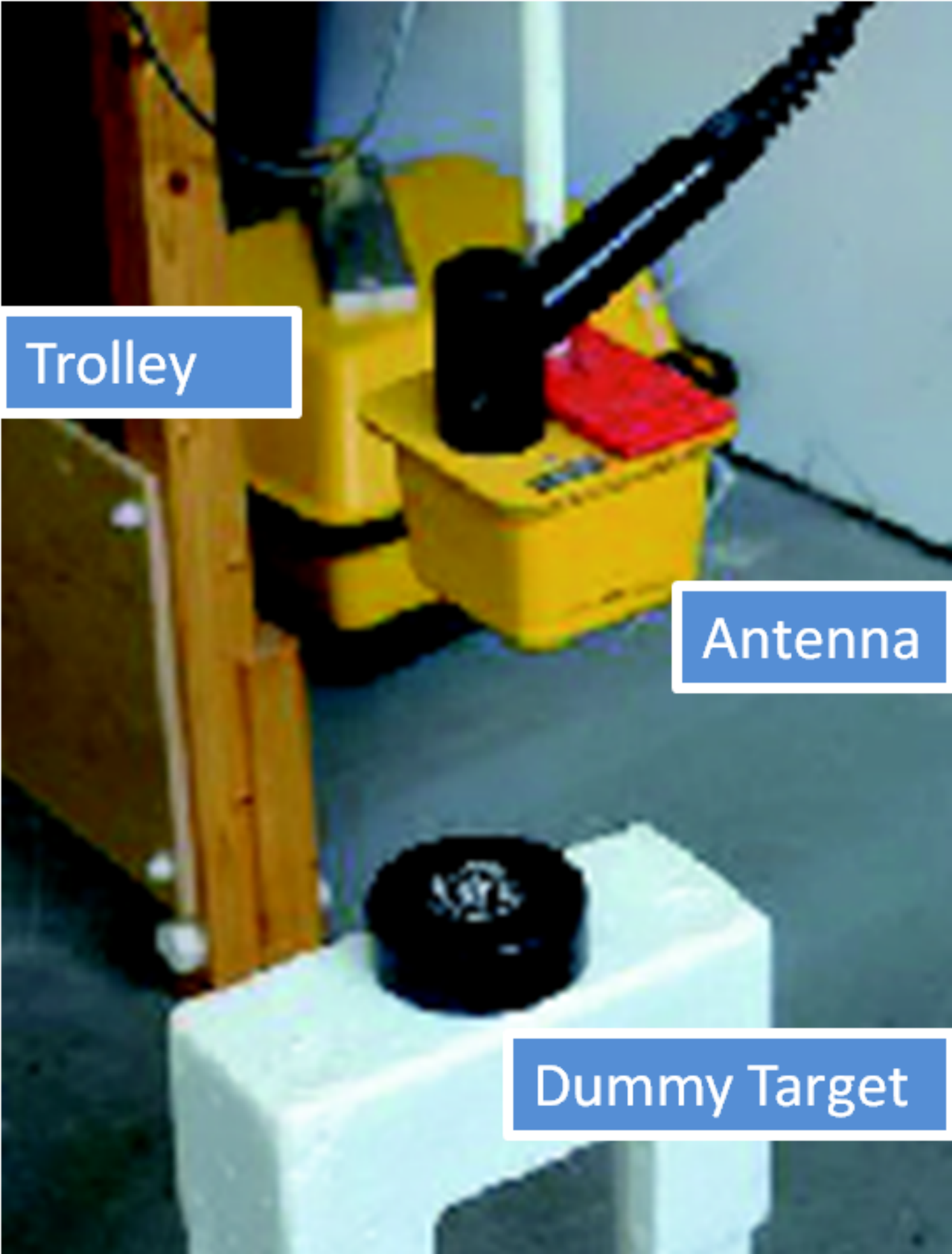}
  	\caption{The L-band GPR system is attached to a movable trolley platform. It is mounted along a rail system and scans the target from above.}
	\label{fig:GPR}
\end{figure}

We employ a commercially available Surface Penetrating Radar unit called \textit{SPRScan} manufactured by ERA Technology. 
The SPRScan is an L-band, impulse waveform, ultra-wideband (UWB) radar that is mounted on a movable trolley platform (see figure \ref{fig:GPR}). Pulsed GPRs are more effective in terms of offering penetration depth and wide bandwidth with respect to the standard Stepped-Frequency Continuous Wave (SFCW) systems. The former is also more robust to electronic interference and does not suffer from unequal balancing of antenna signals.

Table \ref{tbl:techparams} lists the most important operational parameters of the system. The radar uses a $8\times8$ cm dual bow-tie dipole antenna for both transmit (Tx) and receive (Rx) sealed in a metallic shielding filled with an internal absorber. The central frequency of the system ($f_c$) and its bandwidth ($\Delta f$) are $2$ GHz. 

The pulse repetition frequency (PRF) and the sampling of the receiver ADC is $1$ MHz. The scanning system has a resolution of $1$ cm towards the perpendicular broadside (or X direction) and $4$ cm towards the cross-beam (Y direction). In our field campaigns, the SPRScan system moves along the survey area over a rail system which allows accurate positioning of the sensor head in order to obtain the aforementioned resolution in X and Y (see also Section~\ref{sec:GPR}).

Our GPR system employs stroboscopic sampling to reach a pseudo sampling frequency of $f_s=1/T_s=40$ GHz (much above the Nyquist rate) to yield the discrete-time signal $y[n] = y(nT_s)$.
The receiver has the ability to acquire a maximum of 195 profiles per second, each one consisting of 512 time samples. Prior to the A/D conversion, the signal is averaged to improve the SNR. A time-varying gain correction can be applied to compensate for the soil attenuation and increase the overall dynamic range of the system.  The receiver averages 100 range profiles for each antenna position. 

\begin{table}[t]
\centering
    \caption{Technical characteristics of impulse GPR}
	\label{tbl:techparams}
	\begin{tabular}{ p{5.0cm} | p{4cm} }
		\hline
         \noalign{\vskip 1pt}    
         	Parameter & Value\\[1pt]
		\hline
		\hline
        \noalign{\vskip 1pt}    
		   	Operating frequency & 2 GHz\\[1pt]
		   	Pulse repetition frequency & 1 MHz\\[1pt]
		   	Pulse length & 0.5 ns\\[1pt]            
            Sampling time & 25 ps\\[1pt]
            Spatial sampling along the beam & 1 cm\\[1pt]
            Cross-beam resolution & 4 cm\\[1pt]
            Antenna height & 5-9 cm\\[1pt]
            Antenna configuration & Perpendicular broadside\\[1pt]
            Samples/A-scan & 512\\
		\hline
		\hline
        \noalign{\vskip 2pt}
	\end{tabular}
\end{table}

\subsection{Test Field}
\label{subsec:test_field_meas}
\begin{figure}[t]
\centering
  \includegraphics[scale=0.25]{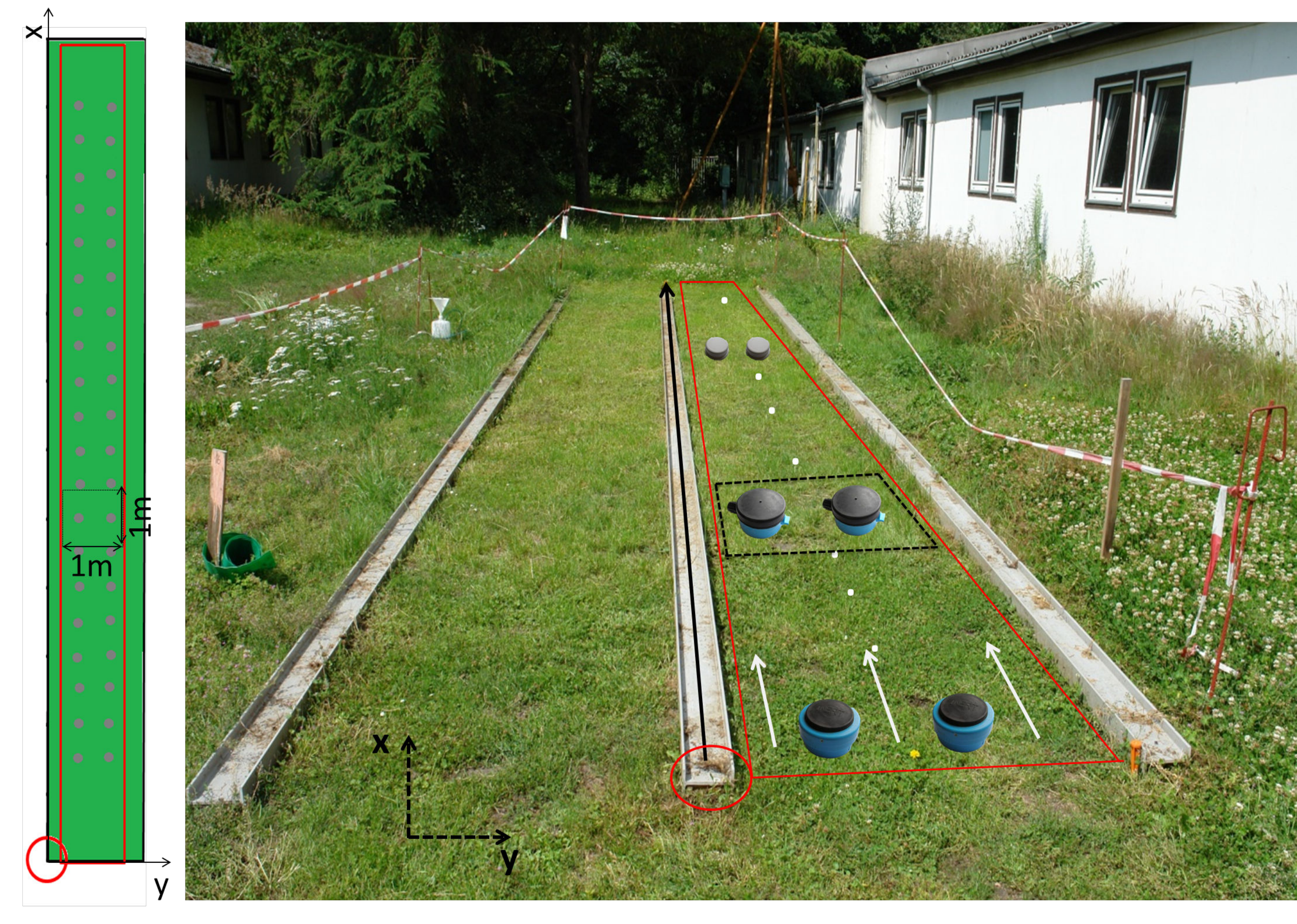}
  \caption{The LIAG test field in Hannover (right) along with its layout (left). The scan directions X and Y of the radar are indicated on the photograph and layout. The radar coverage region is indicated by solid red lines with a red circle showing the origin of the scan. The white arrows in the photograph indicate specific lanes scanned in the X direction that are separated in the Y direction by $4$ cm. In the layout, each gray dot represents the location of a buried test target. An individual survey area unit of $1$ m $\times$ $1$ m that contains 2 targets is also indicated on the layout (solid black lines) and the photograph (dotted black lines). The solid black arrow over the middle rail in the photograph is where the SPRScan was mounted.}
\label{fig:field}
\end{figure}
We use the measurement data from a 2013 field campaign at Leibniz Institute for Applied Geophysics (LIAG) in Hannover (Germany) \cite{gonzalez2013combined}. Fig.~\ref{fig:field} shows the test field, for detailed ground truth informations. The soil texture was sandy and highly inhomogeneous (due to the presence of material such as organic matter and stones), thereby leading to a high variability in the electrical parameters. The dielectric constant at three different locations of the testbed was measured with a Time Domain Reflectometer (TDR) to obtain an estimate of its mean value and variability. The average value oscillated between 4.6 and 10.1 with $15\%$ standard deviation and \textit{correlation length} \cite{gonzalez2013combined} of $20$ cm. These big variations in soil dielectric properties pose difficulties in mine detection.

During the field tests, the SPRScan system moved on two plastic rails with the scan resolution in the X and Y directions being $1$ and $4$ cm, respectively. The entire survey lane was divided in $1\times1$ m sections (see Fig.~\ref{fig:field}), each containing two targets in the center. The targets on the left and right sides of the lane were buried at approximately $10$ and $15$ cm depths, respectively.

\begin{figure}[t]
\centering
  \includegraphics[scale=0.5]{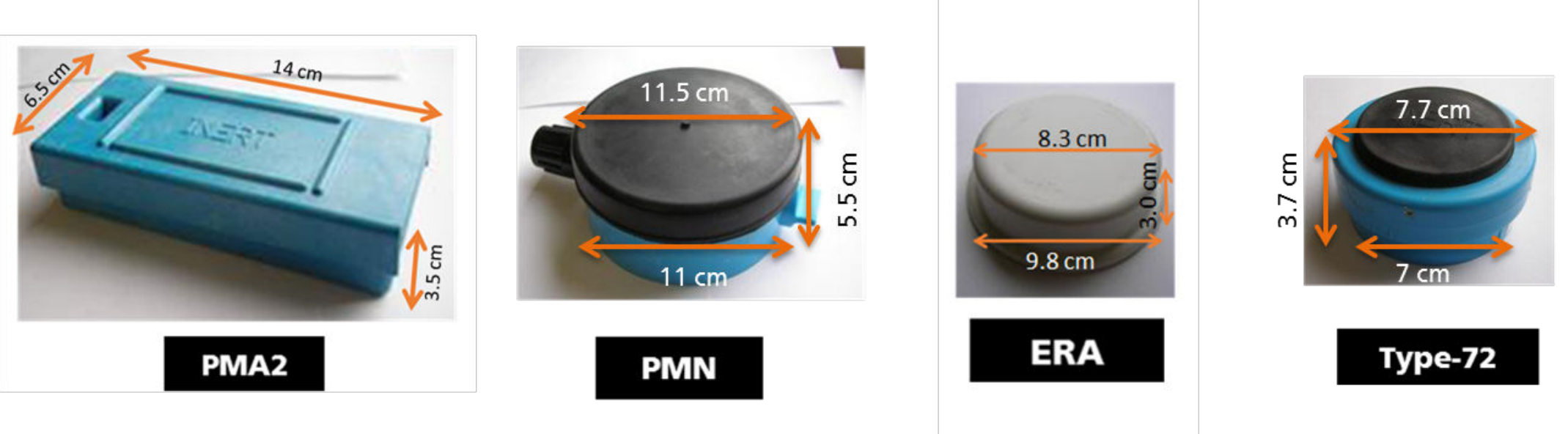}
  \caption{Details of the simulant landmines and the standard test target buried in the test field.}
\label{fig:mines}
\end{figure}

Our testbed contains standard test targets (STT) and simulant landmines (SIM) of different sizes and shapes. An STT is a surrogate target used for testing landmine detection equipment. It is intended to interact with the equipment in an identical manner as a real landmine does. An SIM has the representative characteristics of a specific landmine class although it is not a replica of any specific model. Here, we study three STTs (PMA2, PMN and Type-72) and one SIM (ERA). All of these test objects are buried at a depth of $10$-$15$ cm in the test field \cite{gonzalez2013accurate}. For classification purposes, we group PMN and PMA2 together as the largest targets while T72 mines are the smallest (Fig.~\ref{fig:mines}).
\begin{table}[t]
\centering
 	\caption{Training set classes}
	\label{tbl:tset}
	\begin{tabular}{ l | l }
		\hline
         \noalign{\vskip 1pt}
           	Target class & Number of elements\\
        \noalign{\vskip 1pt}
		\hline
        \hline
        \noalign{\vskip 1pt}
            Clutter & $463$  \\[1pt]
            PMN & $168$ \\[1pt]
            ERA & $167$ \\[1pt]
            T72 & $128$ \\[1pt]
		\hline
		\hline
        \noalign{\vskip 1pt}
	\end{tabular}
\end{table}


\section{Sparse Representation 
Techniques}
\label{sec:SR}
We now consider the SR techniques which are aimed at the solution of a under-determined linear system. The regularization process comprises minimizing a cost function which enforces the sparsity of the vector $\mathbf{x}$ (i.e. the number of its non-zero elements):
\begin{align}
\label{eq:CS}
\hat{\mathbf{x}}=\underset{\mathbf{x}}{\mathrm{argmin}}\phantom{1}||\mathbf{x}||_0  \phantom{1} \text{subject to}\phantom{1} \mathbf{y}=\mathbf{D}\mathbf{x},
\end{align}
where $\mathbf{y} \in \mathbb{C}^{M \times 1}$ represents the measurements of the observed scene, arranged in a $M$ dimensional column vector, $\mathbf{D} \in \mathbb{C}^{M \times K}$ is a $M\times K$ matrix called  \textit{dictionary}, and $\mathbf{x} \in \mathbb{C}^{1 \times K}$ is a coefficient vector which represents the signal information according to the model specified in $\mathbf{D}$.  The dictionary $\mathbf{D}$ contains a collection of realizations of the signal of interest in a certain domain. For image processing and classification purposes, every realization (atom) could represent (for example) a particular patch of an image or a certain signal class associated to a particular target. The vector $\mathbf{x}$ represents the information to be recovered and $||\mathbf{x}||_0$ is its $\ell_0$-norm.

If the observed scene is sparse, then it has been shown that \eqref{eq:CS} has a unique solution i.e. one can find a matrix $\mathbf{D}$ which gives a perfect reconstruction of the measurements with probability close to one \cite{donoho2006compressed}. The number of non-zero elements of the vector $\mathbf{\mathbf{x}}$ is called the \textit{support} of the signal $\mathbf{\mathbf{y}}$; if $\mathbf{\mathbf{y}}$ is mapped on the vector $\mathbf{x}$ using $S$ non-zero elements then it is said to be \textit{S-sparse}. The problem in (\ref{eq:CS}) is NP-hard. However, techniques based on greedy algorithms and convex optimization are shown to be efficient in relaxing this problem.

Assume that matrix $\mathbf{D} \in \mathbb{C}^{M \times K}$ has $M < K$ and that $\mathbf{y}$, the measured signal, is \textit{sparse} or \textit{compressible}. A signal is sparse in a certain domain (for example: space, time, frequency, etc.) if there exists a basis (or frame) in which this signal is represented using a very limited number of elements of the aforementioned basis. A signal is compressible when it can be well-approximated by sparse signals; this is relevant for noisy real-world signals. We define the compressibility of order $p$ of a signal $\mathbf{y}$ as $\sigma_S(\mathbf{y})_p$ by calculating the error in approximation, i.e.,
\begin{align}
\label{eq:compressibility}
\sigma_S(\mathbf{y})_p  = \underset{\hat{\mathbf{y}} \in \Sigma_S }{\mathrm{min}}\phantom{1} ||\mathbf{y} - \hat{\mathbf{y}}||_p, 
\end{align}
where ${\Sigma_S}$ is a set of S-sparse signals, i.e. with ${||x||_0 \leq S}$.

Popular greedy approaches to solve \eqref{eq:CS} such as matching pursuit (MP), orthogonal matching pursuit (OMP), block orthogonal matching pursuit (BOMP), etc. \cite{tropp2007signal}) approximate the $\ell_0$-norm solution in \eqref{eq:CS} using iterative strategies. These methods are computationally fast but may not lead to a global optimum solution. State-of-the-art algorithms for sparse representation such as basis pursuit (BP), basis pursuit de-noising (BPDN) \cite{van2008probing}, least absolute shrinkage and selection operator (LASSO) \cite{tibshirani1996regression} and least-angle regression (LARS) \cite{efron2004least} aim to approximate (\ref{eq:CS}) using less restrictive constraints ($\ell_1$ and $\ell_2$) which also assure a sparse solution for $\mathbf{x}$ \cite{candes2006stable}. These methods based on convex optimization are more robust to noise and, therefore, may lead to a better reconstruction of $\mathbf{y}$ even although they are computationally more demanding. 

In the following, we discuss the following SR techniques: 
OMP along with its batch version (batch-OMP)\cite{rubinstein2008efficient}. We also include convex optimization methods (such as BPDN, LASSO and LARS) because they are used for the dictionary update step of some of the DL algorithms mentioned later in Section~\ref{sec:DL}.

\subsection{OMP}
Greedy strategies for SR are iterative approaches which searches for the best local optimal solution for each iteration (i.e. the \textit{greediest} one) hoping that it coincides to the global best one. They can approximate the solution of the problem in \eqref{eq:CS}. These algorithms choose only the most appropriate elements in $\mathbf{x}$ according to the input constraint (which is the expected sparsity or the reconstruction error) to reconstruct the measurement vector $\mathbf{y}$. 
The OMP problem has a dual formulation:
\begin{align}
\label{eq:OMP1}
\hat{\mathbf{x}}=\underset{\mathbf{x}}{\mathrm{argmin}}\phantom{1} ||\mathbf{y-Dx}||_2 \phantom{1} \text{subject to}\phantom{1} ||\mathbf{x}||_0 \leq S.
\end{align}
Alternatively,
\begin{align}
\label{eq:OMP2}
\hat{\mathbf{x}}=\underset{\mathbf{x}}{\mathrm{argmin}}\phantom{1}||\mathbf{x}||_0  \phantom{1} \text{subject to}\phantom{1} ||\mathbf{y-Dx}||_2 \leq \delta.
\end{align}
The constraint in (\ref{eq:OMP1}) is based on the sparsity $S$ while the one in (\ref{eq:OMP2}) employs the $\ell_2$-norm of the residual by the parameter $\delta$. The OMP is an iterative process, at each iteration it search for an atom $\mathbf{d_i}$ which gives the best projection onto the residual signal $\mathbf{y-Dx}$ and obtains the value of the corresponding coefficient $\mathbf{x_i}$ by least squares. Either $S$ or $\delta$ is used as a stopping criterion.
The least square problem in OMP is computationally demanding and usually computed using a Cholesky factorization or a QR decomposition \cite{rubinstein2008efficient}. To speed up computations we propose a faster variant of OMP (see Algorithm~\ref{alg:OMP2}). The difference in this implementation is that the new found atom is orthogonalized against all the previously found one $\mathbf{d_j}$ with $j = 1, \cdots, i-1$ via Gram–Schmidt process (see step 10) \cite{cheney2011linear} before the coefficient update. Having an orthonormal basis among the new found atoms is beneficial for the coefficient update  where, instead of solving a least square problem, we can directly obtain the updated coefficients $\mathbf{x}$ by projecting them over the residual $\mathbf{r_i}$ (see step 13). After the stopping criterion is met, we adjust the coefficient vector obtained at the last iteration ($\mathbf{x_i}$) solving just one least square problem without orthogonalizing before, this makes possible that the coefficients $\mathbf{x}$ have the right coefficient values with respect to $\mathbf{y}$. We define the set of new atoms to be updated during the iterations as $\mathcal{D}$.
\begin{algorithm}
\LinesNumbered
  \caption{Fast OMP}
  \label{alg:OMP2}
  \SetAlgoLined
  \SetKwFor{Loop}{Loop}{until $i=S$ or $\mathbf{||r_i-Dx||_2} \geq \delta$}{EndLoop}
{\scriptsize
   \KwIn{ Measurement vector ($\mathbf{y} \in \mathbb{C}^{mx1}$), Dictionary ($\mathbf{D} \in \mathbb{C}^{mxn} $), Sparsity number ($S$) or Residual error threshold ($\delta$) }
   \KwOut{ Sparse coefficients vector ($\mathbf{x} \in \mathbb{C}^{1xn}$)}
   \BlankLine
   Initialize iteration count $i = 1$ \par
   Initialize the residual as the measurement vector: $\mathbf{r_i}=\mathbf{y}$ \par
   Initialize the vector of sparse coefficients: $\mathbf{x}=\mathbf{\underline{0}} \in \mathbb{C}^{1xn}$ \par
   Initialize the matrix of new found atoms: $\mathcal{D}_{0}=\mathcal{\varnothing}$ \par
   \BlankLine
   \Loop{}{
        Find the atom $\mathbf{d_i}$ which gives the maximum dot product with $\mathbf{r}$: $\mathbf{di} = \underset{\mathbf{d_i}}{\mathrm{argmax}} \left\langle\mathbf{r_i}, \mathbf{d_i} \right\rangle$ \par
         Normalize the new found atom by the norm-2 : $\mathbf{d_i}=\frac{\mathbf{d_i}}{||\mathbf{d_i}||_2}$ \par        
        Include $\mathbf{d_i}$ in the set of new found atoms: $\mathcal{D}_{i}=\mathcal{D}_{i-1} \cup \mathbf{d_i}$ \par
        \textbf{for} $\phantom{1} j = 1, \cdots, i-1$ \par
        $\phantom{1} \phantom{1} \phantom{1}$ Orthogonalize $\mathbf{d_i}$ against all previously found $\mathbf{d_j}$:  $\phantom{1} \mathbf{d_i} = \mathbf{d_i} -  \left\langle\mathbf{d_j},\mathbf{d_i}\right\rangle \mathbf{d_j}$   \par
         \textbf{end for} \par
        \textbf{for} $\phantom{1} j = 1, \cdots, i$ \par
        $\phantom{1} \phantom{1} \phantom{1}$ Coefficients update $\mathbf{x_i} =\left\langle \mathbf{x_j},\mathbf{r_j} \right\rangle$ \par
         \textbf{end for} \par    
   }
         UPDATE $x_i$ by solving a least square problem: $\mathbf{x_i} = \underset{\mathbf{x}}{\mathrm{argmin}} ||\mathbf{y}-\mathcal{D}_{i}\mathbf{x}||_2$ \par
                  Update residual: $\mathbf{r_i} = \mathbf{r_{i-1}} - \mathcal{D}_{i}\mathbf{x_i}$ \par         
}
\end{algorithm}
\subsection{Batch-OMP} 
Batch-OMP is a variant of the classical OMP and it is used when a large number of signals must be represented over the same dictionary. The intuition behind this method is that it is not required to compute the residual $\mathbf{r}$ at each iteration for the atom selection step, only $\mathbf{D}^T\mathbf{r}$ is required. The atom selection step can be rewritten in order to exploit its dependence only on the dictionary at the iteration $i$ (called $\mathbf{D_i}$) and the measurement vector $\mathbf{y}$ without calculating $\mathbf{D}^T\mathbf{r}$.

Calling $\boldsymbol{\alpha} = \mathbf{D}^T\mathbf{r}$, $\boldsymbol{\alpha}^0 = \mathbf{D}^T\mathbf{y}$ and $\mathbf{G}=\mathbf{D}^T\mathbf{D}$ one can write;
\begin{align}
\label{eq:BatchOMP}
\boldsymbol{\alpha}&=\mathbf{D}^T(\mathbf{y}-\mathbf{D}_I(\mathbf{D}_I)^+\mathbf{y})\\
&=\boldsymbol{\alpha}^0-\mathbf{G}_I(\mathbf{D}_I)^+\mathbf{y}\\
&=\boldsymbol{\alpha}^0-\mathbf{G}_I(\mathbf{D}_I^T\mathbf{D}_I)^{-1}\mathbf{D}_I^T \mathbf{y}\\
&=\boldsymbol{\alpha}^0-\mathbf{G}_I(\mathbf{G}_{I,I})^{-1}\boldsymbol{\alpha}^0_I.
\end{align}
Given the pre-computed $\boldsymbol{\alpha}^0$ and $\mathbf{G}$, one can compute $\boldsymbol{\alpha}$ without explicitly computing $\mathbf{r}$.
However, if the residual is not computed, it is impossible to set a stopping criterion based on the error. In \cite{rubinstein2008efficient} an extention of batch-OMP where the residual $\ell_2$ norm $||\mathbf{r}||_2^2$ at the iteration $i$ is calculated by an incremental formula; this permits the use of the residual as a stopping criterion without explicitly computing it; see \cite{rubinstein2008efficient} for the implementation details. 

\subsection{Basis Pursuit Denoising, LASSO, and Penalized Least Square}
A convex optimization problem deals with miminizing convex functions over a convex set; the convexity makes the minimization easier since the local minimum of the fuction is also the global one.
Convex optimization methods for SR, contrary to greedy approaches, aim to minimize the $\ell_1$-norm of the coefficient vector $x$. 
\begin{align}
\label{eq:BP}
\hat{\mathbf{x}}=\underset{x}{\mathrm{argmin}}\phantom{1} ||\mathbf{x}||_1   \phantom{1} \text{subject to} \phantom{1} \mathbf{D}\mathbf{x}=\mathbf{y}.  
\end{align}
\normalsize
The objective function $||.||_1$ is in fact a convex and tractable, whereas $||.||_0$ is non-convex and generally very difficult to solve. The problem in \eqref{eq:BP} is known as basis pursuit (BP). It can be solved efficiently using linear programming techniques and it is demonstrated \cite{donoho2006compressed} that, under certain conditions, it still leads to a sparse approximation of the underdetermined system solution in \eqref{eq:CS}.

The BPDN accounts for the noise in the data through an additional approximation: 
\begin{align}
\label{eq:BPDN}
\hat{\mathbf{x}}=\underset{x}{\mathrm{argmin}}\phantom{1} ||\mathbf{x}||_1   \phantom{1} \text{subject to} \phantom{1} ||\mathbf{y}-\mathbf{D}\mathbf{x}||_2 \phantom{1}  \leq \delta.
\end{align}
This does not exactly solves \eqref{eq:BP} but makes an approximation based on a positive parameter $\delta$ whose amplitude (between 0 and 1 if the data is normalized) corresponds to the noise level. This parameter plays the same role as $\delta$ in OMP (see Algorithm~\ref{alg:OMP2}). 

Through a different derivation, LASSO  \cite{tibshirani1996regression} solves \eqref{eq:BPDN} as
\begin{align}
\label{eq:LASSO}
\hat{\mathbf{x}}=\underset{x}{\mathrm{argmin}}\phantom{1} ||\mathbf{y}-\mathbf{D}\mathbf{x}||_2    \phantom{1} \text{subject to} \phantom{1} ||\mathbf{x}||_1 \phantom{1}  \leq \tau.
\end{align}
\normalsize
Another technique is penalized least squares which models the problem as
\begin{align}
\label{eq:PLS}
\hat{\mathbf{x}}=\underset{x}{\mathrm{argmin}}\phantom{1} ||\mathbf{y}-\mathbf{D}\mathbf{x}||_2    \phantom{1} + \lambda||\mathbf{x}||_1,
\end{align}
where $\lambda$ is a regularization parameter. For an appropriate choice of $\delta$, $\tau$ and $\lambda$ the solutions of these three approaches coincide.

One of the most popular algorithm for solving the problems in \eqref{eq:BPDN} and \eqref{eq:LASSO} was proposed from Van Der Berg and Friedlander and it is based on the relation between LASSO and BPDN, defined by the Pareto curve. The Pareto curve indicates the optimal trade off between $l_2$-norm of the residual ($||\mathbf{y}-\mathbf{D}\mathbf{x}||_2$) and the one-norm of the solution $\mathbf{x}$. The solution of a BPDN problem using this approach consists on solving a sequence of LASSO problems using spectral projected gradient and use the Newton method applied to the Pareto curve to observe how much the solution of \eqref{eq:LASSO} get close to \eqref{eq:BPDN}; the process stops when a satisfactory threshold is met \cite{van2008probing}.

\subsection{LARS}
Homotopy approaches such as LARS (Least Angle Regression) \cite{efron2004least} solve the BPDN problem by repeatedly solving \eqref{eq:PLS} for all possible values of $\lambda$. LARS is applied for both variable and coefficient selection 
and has been proposed for $\ell_1$ minimization. It is convenient for solving the penalized least square problem in \eqref{eq:PLS} because of its speed (comparable to forward selection algorithms) and producing a full piece-wise linear solution path (i.e. for every $\lambda$). Moreover, LASSO can be derived by LARS. In fact, LASSO is viewed as a variation of a ridge-regression problem 
with the constraint of some coefficients set to zero. Algorithm~\ref{alg:LARS} summarises this LARS-LASSO technique \cite{efron2004least}. 
\begin{algorithm}
\LinesNumbered
  \caption{Least-Angle Regression (LARS)}
  \label{alg:LARS}
  \SetAlgoLined
  \SetKwFor{Loop}{Loop}{until all atoms in $\mathbf{D}$ are considered }{EndLoop}
{\scriptsize
   \KwIn{ Measurement vector ($\mathbf{y} \in \mathbb{C}^{mx1}$), Dictionary ($\mathbf{D} \in \mathbb{C}^{mxn} $), Regularization parameter ($\lambda$) }
   \KwOut{ Sparse coefficients vector ($\mathbf{x} \in \mathbb{C}^{1xn}$)}
   \BlankLine
   Initialize all coefficients $\mathbf{x}_j$ with ($j=1...K$) equal to zero, set the initial residual $\mathbf{r}$ as $\mathbf{y}$\par
   \BlankLine
   \Loop{}{
        Find the atom $\mathbf{d}_j$ which has the highest correlation with $\mathbf{r}$ \par
        Increase the coeficient $\mathbf{x}_j$ in the direction of the sign of its correlation with $\mathbf{y}$, calculate the residual $\mathbf{r}=||\mathbf{y}-\mathbf{D}\mathbf{x}||_2^2$ for each increment and stop when some other atom $\mathbf{d}_k$ has as as much correlation with $\mathbf{r}$ as $\mathbf{d}_j$ has. \par
        Increase $\mathbf{x}_j,\mathbf{x}_k$ in the direction defined by their joint least-square coefficient of the current residual on $\mathbf{d}_j,\mathbf{d}_k$ , until some other $\mathbf{d}_m$ has as much correlation with the residual $\mathbf{r}$       
}
}
\end{algorithm}

\section{Dictionary Learning Framework and Algorithms}
\label{sec:DL}
We now describe the theory and algorithms of some important DL methods that are useful for GPR target recognition \cite{giovanneschi2017online,giovanneschi2019dictionary}. We consider both batch- \cite{rubinstein2008efficient} and online-learning \cite{mairal2009online,naderahmadian2016correlation} approaches. We then analyze in detail a novel online dictionary learning approach proposed in our prior work \cite{giovanneschi2019dictionary}. This is an improvement over the state-of-the-art online dictionary learning approaches.

The dictionary $\mathbf{D}$ may be learned from the data it is going to represent. The DL techniques aim to create adapted dictionaries which provide the sparsest reconstruction for given training sets, i.e., a representation with a minimum number of constituting atoms. DL methods are critical building blocks in many applications such as deep learning, image denoising, and super-resolution; see \cite{elad2010prologue}.


The first step in DL consists of building a training database $\mathbf{Y} = \begin{bmatrix} \mathbf{y}_1 & \cdots & \mathbf{y}_L \end{bmatrix} \in \mathbb{R}^{MxL}$ 
of $L$ vectors, each with $M$ elements. We assume that every vector $\mathbf{y}_i$ is generated by a linear combination of the $K$ atoms of a certain dictionary matrix $\mathbf{D} \in \mathbb{R}^{MxK}$ and an associated sparse vector of coefficients $\mathbf{x_i}$, having $\mathbf{X} = \begin{bmatrix} \mathbf{x}_1 & \cdots & \mathbf{x}_L \end{bmatrix} \in \mathbb{R}^{KxL}$.
The core problem of DL is to find the dictionary $\mathbf{D}$ which give the set of sparsest solution $\mathbf{X}$, we have then:
\begin{flalign}
\label{eq:DL1}
\mathbf{Y}_{(MxL)} = \mathbf{D}_{(MxK)}\phantom{1}\mathbf{X}_{(KxL)},  
\end{flalign}
where the dimension $L$ is usually larger than the number of the dictionary elements ($L > K$).
The DL problem has a dual formulation, whether it constraints sparsity (\ref{eq:DL_S})
\begin{flalign}
\label{eq:DL_S}
	& \mathbf{\hat{D},\mathbf{\hat{X}}}=\underset{\mathbf{D},\mathbf{X}}{\text{argmin}}\phantom{1}\left\Vert \mathbf{Y}-\mathbf{D}\mathbf{X}\right\Vert _{F}\nonumber\\
	& \text{subject to}\phantom{1} \left\Vert\mathbf{x}_i\right\Vert_0 \le S,\: 1\le i \le L,
\end{flalign}
or the model deviation (\ref{eq:DL_eps})
\begin{flalign}
\label{eq:DL_eps}
	& \mathbf{\hat{D},\mathbf{\hat{X}}}=\underset{\mathbf{D},\mathbf{X}}{\text{argmin}}\phantom{1}\left\Vert \mathbf{Y}-\mathbf{D}\mathbf{X}\right\Vert _{F}\nonumber\\
	& \text{subject to}\phantom{1} \left\Vert \mathbf{y}_i - \mathbf{D}\mathbf{x}_i \right\Vert_2 \le \epsilon,\: 1\le i \le L.
\end{flalign}
Where $S$ is the sparsity number, i.e. the desired number of the non-zero elements in the decomposed vector and $\epsilon$ is the residual error.

Since both $\mathbf{D}$ and $\mathbf{X}$ are unknown, a common approach is to use alternating minimization in which we start with an initial guess of $\mathbf{D}$ and then obtain the solution iteratively by alternating between two stages: \textit{sparse representation} and \textit{dictionary update} \cite{elad2006image} - as follows:

\textit{1) Sparse representation}: Obtain $\mathbf{X}_{(t)}$ for each $\mathbf{y}_i$ as:
\begin{align}
\label{eq:spcoding1}
	\mathbf{{X}}_{(t)} &= \underset{\mathbf{X}}{\text{argmin}}\phantom{1}\left\Vert \mathbf{Y}-\mathbf{D}_{(t-1)}\mathbf{X}\right\Vert _{F}\nonumber\\
	& \text{subject to}\phantom{1} \left\Vert\mathbf{x}_{i_{(t-1)}}\right\Vert_p \le S,\: 1\le i \le L,
\end{align}
where $\mathbf{X}_{(t)}$ is the SR in $t^{th}$ iteration. This can be solved using greedy algorithms such as orthogonal matching pursuit (OMP) ($p=0$) or convex relaxation methods like basis pursuit denoising (BPDN) ($p=1$).

\textit{2) Dictionary Update}: Given $\mathbf{X}_{(t)}$, update $\mathbf{D}_{(t)}$ such that
\begin{align}
\label{eq:dl update1}
	\mathbf{D}_{(t)} &= \underset{\mathbf{D}\in\mathcal{D}}{\text{argmin}}\phantom{1}\left\Vert \mathbf{Y}-\mathbf{D}\mathbf{X}_{(t)}\right\Vert _{F},
\end{align}
where $\mathcal{D}$ is a set of all dictionaries with unit column-norms, $\left\Vert\mathbf{d}_j\right\Vert_2 = 1$ for $1\le j\le K$. This subproblem is solved by methods such as singular value decomposition or gradient descent \cite{aharon2006k,mairal2009online}. 

Conventional DL techniques deal with the entire training set in each iteration, making the learning process slow for high dimensional datasets; this approach is called batch Dictionary Learning. Online-DL techniques deal with the training set by considering one element at a time or in mini batches, making the processing of learning faster than batch-DL and adaptive to variations. For batch-DL, we focus on the popular K-times Singular Value Decomposition (K-SVD) \cite{rubinstein2008efficient} and the more recent low-rank shared DL (LRSDL) \cite{vu2017fast}. The latter is capable of generating a class-discriminative dictionary. Among online-DL techniques, we analyze the state-of-the-art Online Dictionary Learning (ODL) \cite{mairal2009online}, recently proposed Correlation-Based Weighted Least Square Update (CBWLSU) \cite{mairal2009online,naderahmadian2016correlation} and our novel Online-DL strategy named Drop-Off Mini-batch Online Dictionary Learning (DOMINODL) \cite{giovanneschi2019dictionary,fabio2020online}.

\subsection{K-SVD}
K-SVD is a batch-DL algorithm which iteratively alternates a sparse decomposition and a dictionary update step solving a minimization over the number of non-zero elements in the set of representation vectors $\mathbf{X}$, and one over $\mathbf{D}$ for updating the dictionary. For the sparse coding step at the iteration $t$, K-SVD solves OMP for each element $\mathbf{y_i}$ in the training set $\mathbf{Y}$: 
\begin{flalign}
\label{eq:omp}
& {\mathbf{x}}_i=\underset{\mathbf{x}_i}{\text{argmin}}\phantom{1}\left\Vert\mathbf{x}_i\right\Vert_0\nonumber\\
	& \text{subject to}\phantom{1} \left\Vert \mathbf{y}_i-\mathbf{D_{(t-1)}}\mathbf{x}_i\right\Vert^2_{2} \le \delta,\:\forall 1\le i \le L,
\end{flalign}
\normalsize
where $i$ is the index which represent the training set elements, $\mathbf{D_{(t-1)}}$ is the dictionary computed at the previous iteration ($t-1$) and $\delta$ is the maximum residual error.

K-SVD shares the same strategy for the sparse decomposition step with another batch-DL method called Method of Optimal Directions (MOD) \cite{engan1999method} but their respective dictionary update rules are different. 
Let $K$ be an input parameter of K-SVD which indicates the number of columns of the learned dictionary. For the dictionary update step, K-SVD solves the global minimization problem in (\ref{eq:dl update1}) via $K$ sequential minimization
problems, wherein every column $\mathbf{d}_k$ of $\mathbf{D}$ and its corresponding row of coefficients $\mathbf{X}_{\text{row},k}$ of $\mathbf{X}$ are updated.  
Let's assume that we are at the iteration $t$ and we already peformed the sparse decomposition step using the dictionary which has been obtained in the iteration $t-1$. For each $k$th dictionary atom, we calculate the error term $\mathbf{Y}_{r}=\mathbf{Y}- \sum\limits_{l\neq k}\mathbf{d}_{l_{t-1}}\mathbf{X}_{\text{row},l_{(t-1)}}$ and extract a subset $\mathcal{Y}_k$ of it which comprises the elements of $\mathbf{Y_r}$ which use the selected atom ($k$). Then we use SVD to find the closest rank-1 approximation of $\mathcal{Y}_k$ to obtain $\mathbf{d}_k$  subjected to the constraint $\Vert\mathbf{d}_{k_{(t)}}\Vert_2 = 1$. K-SVD global optimization will terminate after a series of sparse decomposition plus dictionary update steps depending on the changes on $\left\Vert\mathbf{Y-DX_{(t)}}\right\Vert^2_{F}$.  

%

In this chapter, a particular K-SVD implementation was used \cite{rubinstein2008efficient}. This version employs a faster approximation for the SVD step in the dictionary update and uses batch-OMP for the sparse decomposition step, making it more feasible to deal with large sets of signals. 
The performance of K-SVD can be improved  if the learning process enforces constraints such as hierarchical tree sparsity \cite{varshney2008sparse}, structured group sparsity (StructDL) \cite{suo2014structured}, Fisher discrimination (FDDL) \cite{yang2014sparse}, and low-rank-and-Fisher (D$^2$L$^2$R$^2$) \cite{li2014learning}. 
\subsection{LRSDL}
LRSDL \cite{vu2017fast} is one of the latest evolution of (batch) discriminative DL algorithms; we tested it on our classification approach due to its promising capabilities for class recognition. 
Here, we provide the basic idea of discriminative-DL and briefly describe the theoretical background on which LRSDL is based of.

Discriminative DL algorithms, as D-KSVD \cite{zhang2010discriminative} and LC-KSVD \cite{jiang2013label}, employ a learning strategy which promotes the generation of a dictionary $\mathbf{D}$ which is separated in blocks of atoms associated to different classes as $\mathbf{D} = [\mathbf{D}_1, \cdots, \mathbf{D}_C] \in \mathbb{R}^{M\times K}$ where $C$ is the number of classes present in the training set $\mathbf{Y}$. The resultant coefficient matrix $\mathbf{X}$ is close to be sparse with all non-zeros being one while satisfying a block diagonal structure.

However, objects belonging to different classes often have common features, therefore the assumption of non-overlapping subspaces done by such algorithms is often unrealistic in practice. Techniques such as DL with structured incoherence and shared features (DLSI) \cite{ramirez2010classification}, separating the commonality and the particularity (COPAR) \cite{kong2012dictionary} and convolutional sparse DL (CSDL) \cite{gao2014learning} exploit common patterns among different classes even though different objects possess distinct class-specific features. These methods produce am additional constituent $\mathbf{D_0}$ which is shared among all classes so that $\mathbf{D} = [\mathbf{D}_1, \cdots, \mathbf{D}_C, \mathbf{D}_0] \in \mathbb{R}^{M\times K}$. The drawback of these strategies is that the shared dictionary may also contain class-discriminative features.

To avoid this problem, LRSDL \cite{vu2017fast} requires that the shared dictionary must have a low-rank structure and that its sparse coefficients have to be almost similar. LRSDL learns both $\mathbf{D}$ and $\mathbf{X}$ by solving a minimization problem with a  cost function which is closely related to the one of another DL algorithm called Fisher Discriminative Dictionary Learning (FDDL) \cite{yang2011fisher}.
The LRSDL dictionary update step employs ADMM \cite{boyd2011distributed} and Fast Iterative Shrinkage-Thresholding Algorithm (FISTA) \cite{beck2009fast} for the sparse decomposition step.
Once the data is sparsely represented with such dictionaries, a sparse-representation-based classifier (SRC) \cite{giovanneschi2015preliminary} is used to predict the class of new data. 

\subsection{ODL}
The ODL is an interesting alternative for inferring a dictionary from large training sets or ones which change over time \cite{mairal2009online}, like K-SVD this algorithm also updates the entire dictionary sequentially, but draws one element of training data at a time for the dictionary update. ODL assumes that the training set $\mathbf{Y}$ is composed of i.i.d. samples of a distribution $\mathbf{p(x)}$. In many practical applications, this condition is usually not fulfilled and $\mathbf{Y}$ is obtained by a collection of measurements that has been randomly permuted and drawn consecutively for each iteration (we follow a similar procedure in Section~\ref{sec:parameval}).

The input parameters of ODL are: the initial dictionary 
$\mathbf{D}_{{0}} \in \mathbb{C}^{{M \times K}}$, the regularization parameter $\lambda$, the dimension of the learned dictionary $K$, and the number of iterations $T$ (which will also correspond to the number of training set elements that will be used for learning $\mathbf{D}$). The first step at the iteration $t$, with with $t=1...T$, is to draw an example of the training set $\mathbf{y}_{t}$ from $\mathbf{Y}$. The sparse decomposition step is done using the dictionary obtained at the previous operation $\mathbf{D}_{(t-1)}$ via a Cholesky-based implementation of the LARS-Lasso algorithm which solves a $\ell_1$-regularized least-squares problem as indicated in \ref{LARSODL}. In the dictionary update we consider all the training set elements analyzed so far: $\mathbf{y}_i \phantom{1} \text{with} \phantom{1} i = 1 ... t$.
\par\noindent\small
\begin{align}
\label{LARSODL}
\hat{\mathbf{x}}_{(t)} = \underset{\mathbf{x} \in \mathbb{C}^n }{\mathrm{argmin}}\phantom{1} \frac{1}{2}||\mathbf{y}_t - \mathbf{D_{(t-1)}}\mathbf{x}||^2_2 + \lambda||\mathbf{x}||_1,
\end{align}
\normalsize
The next step is the dictionary update, this step requires the input dictionary:
\begin{align}
{\mathbf{D}=[\mathbf{d}_1,\mathbf{d}_2,\cdots,\mathbf{d}_k]}  \in {\mathbb{R}^{M \times K}}
\end{align}
\normalsize
and two matrices:
\begin{align}
&{\mathbf{A}=[\mathbf{a}_1,\mathbf{a}_2,\cdots,\mathbf{a}_k]}   = \sum_{i=1}^{t} \mathbf{x}_{i}\mathbf{x}_{i}^t \in  {\mathbb{R}^{K \times K}} \\
&{\mathbf{B}=[\mathbf{b}_1,\mathbf{b}_2,\cdots,\mathbf{b}_k]}  = \sum_{i=1}^{t} \mathbf{y}_{i}\mathbf{x}_{i}^t  \in {\mathbb{R}^{M \times K}}
\end{align}
\normalsize
This algorithm updates each column of $\mathbf{D}$ sequentially using block coordinate descent with a “warm restart” (which consists in the dictionary calculated at the previous step ${\mathbf{D}_{t-1}}$). The following equations are used for updating the $j$th column of $\mathbf{D}$ while keeping the other ones fixed:
\par\noindent\small
\begin{align}
\mathbf{u}_{j} \leftarrow \frac{1}{\mathbf{A}_{jj}} {\left( \mathbf{b}_{j} -\mathbf{D}\mathbf{a}_{j} \right) +\mathbf{d}_{j}}
\end{align}
\normalsize
\par\noindent\small
\begin{align}
\mathbf{d}_{j} \leftarrow \frac{1}{max\left( \vert\vert \mathbf{u}_{j} \vert\vert_{2} \right)} \mathbf{u}_{j} 
\end{align}
\normalsize
This approach gives the solution to the dictionary update problem:
\par\noindent\small
\begin{align}
\normalsize
\mathbf{D}_{t} = \underset{{\mathbf{D}} \in C }{\mathrm{argmin}}\phantom{1} \frac{1}{t} \left( \sum_{i=1}^{t} \vert\vert \mathbf{y}_{i} - \mathbf{D}\mathbf{x}_{i} \vert\vert ^2_2 + \lambda \vert\vert \mathbf{x}_i \vert\vert _1 \right) = \\ \underset{{\mathbf{D}} \in C }{\mathrm{argmin}}\phantom{1} \frac{1}{t} \left( \frac{1}{2} \mathrm{Tr} \left( \mathbf{D}^T \mathbf{D} \mathbf{A}_t \right) - \mathrm{Tr} \left( \mathbf{D}^\mathbf{T} \mathbf{B}_t \right) \right)
\end{align}
\normalsize
Where $\mathbf{C}$ is a convex set, i.e. we have to impose that the column of the dictionary matrix have an $\ell_2$-norm less than or equal to one, then:
\begin{align}
C \overset{\Delta}{=} \left\lbrace \mathbf{D} \in \mathbb{R}^{M \times K} \phantom{1} \mathrm{subject \phantom{1} to} \phantom{1} \forall \phantom{1} j = 1,...,k \phantom{1}\mathbf{d}_{j}^T \mathbf{d}_t \le 1 \right\rbrace  
\end{align}
\normalsize
The ODL algorithm is faster than the batch K-SVD and, since it uses $\mathbf{\mathbf{D}_{(t-1)}}$  as warm restart for computing $\mathbf{\mathbf{D}_t}$, few iterations could be enough for the correct reconstruction. 

\subsection{CBWLSU}
A recent study \cite{naderahmadian2016correlation} notes that even though online processing reduces computational complexity compared to batch methods, ODL performance can be further improved if only useful information from previous data is used for updating the atoms. In this study, a new online DL called Correlation-Based Weighted Least Square Update (CBWLSU) was proposed. CBWLSU is an online method that introduces an interesting alternative for the dictionary update step. Like ODL, CBWLSU evaluates one new training data at a time, $\mathbf{y}_t$. However, to update the dictionary, it searches among all previous training data and uses only the ones which share the same atoms with $\mathbf{y}_{t}$. 
Let $\mathbf{Y}_{\mathcal{Q}_t} = \begin{bmatrix}\mathbf{y}_{l_1} & \cdots & \mathbf{y}_{l_{|\mathcal{Q}_t|}} \end{bmatrix}$, where $l_1,\cdots,l_{|\mathcal{Q}_t|}\in \mathcal{Q}_t$.
be the set of previous training elements at iteration $t$. Define $\mathcal{N}_t = \left\lbrace l : 1 < l < t, \left\langle \mathbf{x}_l^T,\mathbf{x}_t \right\rangle \neq 0 \right\rbrace \subset \mathcal{Q}_t$ as the set of indices of all previous training elements that are correlated with the new element such that $|\mathcal{N}_t| = N_{p_t}$. The new index set is $\mathcal{N}_t = \mathcal{N}_t \cup \{t\}$ so that the training set becomes 
$\mathbf{Y}_{\mathcal{N}_t} = \begin{bmatrix}\mathbf{y}_{l_1} & \cdots & \mathbf{y}_{l_{|\mathcal{N}_t|}} \end{bmatrix}$, where $l_1,\cdots,l_{|\mathcal{N}_t|}\in \mathcal{N}_t$. CBWLSU then employs a weighting matrix $\mathbf{W}(\mathbf{y}_t)$ to evaluate the influence of the selected previous elements for the dictionary update step and solves the optimization problem therein via weighted least squares (WLS). 
The sparse coding in CBWLSU is achieved via batch OMP.

\subsection{DOMINODL}
We now introduce our DOMINODL approach for online DL which not only leads to a dictionary ($\mathbf{D}$) that is tuned to sparsely represent the training set ($\mathbf{Y}$) but is also faster than other online algorithms. The key idea of DOMINODL is as follows: When sequentially analyzing the training set, it is pertinent to leverage the memory of previous data in the dictionary update step. However, algorithms such as CBWLSU consider \textit{all} previous elements. Using all previous training set samples is computationally expensive and may also slow down convergence. The samples which have already contributed in the dictionary update do not need to be considered again. Moreover, in some real-time applications (such as highly correlated range profiles of GPR), their contribution may not be relevant anymore for updating the dictionary.

In DOMINODL, we save computations by considering only a small batch of previous elements that are correlated with the new elements. The two sets are defined correlated if, in their sparse decomposition, they have at least one common non-zero element. The time gained from considering fewer previous training elements is used to consider a mini-batch of new training data (instead of a single element as in ODL and CBWLSU).

The sparse coding step of DOMINODL employs batch OMP, selecting the maximal residual error $\delta$ in (\ref{eq:omp}) using a data-driven entropy-based strategy as described later in this section. At the end of each iteration, DOMINODL also drops-off those previous training set elements that have not been picked up after a certain number of iterations, $N_u$. The mini-batch drawing combined with dropping off training elements and entropy-based criterion to control sparsity results in an extremely fast online DL algorithm that is beneficial for real-time radar operations.

We initialize the dictionary $\mathbf{D}$ using a collection of $K$ training set samples that can be randomly chosen from $\mathbf{Y}$ (alternatively one can also use random vectors with a given distribution) and then perform a sparse decomposition of $\mathbf{Y}$ with the dictionary $\mathbf{D}$; the algorithm then scans the entire training set sequentially. 
Define the mini-batch of $N_b+1$ 
new training elements as $\mathbf{Y}_{\mathcal{B}_t} = \begin{bmatrix}\mathbf{y}_{l_1} & \cdots & \mathbf{y}_{l_{N_b+1}} \end{bmatrix}$ with $l_1,\cdots,l_{N_b+1}\in \mathcal{B}_t$ such that the index set $\mathcal{B}_t = \{l: t \le l < t+N_b\}$. 
When $t>L-N_b$, we simply take the remaining new elements to constitute this mini-batch\footnote{In numerical experiments, we observed that the condition $t>L-N_b$ rarely occurs because DOMINODL updates the dictionary and converges in very few iterations. The algorithm also ensures that the number of previous samples $\geq 2N_r$ before the dictionary update. If this condition is not fulfilled, then it considers all previous training samples.}. 
We store the set of dictionary atoms participating in the SR of the signals in $\mathbf{Y}_{\mathcal{B}_t}$ as $\mathbf{D}_{\mathcal{B}_t}$. Let the coefficient vectors associated with the SR of $\mathbf{Y}_{\mathcal{B}_t}$ are indicated with $\mathbf{X}_{\mathcal{B}_t}$ and 
\par\noindent\small
\begin{align}
\normalsize
\mathbb{I}_{\mathbf{X}_{\mathcal{B}_t}}=\sum_{p \in \mathcal{B}_t} \mathbf{x}_{p},
\end{align}
\normalsize
being an indicator vector whose non-zero elements indicate the atoms of $\mathbf{D}$ being used by $\mathbf{Y}_{\mathcal{B}_t}$.

Define 
$\mathbf{Y}_{\mathcal{Q}_t} = \begin{bmatrix}\mathbf{y}_{l_1} & \cdots & \mathbf{y}_{l_{t}} \end{bmatrix}$ with $l_1,\cdots,l_{t}\in \mathcal{Q}_t$ as the collection of previous training elements with the index set $\mathcal{Q}_t = \{l: 1 \le l < t-1\}$. Consider 
$\mathbf{Y}_{\mathcal{M}_t} = \begin{bmatrix}\mathbf{y}_{l_1} & \cdots & \mathbf{y}_{l_{N_r}} \end{bmatrix}$ with  $l_1,\cdots,l_{N_r}\in \mathcal{M}_t \subset \mathcal{Q}_t$ as a randomly selected mini-batch of $N_r$ previous elements. 
The coefficient vectors associated with the SR of $\mathbf{Y}_{\mathcal{M}_t}$ are indicated with $\mathbf{X}_{\mathcal{M}_t}$ and
\par\noindent\small
\begin{align}
\normalsize
\mathbb{I}_{\mathbf{X}_{\mathcal{M}_t}}=\sum_{l \in \mathcal{M}_t} \mathbf{x}_{l},
\end{align}
\normalsize
is an indicator vector whose non-zero elements indicate the atoms of $\mathbf{D}$ being used by $\mathbf{Y}_{\mathcal{M}_t}$.

Define 
$\mathbf{Y}_{\mathcal{A}_t} = \begin{bmatrix}\mathbf{y}_{l_1} & \cdots & \mathbf{y}_{l_{|\mathcal{A}_t|}} \end{bmatrix}$ with $l_1,\cdots,l_{|\mathcal{A}_t|}\in \mathcal{A}_t$ where
\begin{align}
\normalsize
\mathcal{A}_t = \left\lbrace l : l \in \mathcal{M}_t, \left\langle {\mathbb{I}_{\mathbf{X}_{\mathcal{B}_t}}}^T,{\mathbb{I}_{\mathbf{X}_{\mathcal{M}_t}}} \right\rangle \neq 0 \right\rbrace \subset \mathcal{M}_t,
\end{align}
is a subset of previous training elements that are correlated with the mini-batch of new elements. In order to avoid multiple occurrences of the same element in consecutive mini-batches, DOMINODL ensures that $\mathcal{M}_t \cap \mathcal{M}_{t-1} = \varnothing$ providing that a sufficient number of previous training set elements is available. 
Our new training set is $\mathbf{Y}_{\mathcal{C}_t} = \mathbf{Y}_{\mathcal{A}_t} \cup \mathbf{Y}_{\mathcal{B}_t}$. Both mini-batches of new and previous elements are selected such that the entire training set size ($N_b+N_r$) is still smaller than that of CBWLSU where it is $N_{p_t}+1$ (see above on CBWLSU explanation).

The dictionary update subproblem then reduces to considering only the sets $\mathbf{Y}_{\mathcal{C}_t}$, $\mathbf{D}_{\mathcal{C}_t}$ and $\mathbf{X}_{\mathcal{C}_t}$:
\begin{align}
\label{eq:DL_DOMINODL}
\mathbf{\hat{D}}_{\mathcal{C}_t} = \underset{\mathbf{D}_{\mathcal{C}_t} \in \mathcal{D}}{\text{argmin}} \phantom{1} || \mathbf{Y}_{\mathcal{C}_t}-\mathbf{D}_{\mathcal{C}_t}\mathbf{X}_{\mathcal{C}_t}||_{F}^2.
\end{align}
\normalsize
Assume that the sparse coding for each example is known and define the errors as
\begin{align}
\label{eq:Error_DOMINODL}
\mathbf{E}_{\mathcal{C}_t} = \mathbf{Y}_{\mathcal{C}_t} -\mathbf{D}_{\mathcal{C}_t}\mathbf{X}_{\mathcal{C}_t} = [\mathbf{e}_1, \cdots,\mathbf{e}_{N_r}].
\end{align}
\normalsize

We can update $\mathbf{D}_{\mathcal{C}_t}$, such that the above error is minimized, with the assumption of fixed $\mathbf{X}_{\mathcal{C}_t}$. A similar problem is considered in MOD where error minimization is achieved through least squares. Here, we employ weighted least squares inspired by the fact that it has shown improvement in convergence over standard least squares \cite{naderahmadian2016correlation}. We compute the weighting matrix $\mathbf{W}_{\mathcal{C}_t}$ using the sparse representation error $\mathbf{E}_{\mathcal{C}_t}$
\begin{align}
\label{eq:boh}
\mathbf{W}_{\mathcal{C}_t} = \text{diag}\left(\frac{1}{||\mathbf{e}_1||_2^2},...,\frac{1}{||\mathbf{e}_{N_r}||_2^2}\right),
\end{align}
\normalsize
and then solve the following optimization problem
\begin{align}
\label{eq:DL_DOMINODLW}
\mathbf{\hat{D}}_{\mathcal{C}_t} = \underset{\mathbf{D}_{\mathcal{C}_t} \in \mathcal{D}}{\text{argmin}} \phantom{1} ||(\mathbf{Y}_{\mathcal{C}_t}-\mathbf{D}_{\mathcal{C}_t}\mathbf{X}_{\mathcal{C}_t})\mathbf{W}_{\mathcal{C}_t}^{\frac{1}{2}}||_{F}^2.
\end{align}
\normalsize
This leads to the weighted least squares solution
\begin{align}
\label{eq:mah}
\mathbf{\hat{D}}_{\mathcal{C}_t} = \mathbf{{Y}}_{\mathcal{C}_t} \mathbf{{W}}_{\mathcal{C}_t} \mathbf{Y}^T_{\mathcal{C}_t} (\mathbf{{Y}}_{\mathcal{C}_t} \mathbf{{W}}_{\mathcal{C}_t} \mathbf{Y^T}_{\mathcal{C}_t})^{-1}.
\end{align}
\normalsize
The dictionary $\mathbf{D}$ is then updated with the atoms $\mathbf{\hat{D}}_{\mathcal{C}_t}$ and its columns are normalized by their $\ell_2$-norms. The $\mathbf{D}$ is then used for updating the sparse coding of $\mathbf{{Y}}_{\mathcal{C}_t}$ using batch OMP. Algorithm~\ref{alg:dominodl} summarizes all major steps of DOMINODL. 
\begin{algorithm}[ht]
\LinesNumbered
  \caption{Drop-Off MINi-Batch Online Dictionary Learning (DOMINODL)}
  \label{alg:dominodl}
  \SetAlgoLined
  \SetKwFor{Loop}{Loop}{}{EndLoop}
{\scriptsize
   \KwIn{ Training set ($\mathbf{Y}$), number of trained atoms ($K$), mini-batch dimension for new training data ($N_b$), mini-batch dimension for previous training data ($N_r$), drop-off value ($N_u$), convergence threshold ($\chi\in\mathbb{R}$) and Residual error threshold ($\delta$) for SR }
   \KwOut{ Learned dictionary ($\mathbf{D}$), sparse decomposition of the training set ($\mathbf{X}$) }
   \BlankLine
   Generate the initial dictionary $\mathbf{D}$ of dimension $K$ using training samples \par
   Normalize the columns of $\mathbf{Y}$ and $\mathbf{D}$ by their $\ell_2$-norms\par
   Sparsely decompose $\mathbf{Y}$ with the initial dictionary using batch OMP\par
   \Loop{}{
        Gather a mini-batch of new training set elements $\mathbf{Y}_{\mathcal{B}_t} = \begin{bmatrix}\mathbf{y}_{l_1} & \cdots & \mathbf{y}_{l_{N_b+1}} \end{bmatrix}$ with $l_1,\cdots,l_{N_b+1}\in \mathcal{B}_t$ such that the index set $\mathcal{B}_t = \{l: t \le l < t+N_b\}$ \par
        SR of $\mathbf{Y}_{\mathcal{B}_t}$ with the dictionary $\mathbf{D}$ using entropy-thresholded batch OMP \par
        Store the set of atoms $\mathbf{D}_{\mathcal{B}_t}$ participating in the SR of $\mathbf{Y}_{\mathcal{B}_t}$ \par
                $\mathbf{X}_{\mathcal{B}_t} \leftarrow$ coefficient vectors associated with the SR of $\mathbf{Y}_{\mathcal{B}_t}$ and indicator vector $\mathbb{I_{\mathbf{X}_{\mathcal{B}_t}}}\leftarrow\sum_{p \in \mathcal{B}_t} \mathbf{x}_{p}$ \par
        $\mathbf{Y}_{\mathcal{Q}_t} \leftarrow \begin{bmatrix}\mathbf{y}_{l_1} & \cdots & \mathbf{y}_{l_{t}} \end{bmatrix}$ with $l_1,\cdots,l_{t}\in \mathcal{Q}_t$ as the collection of previous training elements with the index set $\mathcal{Q}_t = \{l: 1 \le l < t-1\}$\par
        Randomly select a mini-batch of previous training set elements $\mathbf{Y}_{\mathcal{M}_t} = \begin{bmatrix}\mathbf{y}_{l_1} & \cdots & \mathbf{y}_{l_{N_r}} \end{bmatrix}$ with  $l_1,\cdots,l_{N_r}\in \mathcal{M}_t \subset \mathcal{Q}_t$ \par 
        
$\mathbf{X}_{\mathcal{M}_t} \leftarrow$ the coefficient vectors associated with the SR of $\mathbf{Y}_{\mathcal{M}_t}$ and indicator vector $\mathbb{I_{\mathbf{X}_{\mathcal{M}_t}}}\leftarrow\sum_{l \in \mathcal{M}_t} \mathbf{x}_{l}$        \par
        
        $\mathbf{Y}_{\mathcal{A}_t} \leftarrow \begin{bmatrix}\mathbf{y}_{l_1} & \cdots & \mathbf{y}_{l_{|\mathcal{A}_t|}} \end{bmatrix}$ with $l_1,\cdots,l_{|\mathcal{A}_t|}\in \mathcal{A}_t$, where
        $\mathcal{A}_t = \left\lbrace l : l \in \mathcal{M}_t, \left\langle {\mathbb{I}_{\mathbf{X}_{\mathcal{B}_t}}}^T,{\mathbb{I}_{\mathbf{X}_{\mathcal{M}_t}}} \right\rangle \neq 0 \right\rbrace \subset \mathcal{M}_t,$  \par
        $\mathbf{Y}_{\mathcal{C}_t} \leftarrow \mathbf{Y}_{\mathcal{A}_t} \cup \mathbf{Y}_{\mathcal{B}_t}$ and store $\mathbf{D}_{\mathcal{C}_t}$ the atoms of $\mathbf{D}$ shared by $\mathcal{B}_t$ and $\mathcal{M}_t$\par

        $\mathbf{E}_{\mathcal{C}_t} \leftarrow \mathbf{Y}_{\mathcal{C}_t} -\mathbf{D}_{\mathcal{C}_t}\mathbf{X}_{\mathcal{C}_t} = [\mathbf{e}_1, \cdots,\mathbf{e}_{N_r}].$ \par
        
        $\mathbf{W}_{\mathcal{A}_t} \leftarrow \text{diag}\left(\frac{1}{||\mathbf{e}_1||_2^2},...,\frac{1}{||\mathbf{e}_{N_r}||_2^2}\right)$ \par
        $\mathbf{\hat{D}}_{\mathcal{C}_t} \leftarrow \mathbf{\hat{D}}_{\mathcal{C}_t} = \mathbf{{Y}}_{\mathcal{C}_t} \mathbf{{W}}_{\mathcal{C}_t} \mathbf{Y}^T_{\mathcal{C}_t} (\mathbf{{Y}}_{\mathcal{C}_t} \mathbf{{W}}_{\mathcal{C}_t} \mathbf{Y}^T_{\mathcal{C}_t})^{-1}$ and normalize its columns \par
        Replace the updated atoms $\mathbf{D}_{\mathcal{C}_t}$ into $\mathbf{D}$ and normalize its columns \par
        Perform SR of selected signals used in the previous step using entropy-thresholded batch OMP \par
        Eliminate previous training set elements which have not been used for the last $N_u$ iterations  \par
       \textbf{if} { ${\vert\vert\left(\mathbf{Y}_{\mathcal{C}_t}- \mathbf{D}_{t}\mathbf{X}_{\mathcal{C}_t}\right) \left( \mathbf{W}_{i}\right)^{0.5} \vert\vert_F^2 < \chi}$ } \textbf{then} break
   }
}
\end{algorithm}

\subsection{Comparison of DL algorithms}
Table~\ref{tbl:dlcomparison} summarizes the important differences between DOMINODL and other related algorithms. 
Like MOD and CBWLSU, DOMINODL uses a weighted least squares solution in the dictionary update. The proof of convergence for the alternating minimization method in MOD was provided in \cite{agarwal2014learning} where it is shown that alternating minimization converges linearly as long as the following assumptions hold true: sparse coefficients have bounded values, sparsity level is on the order of $\mathcal{O}(M^{1/6})$ 
and the dictionary satisfies the RIP property. In \cite{naderahmadian2016correlation}, these assumptions have been applied for CBWLSU convergence. Compared to CBWLSU, the improvements in DOMINODL include mini-batch based data selection and data reduction via drop-off strategy but the update algorithms remain the same. Numerical experiments in \ref{sec:target_recog} suggest that DOMINODL usually converges in far fewer iterations than CBWLSU.
\begin{table*}
\scriptsize
\centering
 	\caption{{Comparison of DL steps}} 
	\label{tbl:dlcomparison}
	\begin{tabular}{ l l l l l l }
		\hline
         \noalign{\vskip 1pt}    
         	 DL step & K-SVD &  LRSDL & ODL & CBWLSU   & DOMINODL\\[1pt]
		\hline
		\hline
        \noalign{\vskip 1pt}    
		   Training method & Batch & Batch & Online & Online  & Online\\[1pt]
            Sparse coding method & OMP & FISTA & LARS & Batch OMP & batch OMP\\[1pt]  
            Dictionary update & Entire $\mathbf{D}$ & Entire $\mathbf{D}$ & Entire $\mathbf{D}$ & Entire $\mathbf{D}$   & Partial $\mathbf{D}$ adaptively \\[1pt]
            Samples per iteration & Entire $\mathbf{Y}$ & Entire $\mathbf{Y}$ & $\mathbf{Y}_{t}$ & $\mathbf{Y}_{\mathcal{N}_t}$  & $\mathbf{Y}_{\mathcal{C}_t}$\\[1pt]
            Optimization method &  SVD & ADMM &  Gradient descent & WLS & WLS\\[1pt]
            Dictionary pruning &  Yes & No &  Yes & No & No\\[1pt]
             Training-set drop-off &  No &  No &  No & No & Yes\\[1pt]
		\hline
		\hline
        \noalign{\vskip 2pt}
	\end{tabular}
\end{table*} 

Computational complexity of DOMINODL has a very low order compared to other online approaches. Let us indicate the order of complexity as $OC_{DL}$, with the subscript $DL$ indicating the employed DL algorithm. As mentioned earlier, there are $N$ atoms in the dictionary. Assume that every signal is represented by a linear combination of $K_s$ atoms, $K_s \ll N$. Empirically, among all possible combinations of $K_s$ atoms from $N$, the probability to have a common atom in the sparse representation is $K_s/N$. Given $L$ training elements, the number of training data which have a specific atom in their representation is proportional to $LK_s/N$. Suppose our mini-batch has elements that reduce the number of training data by a factor $\beta < 1$ (depending on the value of $N_b$ and $N_r$. Further, assume that the dropping off step reduces the training set elements by a factor $\rho < 1$. The number of used training data $L_t$ in the $t^{\text{th}}$ iteration is proportional to $\beta\rho tK_s/ N$. Then, the worst estimate of DOMINODL's computational complexity is due to the sparse coding batch OMP which is of order $OC_{DOMINODL}=\mathcal{O}(L_tN^2) = \mathcal{O}(\beta\rho tK_sN) \approx \mathcal{O}(\beta\rho tN)$. This is much smaller than the complexity of ODL ($OC_{ODL}=\mathcal{O}(N^3)$) or CBWLSU ($OC_{CBWLSU}=\mathcal{O}(tN)$) \cite{naderahmadian2016correlation,mairal2009online}.

\begin{figure}[t]
\centering
  \includegraphics[width=0.85\textwidth]{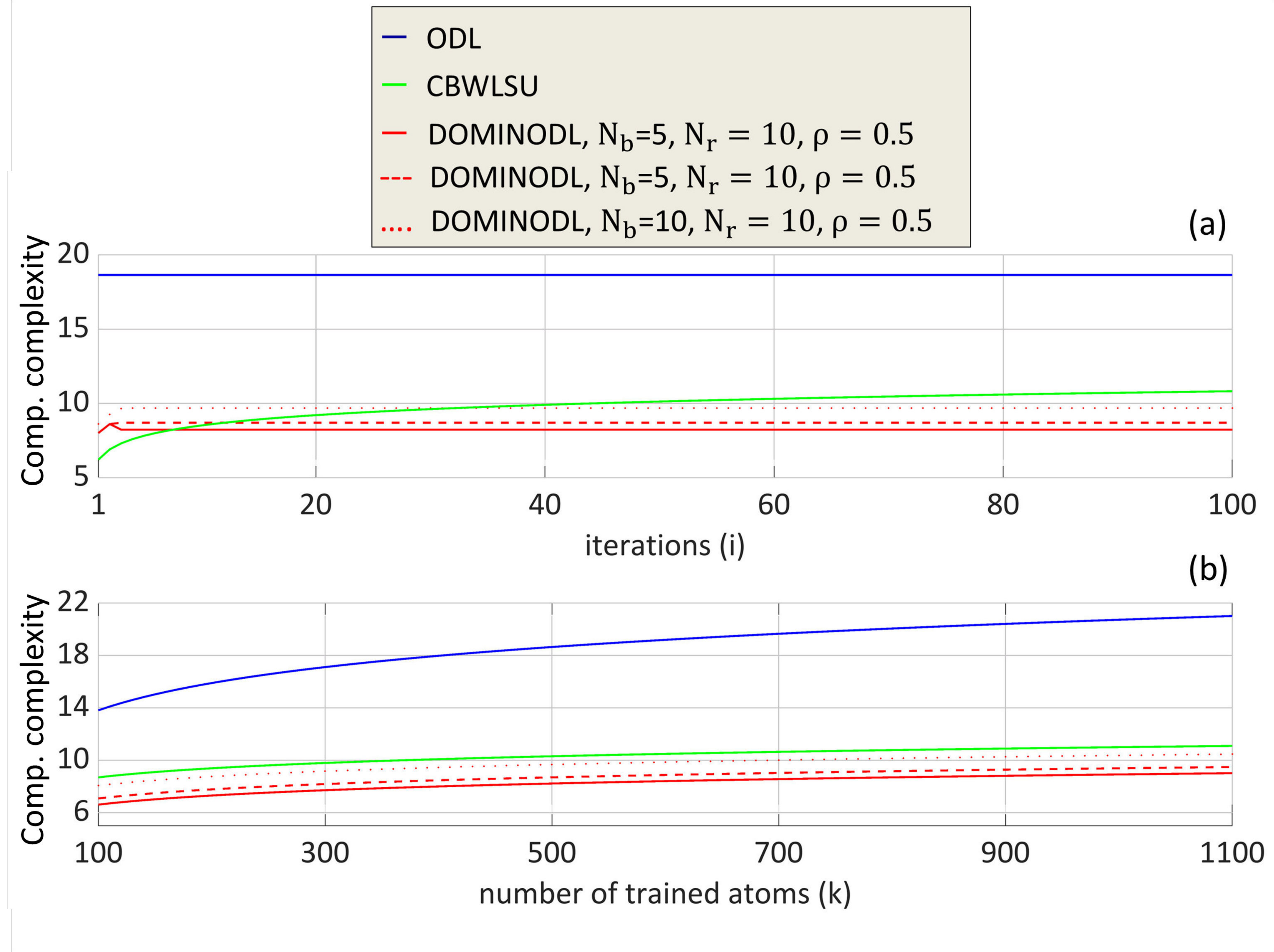}
  \caption{Computational complexity of online DL strategies for increasing number of (a) iterations and (b) trained atoms.}
\label{fig:computational_complexity_vert}
\end{figure}

Figure \ref{fig:computational_complexity_vert} illustrates the computational complexity of online DL approaches compared to DOMINODL. Here, we generally indicate with $OC_{DL}(t)$ and $OC_{DL}(k)$ the order of complexity in function of the number of iterations and atoms of the dictionary. Figure \ref{fig:computational_complexity_vert}(a) shows that, for fixed number of iterations ($t=60$), the general trend of complexity with respect to the increase in the number of atoms ($K$) is similar for all algorithms. However, the complexity of ODL is higher than CBWLSU and DOMINODL; the latter being the least complex. When the number of iterations is increased, the complexity of ODL and CBWLSU have a similar increasing trend (see Fig. \ref{fig:computational_complexity_vert}(b)). In case of DOMINODL, its complexity is similar to the increasing trend of CBWLSU and determined largely by $N_b$. When DOMINODL iterations begin accounting for $N_r$ previous elements, its complexity stays constant. The value of $\beta$ changes for every iteration, while $\rho$ depends on the data itself. In general, after a few dozen of iterations, DOMINODL's complexity always stays lower than CBWLSU.

The proposed DL methods are heuristic, i.e., they do yield over-complete dictionaries but with no theoretical guarantee of obtaining coherent dictionaries. Several state-of-the-art results that outline DL algorithms with concrete performance guarantees require stronger assumptions on the observed data.

An efficient sparse representation (SR) which accurately represents the scattering behaviors related to soil type and targets serves a twofold purpose. First, it allows application of CS in reducing sensing data and associated scan time. Second, the literature \cite{wright2010sparse} indicates that SR is effective in extracting the mid- or high-level features in image classification. For example, in the context of landmines classification with Ground Penetrating Radar, our prior work \cite{giovanneschi2015preliminary,giovanneschi2017online} has shown that frameworks based on sparse representation improve the classifier performance. SR has been frequently applied to data from synthetic aperture radar (SAR) \cite{cetin2014sparsity}), inverse SAR (ISAR) \cite{ender2010compressive}) and interferometric SAR (InSAR) \cite{hongxing2015interferometric} because these are not naturally sparse in the range-time domain. In the particular case of GPR, \cite{gurbuz2012compressive,gurbuz2013sparse,suksmono2010compressive,soldovieri2011sparse} have proposed CS-based imaging for various GPR waveforms.

Techniques to reduce the aquired radar samples using CS have been previously used to address the bottleneck of long scan times in conventional radar systems. The estimation accuracy of target parameters is greatly affected by radar's dwell time \cite{skolnik2008radar}, i.e., the time duration a directional radar beam spends hitting a particular target. But, at the same time, this negatively affects the ability of the radar to look at targets in other directions thereby prolonging the total scan time. For example, \cite{mishra2014compressed} employs matrix completion to reduce scan time in a weather radar. 
Similar techniques have also been shown to be useful in radar imaging applications \cite{akhtar2017compressed}.

\section{Adaptive Statistical DL Evaluation}
\label{sec:parameval}
Adaptive processing is crucial to enhance the radar performance. In the specific context of DL-based classification, we optimize the learned dictionary to sparsely represent GPR measurements. This optimization is based on computing certain statistical properties of the radar data. For example, in our mine detection application, this procedure is applied to the experimental measurements from a GPR test field with buried landmine-simulants (see \ref{subsec:test_field_meas}). The sparse vectors obtained by DL are fed to a Support Vector Machine (SVM) classifier \cite{chang2011libsvm}, which then discriminates between different types of mines and clutter (shown later in Section~\ref{sec:target_recog}). 

The selection of different input parameters for the aforementioned DL algorithms determines their success in sparsifying the data. Hence, it is not useful to directly apply DL with arbitrary parameter values. Prior works set these parameters through hit-and-trial or resorting to metrics that are unable to discriminate the influence of different parameters \cite{shao2013sparse}. Here, we propose a method to investigate the effect of the various input parameters on learning performance and then preset the parameters to \textit{optimal} values that yield the dictionary $\mathbf{D}$ (for each DL method) optimized to sparsely represent our GPR data, therefore improving the quality of the features for classification (i.e. the sparse coefficients).

Table~\ref{tbl:inputparam} lists these input parameters. The K-SVD and ODL admit the number of iterations $N_t$ and number of trained atoms $K$ as input parameters while CBWLSU uses only $K$. The DOMINODL parameters are the dimension of the mini-batch of new ($N_b$) and previous ($N_r$) training elements for each iteration and the drop-off value $N_u$ which indicates after how many iteration the algorithm should discard each unused training set element. We applied K-SVD, ODL, CBWLSU and DOMINODL separately on the training set for different combinations of parameter values. 

In the next subsection, we provide details of the training set obtained from the LIAG survey and follow it by various criteria used for our adaptive DL evaluation such as similarity measure and statistical metrics.

\begin{table}
\centering
 	\caption{DL parameters}
	\label{tbl:inputparam}
	\begin{tabular}{ l | l }
		\hline
         \noalign{\vskip 1pt}
           	DL algorithm & Input parameters\\
        \noalign{\vskip 1pt}
		\hline
        \hline
        \noalign{\vskip 1pt}
            K-SVD & $N_{t}$, $K$  \\[1pt]
            ODL & $N_{t}$, $K$ \\[1pt]
            CBWLSU & $K$ \\[1pt]
            DOMINODL & $K$, $N_b$, $N_r$, $N_u$ \\[1pt]
		\hline
		\hline
        \noalign{\vskip 1pt}
	\end{tabular}
\end{table}

\subsection{Training set generation}
\label{subsec:data_organization}
The entire LIAG data-set described in Section~\ref{sec:landmine_meas} consists of 27 survey sections (or simply, ''surveys'', see \ref{fig:field}) of size $1\times1$ m. Every survey consists of $2500$ range profiles. We divided the data into two sets: \textit{training set} ($\mathbf{Y}$) to be used for both DL and classification, and the \textit{test set} ($\mathbf{Y}_{\text{TEST}}$) to evaluate the performance of the classification.

The training set $\mathbf{Y} \in \mathbb{R}^{M \times L}$ is a matrix whose $L$ columns $\{\mathbf{y}_i\}_{i=1}^L$ consist of sampled range profiles $\mathbf{y}_i = \big[y[0],\cdots,y[M-1]\big]^T$ of $M$ range samples each. The profiles are selected from different surveys and contain almost exclusively either a particular class of landmine or clutter. In total, we have $463$, $168$, $167$ and $128$ range profiles for clutter, PMA2/PMN, ERA and Type-72, respectively, see Table~\ref{tbl:tset}. An accurate separation of these classes was very challenging because of the contributions from the non-homogeneous soil clutter that often masked the target responses completely. A poor selection would lead the DL to learn a dictionary that is appropriate for sparsely representing clutter, instead of landmines. The test set $\mathbf{Y}_{\text{TEST}} \in \mathbb{R}^{M \times J}$ is a matrix with $J=15000$ columns $\{\mathbf{y}_{\text{TEST}_i}\}_{i=1}^J$ that correspond to sampled range profiles from 6 surveys, two for each target class. The test and training sets contain data from separate surveys to enable fair assessment of the classification performance. We denote by the matrices $\mathbf{X} \in \mathbb{R}^{N\times L}$ and $\mathbf{X_{\text{TEST}}} \in \mathbb{R}^{N\times J}$ as the sparse representations of $\mathbf{Y}$ and $\mathbf{Y}_{\text{TEST}}$, respectively and by $K$ the number of atoms of the learned dictionary $\mathbf{D} \in \mathbb{R}^{M\times K}$.

\subsection{Similarity Measure}
In order to compare the dictionaries obtained from various DL algorithms, we use a \textit{similarity measure} that quantifies the closeness of the original training set $\mathbf{Y}$ with the reconstructed set $\hat{\mathbf{Y}}$ obtained using the sparse coefficients of the learned dictionary $\mathbf{D}$. From these similarity values, empirical probability density functions (EPDFs) for any combination of parameter values are obtained; we evaluate these EPDFs using statistical metrics described in Section~\ref{subsec:stat_met}. These metrics efficiently characterize the similarity between $\mathbf{Y}$ and $\hat{\mathbf{Y}}$ and lead us to an optimal selection of various DL input parameters for our experimental GPR dataset.
Consider the cross-correlation between a training set vector $\mathbf{y}_i$ and its reconstruction using a learned dictionary $\hat{\mathbf{y}}_i$: The cross-correlation between $\mathbf{y}_i$ and $\hat{\mathbf{y}}_i$ can be defined as:
\begin{align}
\mathbf{r}_{\mathbf{y}_i,\hat{\mathbf{y}}_i}[l] = \sum\limits_{n=-\infty}^{+\infty}\mathbf{y}_i[n]\hat{\mathbf{y}}_i[n+l]
\end{align}
whereas its normalized version (normalized cross-correlation) is given by:
\begin{align}
\overline{\mathbf{r}_{\mathbf{y}_i,\hat{\mathbf{y}}_i}}[m] = \frac{\mathbf{r}_{\mathbf{y}_i,\hat{\mathbf{y}}_i}[m]}{\sqrt{\mathbf{r}_{\mathbf{y}_i,\mathbf{y}_i}[0]\mathbf{r}_{\hat{\mathbf{y}}_i,\hat{\mathbf{y}}_i}[0]}}.
\end{align}
For the vector $\mathbf{y}_i$, we finally define the similarity measure $s_i$ as
\begin{align}
s_i = \underset{m}{\mathrm{max}}|\overline{\mathbf{r}_{\mathbf{y}_i,\mathbf{\hat{y}}_i}(m)}|,
\end{align}
where a value of $s_i$ closer to unity demonstrates greater similarity of the reconstructed data with the original training set element (i).
\begin{figure}[t]
\centering
  \includegraphics[width=0.5\textwidth]{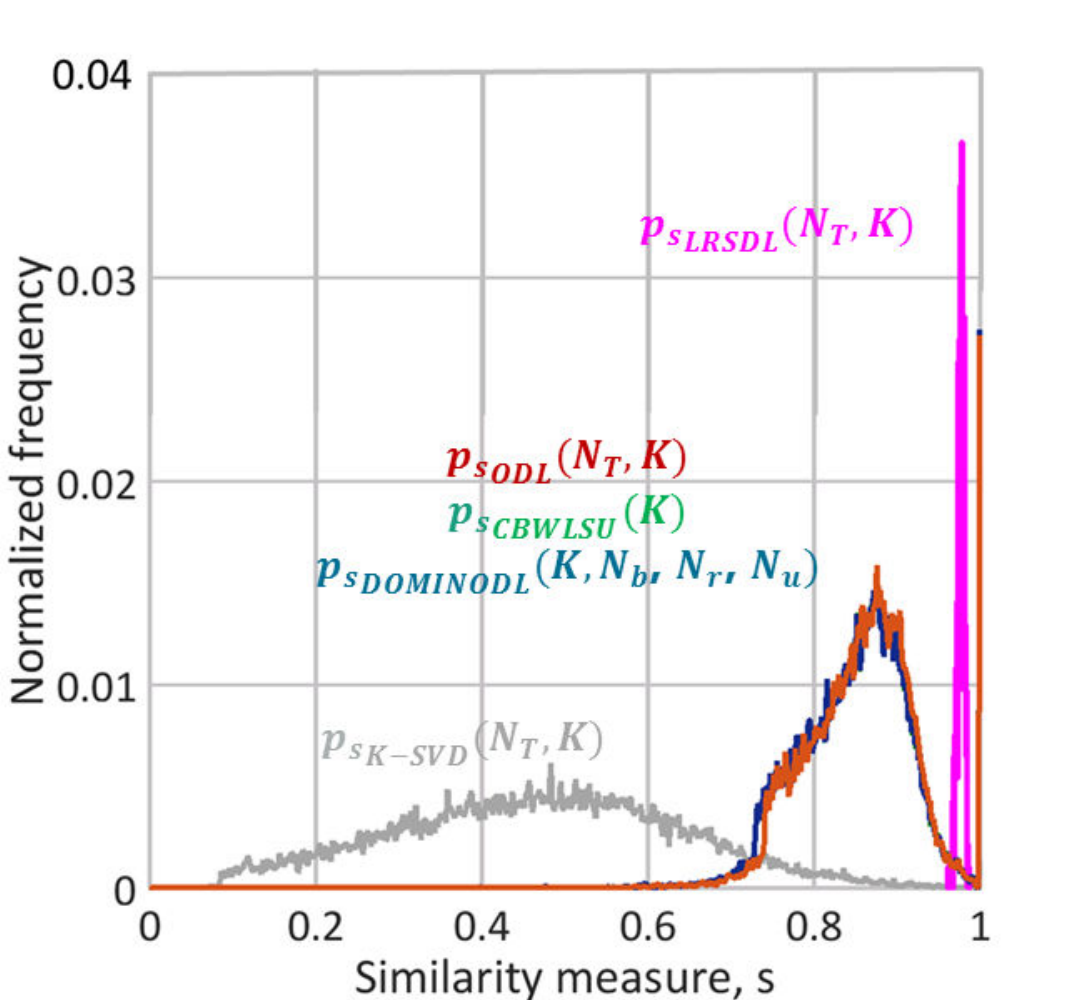}
  \caption{Normalized histograms of similarity measure using the following optimal parameters for the DL algorithm: $N_t=100$, $K=640$, $N_b=30$, $N_r=10$, and $N_u=10$. See Section~\ref{subsec:param_eval_dl} on the process to select these optimal values.}
\label{fig:odlKsvdDist}
\end{figure}

We compute $\{s_i\}_{i=1}^{L}$ for all vectors $\{\mathbf{y}_i\}_{i=1}^L$, and then obtain the normalized histogram or empirical probability density function (EPDF) of all similarity measures. In statistics, an EPDF is defined as the distribution function associated with the empirical measure of a set of data (in our case a set of similarity measures). We indicate the EPDF of a particular DL algorithm as $p_{s_{DL}}$. Here, the subscript DL represents the algorithm used for learning $\mathbf{D}$ e.g. ``K-SVD", ``LRSDL", ``ODL", ``CBWLSU" and ``DOMINODL", respectively.

Various parameter combinations for a specific DL method result in a collection of EPDFs. For a given DL method, our goal is to compare the EPDFs by varying these parameters, and arrive at the thresholds of parameter values after which the changes in $p_{s_{DL}}$ are only incremental. For instance, Fig.~\ref{fig:odlKsvdDist} shows the EPDFs of $\{s_i\}_{i=1}^{L}$ obtained from the GPR mines data where optimal parameters for different DL methods were determined using the statistical methods described in the following subsection. In the figure we also indicate which parameters were affecting each particular distributions (in parenthesis). We note that the online DL approaches ($p_{s_{ODL}}$, $p_{s_{CBWLSU}}$ and $p_{s_{DOMINODL}}$) yield distributions that are more skewed towards unity than K-SVD ($p_{s_{K-SVD}}$). The distribution associated to LRSDL is concentrated on a single peak with high values of similarity, nevertheless, it does not exibith the extremely close to unity values of Online-DL distributions.

\subsection{Statistical metrics}
\label{subsec:stat_met}
Our goal is to find parameter values for which $p_{s_{DL}}$ is skewed towards unity and has small variance. The individual comparisons of mean ($\mu$) and standard deviation ($\sigma$) of EPDFs, as used in previous GPR DL studies \cite{shao2013sparse}, are not sufficient to quantify the observed dispersion in the EPDFs obtained by varying any of the parameter values. Some DL studies \cite{shao2013sparse,chen2016unsupervised} rely on bulk statistics such as the Normalized Root Mean Square Error (NRMSE) but these quantities are less sensitive to changes in parameter values and, therefore, not so helpful in fine-tuning the algorithms. For this evaluation, we will use three different metrics: the coefficient of variation, the Two-sample Kolmogorov-Smirnov (K-S) distance and the Dvoretzky-Kiefer-Wolfowitz (DKW) inequality.
\paragraph{Coefficient of variation}
We choose to simultaneously compare both mean ($\mu$) and variance ($\sigma$) of a single EPDF by using the \textit{coefficient of variation}:
\begin{align}
CV = \sigma/\mu
\end{align}
In our analysis, it represents the extent of variability in relation to the mean of the similarity values.
\paragraph{Two-sample Kolmogorov-Smirnov (K-S) distance}
In the context of our application, it is more convenient to work with the cumulative distribution functions (CDFs) rather than with EPDFs because the well-developed statistical inference theory allows for convenient comparison of CDFs. Therefore, our second metric to compare similarity measurements obtained by successive changes in parameter values is the \textit{two-sample Kolmogorov-Smirnov (K-S) distance}, which is the maximum distance between two given empirical cumulative distribution functions (ECDF). Larger values of this metric indicate that samples are drawn from different underlying distributions. Given two distributions $s_1$ and $s_2$ taken at L discrete points, suppose $\hat{F}_{s_1}$ and $\hat{G}_{s_2}$ are their ECDFs of the same length and correspond to their EPDFs $\hat{f}_{s_1}$ and $\hat{g}_{s_2}$, respectively.\\
Denote $\Omega$ as the set of $L$ observations used to evaluate both distributions. Then, the discrete two sample K-S distance is
\begin{align}
d_{ks}(\hat{F}_{s_1}, \hat{G}_{s_2}) = \sup_{ i \in \Omega}|\hat{F}_{s_1}(i) - \hat{G}_{s_2}(i)|,
\end{align}
where $\sup(\cdot)$ denotes the supremum over all distances. 
In our case, $L$ is the number of range profiles in the training set.\\
We first compute a reference ECDF ($\hat{G}_{s_{\text{ref}}}$) for each DL algorithm with fixed parameter values. For our purposes, this reference ECDF will be obtained by a particular combination of input parameters of the selected DL algorithm. Then, we vary parameter values from this reference and obtain the corresponding ECDF $\hat{F}_{s_{\text{test}}}$ of similarity measure. Finally, we calculate the K-S distance $d_{ks}$ of each $\hat{F}_{s_{\text{test}}}$ with respect to $\hat{G}_{s_{\text{ref}}}$ as
\begin{align}
d_{ks} = d_{ks}(\hat{F}_{s_{\text{test}}}, \hat{G}_{s_{\text{ref}}}) = \sup_{1 \le i \le L}|\hat{F}_{s_{\text{test}}}(i) - \hat{G}_{s_{\text{ref}}}(i)|.
\label{eq:kstest2}
\end{align}
For our evaluation, $d_{ks}$ states how much the selection of certain input parameters of DL changes the ECDFs of similarity values (i.e. how different is the result of DL) w.r.t. the reference one. In other words, using this metric in combination with CV gives us information regarding the variation and the quality of the obtained dictionary according to the selected input parameters.  
\paragraph{Dvoretzky-Kiefer-Wolfowitz (DKW) inequality}
As a third metric, we exploit the \textit{Dvoretzky-Kiefer-Wolfowitz inequality (DKW)} \cite{massart1990tight} which precisely characterizes the rate of convergence of an ECDF to a corresponding exact CDF (from which the empirical samples are drawn) for any finite number of samples.

Let $d_{ks}(\hat{G}_s, F_s)$ be the K-S distance between ECDF $\hat{G}_s$ and the continuous CDF $F_s$ for a random variable $s$ and $L$ samples. Since $\hat{G}_s$ changes with the change in the $L$ random samples, $d_{ks}(\hat{G}_s, F_s)$ is also a random variable. We are interested in the conditions that provide desired confidence in verifying if F and G are the same distributions for a given finite $L$. If the two distributions are indeed identical, then the DKW inequality bounds the probability that $d_{ks}$ is greater than any number $\epsilon$, with $0 < \epsilon < 1$ as follows:
\begin{align}
\label{eq:DKW}
\text{Pr}\left\lbrace d_{ks}\left( \hat{G}_s, F \right) > \epsilon \right\rbrace \le 2e^{-2L\epsilon^2}.
\end{align}
The corresponding asymptotic result that as $L\rightarrow \infty$, $d_{ks} \rightarrow 0$ with probability $1$ is due to the Glivenko-Cantelli theorem \cite{glivenko1933sulla}.

Consider a binary hypothesis testing framework where we use (\ref{eq:DKW}) to test the null hypothesis ${\mathcal{H}_0: F = \hat{G}}$ for a given ${\epsilon}$. The DKW inequality bounds the probability of rejecting the null hypothesis when it is true, i.e., the Type I statistical error.

The probability of rejecting the null hypothesis when it is true is called the p-value of the test and is bounded by the DKW inequality. Assuming the p-value is smaller than a certain confidence level $\alpha$, the following inequality must hold with probability at least $1-\alpha$ \cite{lillacci2011model}:
\begin{align}
\label{eq:epsilon_DKW}
d_{ks}\left( \hat{G}_s, F \right) \leq \sqrt{-\frac{1}{2L}\text{ln}\left(\frac{\alpha}{2}\right)}.
\end{align}

Our goal is to use the DKW inequality to compare two ECDFs $\hat{F}_{s_{\text{test}}}$ and $\hat{G}_{s_{\text{ref}}}$ as in (\ref{eq:kstest2}), to verify if they are drawn from the same underlying CDF. By the triangle inequality, the K-S distance
\begin{align}
\label{eq:triangleinequ}
d_{ks}(\hat{F}_{s_{\text{test}}},\hat{G}_{s_{\text{ref}}}) = d_{ks}(\hat{F}_{s_{\text{test}}},{F}) + d_{ks}(\hat{G}_{s_{\text{ref}}},{F}),
\end{align} 
where $G$ an $F$ are the underlying CDFs corresponding to $\hat{G}$ and $\hat{F}$. We now bound the right side using DKW
\begin{align}
\label{eq:DKW2}
d_{DKW}(\hat{F}_{s_{\text{test}}},\hat{G}_{s_{\text{ref}}}) &\leq \sqrt{-\frac{1}{2L}\text{ln}\left(\frac{\alpha}{2}\right)} + \sqrt{-\frac{1}{2L}\text{ln}\left(\frac{\alpha}{2}\right)}\nonumber\\
&=\sqrt{-\frac{2}{L}\text{ln}\left(\frac{\alpha}{2}\right)},
\end{align} 
which is the maximum distance for which $\hat{F}_{s_{\text{test}}}$ and $\hat{G}_{s_{\text{ref}}}$ are identical with probability $1-\alpha$. The \textit{DKW metric} is the difference
\begin{align}
\label{eq:DKW3}
d_{DKW} = \sqrt{-\frac{2}{L}\text{ln}\left(\frac{\alpha}{2}\right)} - 
d_{L}(\hat{F}_{s_{\text{test}}},\hat{G}_{s_{\text{ref}}}) .
\end{align}
Larger values of this metric imply greater similarity betweem the two ECDFs; a negative value implies that the null hypothesis is not true. For our purposes, $d_{DKW}$ is an alternative way to state the variation of ECDFs respect to $d_{KS}$, however, $d_{DKW}$ also tell us if two distribution are coming from the same underlying CDFs or not (wether the metric is positive or negative, i.e. fulfilling or not the null hypotesis) thus giving a stronger indication on how the input parameters affect the final result of DL.

\subsection{Evaluation results}
\label{subsec:param_eval_dl}
We evaluated the performance of all DL algorithms using the metrics explained in \ref{subsec:stat_met} for the reconstruction of the training set $\mathbf{Y}$. We refer to the Table~\ref{tbl:inputparam} to indicate what are the parameters which affect the presented DL approaches. In particular, the number of iterations $N_t$ is not relevant to CBWLSU and DOMINODL while the latter requires additional parameters for the mini-batch dimensions and the iterations required to drop-off unused training set elements. The role of these parameters in providing an optimized dictionary are summarized in Table~\ref{tbl:paramresume}.
\begin{table}
\centering
	\begin{threeparttable}[b]
    	\caption{Outlook on the influence of the different input parameters for the proposed DL approaches. I = important, S.I. = slightly important, N.U. = not used}
		\label{tbl:paramresume}
		\begin{tabular}{ l | l | l | l | l | l}
			\hline
         	\noalign{\vskip 1pt}
            	& $N_t$ & $K$ & $N_b$ & $N_r$ & $N_u$\\[1pt]
            	
			\hline
        	\hline
        	\noalign{\vskip 1pt}
           		K-SVD & \cellcolor{red!5}N.I. & \cellcolor{blue!25}I. & N.U. & N.U. &  N.U. \\[1pt]
			\hline
			
        	\noalign{\vskip 1pt}
           		ODL & \cellcolor{blue!10}S.I. & \cellcolor{blue!25}I. & N.U. & N.U. &  N.U. \\[1pt]
			\hline

        	\noalign{\vskip 1pt}
           		CBWLSU & N.U. & \cellcolor{blue!25}I. & N.U. & N.U. &  N.U. \\[1pt]
			\hline

        	\noalign{\vskip 1pt}
           		DOMINODL & N.U. & \cellcolor{blue!25}I. & \cellcolor{blue!10}\cellcolor{blue!10}S.I. & \cellcolor{blue!10}S.I. &  \cellcolor{blue!25}I. \\[1pt]
			\hline
			
        	\noalign{\vskip 1pt}
		\end{tabular}
    \end{threeparttable}
\end{table}
Note that we compute the K-S distance and the DKW metric for all methods with respect to a reference distribution $p_{\text{ref}}$, as explained in \ref{subsec:stat_met}. This reference is obtained using the following parameters, as applicable: $N_t = 1$, $K=300$, $N_b = 30$, $N_r=10$ and $N_u=10$.
 
\begin{figure}[t]
\centering
  \includegraphics[width=1.0\textwidth]{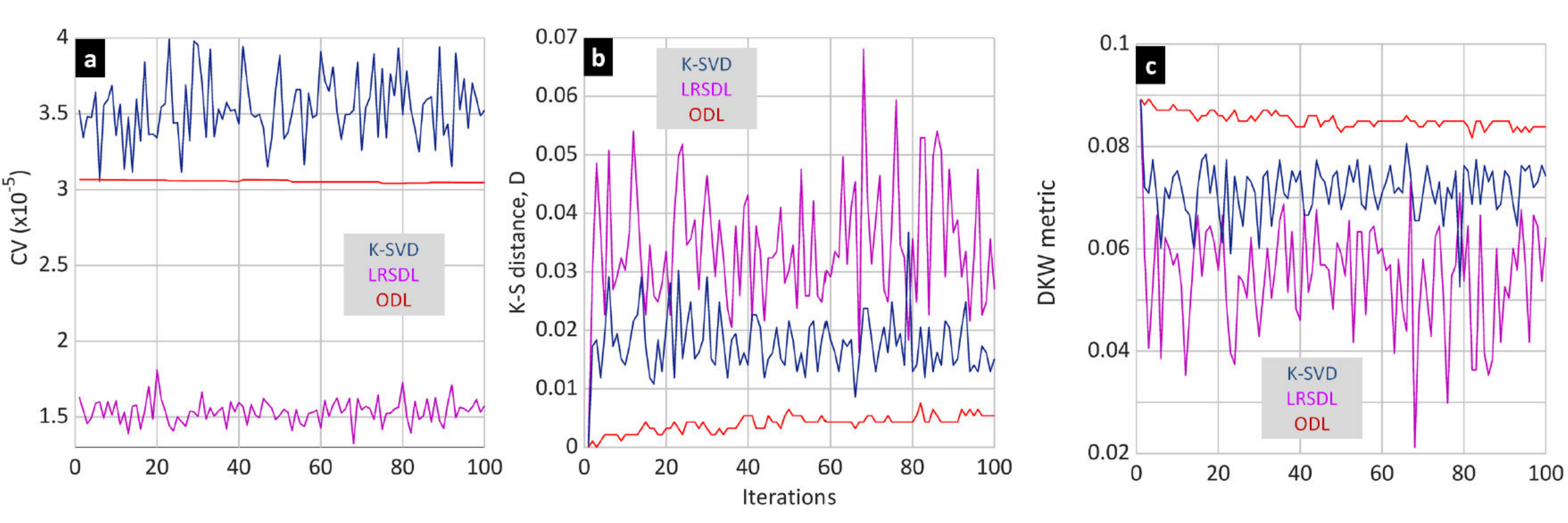}
  \caption{(a) $CV$, (b) K-S distance, and (c) DKW metric for K-SVD, LRSDL, and ODL parameter analyses as a function of the number of iterations $N_t$.}
\label{fig:cv_nit}
\end{figure}

\paragraph{Influence of the number of iterations}
Figures~\ref{fig:cv_nit} (a,b,c) show the effect of $N_t$ on the $CV$, K-S test distance ($d_{ks}$) and the DKW metric ($d_{dkw}$) for K-SVD and ODL and LRSDL. We have skipped CBWLSU and DOMINODL from this analysis because they do not accept $N_t$ as an input. For ODL, the CV remains relatively unchanged with an increase in $N_t$. However, the K-SVD $CV$ exhibits an oscillating behavior and generally high values. In case of the K-S distance, ODL shows slight increase in $d_{ks}$ while K-SVD oscillates around a mean value that is higher than ODL. The DKW metric provides better insight: even though the ODL distributions differ from $p_{\text{ref}}$ with increase in the iterations, the null hypothesis always holds because $d_{DKW}$ remains positive. The $d_{DKW}$ for K-SVD is also positive but much smaller than ODL. It also does not exhibit any specific trend with an increase in iterations. We also observed a similar behavior with the mean of similarity values. The influence of the number of iterations in LRSDL had the same oscillating behaviour as in K-SVD but with larger variation. We conclude that the number of iterations $N_t$ does not significantly influence the metrics for both algorithms, and choose $N_t=100$.

\begin{figure}[t]
\centering
  \includegraphics[width=1.0\textwidth]{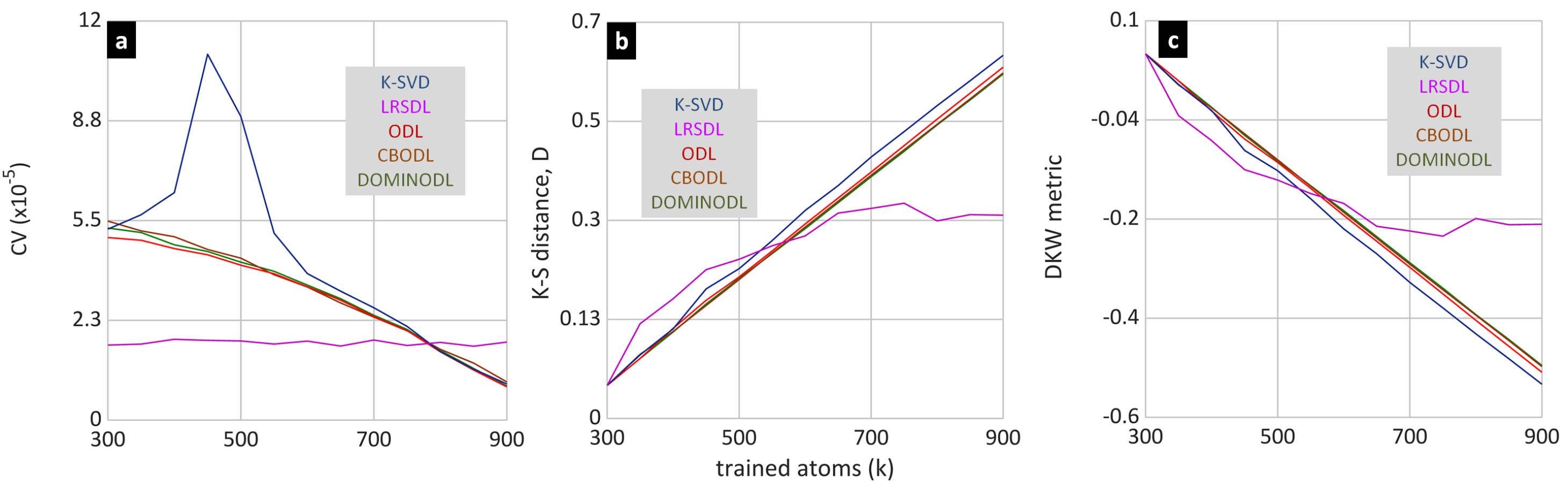}
  \caption{(a) $CV$, (b) K-S distance, and (c) DKW metric for various DL algorithms as a function of the number of trained atoms $K$.}
\label{fig:cv_k}
\end{figure}

\paragraph{Influence of the number of trained atoms}
Figs.~\ref{fig:cv_k} (a,b,c) compare all three metrics with change in the number of trained atoms $K$, a parameter that is common to all DL methods. We observe that $CV$ generally decreases with an increase in $K$. This indicates an improvement in the similarity between the reconstructed and the original training set. K-SVD shows an anomalous pattern for lower values of $K$ but later converges to a trend that is identical to other DL approaches. The K-S distance exhibits a linear change in the the distributions with respect to the reference. Since $d_{ks}$ quantifies the difference between the distributions rather than stating which one is better, combining its behavior with $CV$ makes it evident that an increase in $K$ leads to better distributions of similarity values. The DKW metric $d_{DKW}$, calculated with the same reference, expectedly also shows a linear change. It is clear that, even a slight change in $K$ leads to more negative values of $d_{DKW}$ implying that the null hypothesis does not hold true. This shows the significant influence of the parameter $K$ on the distributions. It was interesting to see a slight improvement for the coefficient of variation when using LRSDL with respect to the other strategies. However, KS-distance and DKW metric indicated that the distributions of similarity values for LRSDL were sensitive to the number of trained atoms only up to a certain value. The value of $K$ is finally chosen such that the dictionary is consistently overcomplete e.g. the number of atoms is three times greater than the number of samples ($K=640$ vs $M=211$).

\paragraph{DOMINODL input parameters selection}
It is difficult to evaluate DOMINODL EPDFs by varying all four parameters together. Instead, we fix the parameter that is common to all algorithms, i.e. the number of trained atoms $K$, and then determine optimal values of $N_b$, $N_r$ and $N_u$.

\begin{figure}[!t]
\centering
  \includegraphics[scale=0.5]{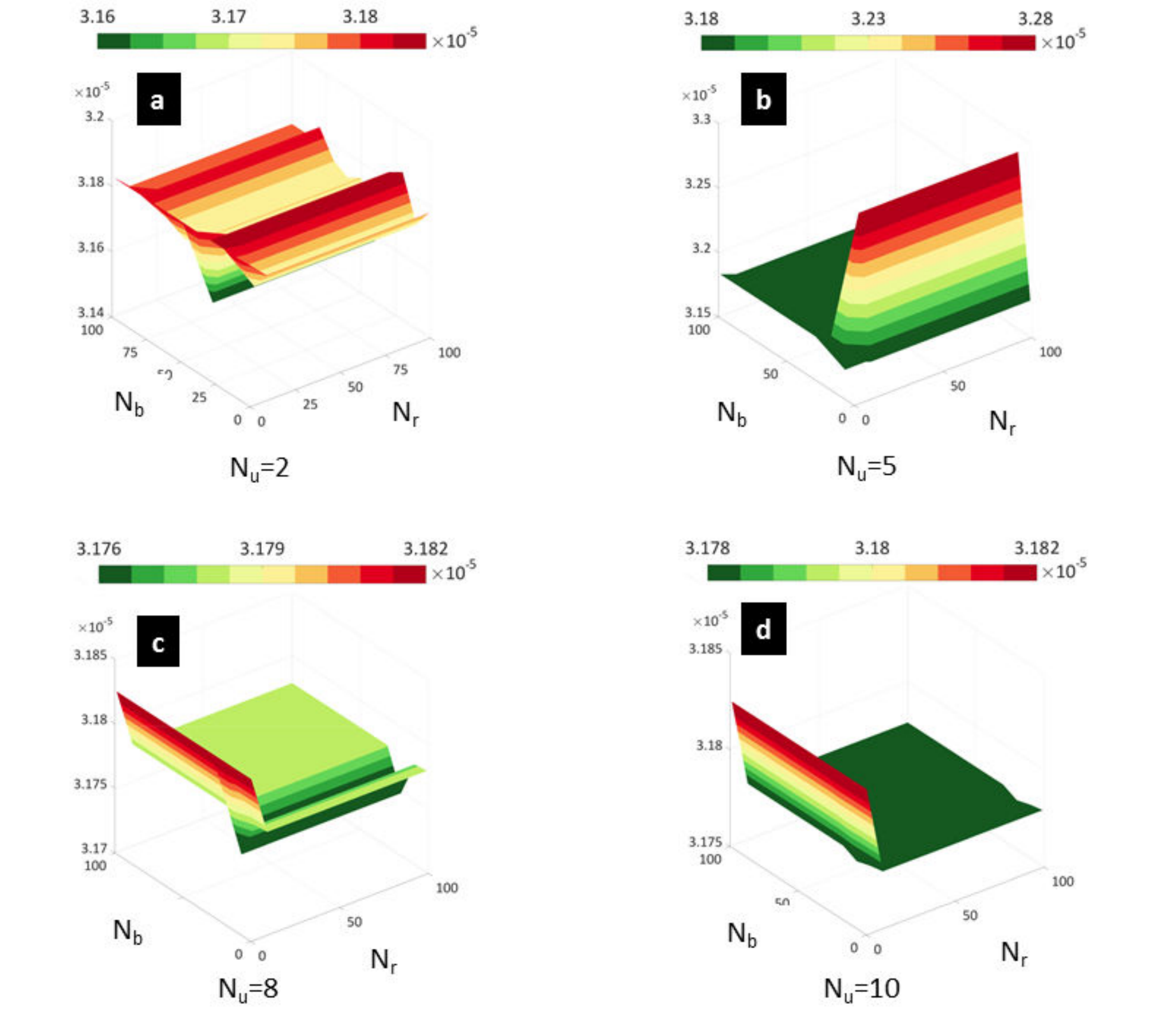}
  \caption{$CV$ as a function of DOMINODL input parameters for $k=640$ and $N_u$ as (a) 2, (b) 5, (c) 8, and (d) 10.}
\label{fig:DOMINOPAR}
\end{figure}

Figure~\ref{fig:DOMINOPAR} shows the coefficient of variation $CV$ of the distribution of similarity values as a function of DOMINODL parameters. The drop-off value $N_u$ appears to have a greater influence with respect to the mini-batch dimensions $N_b$ and $N_r$. To select these parameters we made considerations based on the computational time and the way the algorithm is initialized. The computational time of DOMINODL is essentially independent of $N_r$ and $N_u$ but slightly increases with $N_b$. This was verified and expected because, with $N_b$ we are also increasing the number of SR steps (see Algorithm~\ref{alg:dominodl}) at every iteration which is the source of bulk of computations in DL algorithms \cite{naderahmadian2016correlation}. Further, in order to ensure that the correlation and the drop-off steps kick off from the very first iteration, DOMINODL should admit several new samples for each iteration thereby increasing $N_b$ as well as the number of previous elements accordingly. Taking into account these observations, we choose $N_b=30$ $N_r=10$ and $N_u=10$.


According to the results of the parametric evaluation, we choose the following combination of "optimal" parameters for testing our DL strategies: $N_t=100$, $K=640$, $N_b=30$, $N_r=10$, and $N_u=10$. 

\paragraph{Considerations on computational efficiency of DL algorithms} 
\label{subsec:comp_times}
We used a MATLAB platform on an 8-Core CPU Windows 7 desktop PC to clock the times for DL algorithms. The ODL algorithm from \cite{mairal2009online} is implemented as \texttt{mex} executable, and therefore already fine-tuned for speed. For K-SVD, we employed the efficient implementation from \cite{rubinstein2008efficient} 
to improve computational speed. Table~\ref{tbl:times} lists the execution times of the four DL approaches when using optimal input parameters. The LRSDL is the slowest of all while ODL is more than 4 times faster than K-SVD. The CBWLSU provided better classification results but is three times slower than ODL. This could be because the dictionary update step always considers the entire previous training set elements that correlate with only one new element (i.e. there is no mini-batch strategy). This makes the convergence in CBWLSU more challenging.

The DOMINODL is the fastest DL method clocking 3x speed than ODL and 15x than K-SVD. This is because the DOMINODL updates the dictionary by evaluating only a mini-batch of previous elements (instead of all of them as in CBWLSU) that correlate with a mini-batch of several new elements (CBWLSU uses just one new element). Further, DOMINODL drops out the unused elements leading to a faster convergence. We note that, unlike ODL and K-SVD implementations, we did not use \texttt{mex} executables of DOMINODL which can further shorten current execution times. From Table~\ref{tbl:times}, the reduction in DOMINODL computational time over K-SVD is $((25.8-1.75)\times100)/25.8 \approx93$\%. The reduction for ODL and CBWLSU are computed similarly as $8$\% and $36$\%, respectively.

 The computational bottleneck of mines classification lies in the training times. In comparison, the common steps of sparse decomposition and SVM-based classification during testing take just 0.4 s and 1 s, respectively, for an entire survey (1 m $\times$ 1 m area with 2500 range profiles). Thus, time taken per range profile in ca. 0.59 ms. The average scan rate of our GPR system is 0.19 m/s (or 1 cm/52.1 ms). This can go as high as 2.7 m/s (or 1 cm/3.61 ms) in other GPRs used for landmines application. Therefore, the test times do not impose much computational cost.
\begin{table}
\centering
	\begin{threeparttable}[b]
    	\caption{Computational times for DL algorithms}
		\label{tbl:times}
		\begin{tabular}{ l | l | l | l | l}
			\hline
         	\noalign{\vskip 1pt}
            	& DOMINODL & CBWLSU & ODL & K-SVD\\[1pt]
			\hline
        	\hline
        	\noalign{\vskip 1pt}
           		Time (seconds) & \cellcolor{blue!25}1.75\tnote{1} & 16.49 & 5.75 & 25.8 \\[1pt]
			\hline
			\hline
        	\noalign{\vskip 1pt}
		\end{tabular}
       	\begin{tablenotes}
     		\item[1] Blue denotes the best performance among all DL algorithms
		\end{tablenotes}
    \end{threeparttable}
\end{table}    
\section{Advanced GPR Target Recognition}
\label{sec:target_recog}
We now show the combination of SR, DL, and adaptive processing on GPR target recognition for the case of landmine detection. The general procedure for this target recognition has been illustrated earlier in Section~\ref{subsec:mod_gpr}. Here, we also compare DL-based classification with CNN and evaluate effect of reduced samples. This is reflective of modern adaptive GPR processing which adopts a \textit{swiss-knife} approach based on blending many techniques. 

\begin{figure}[t]
\centering
  \includegraphics[width=1.0\textwidth]{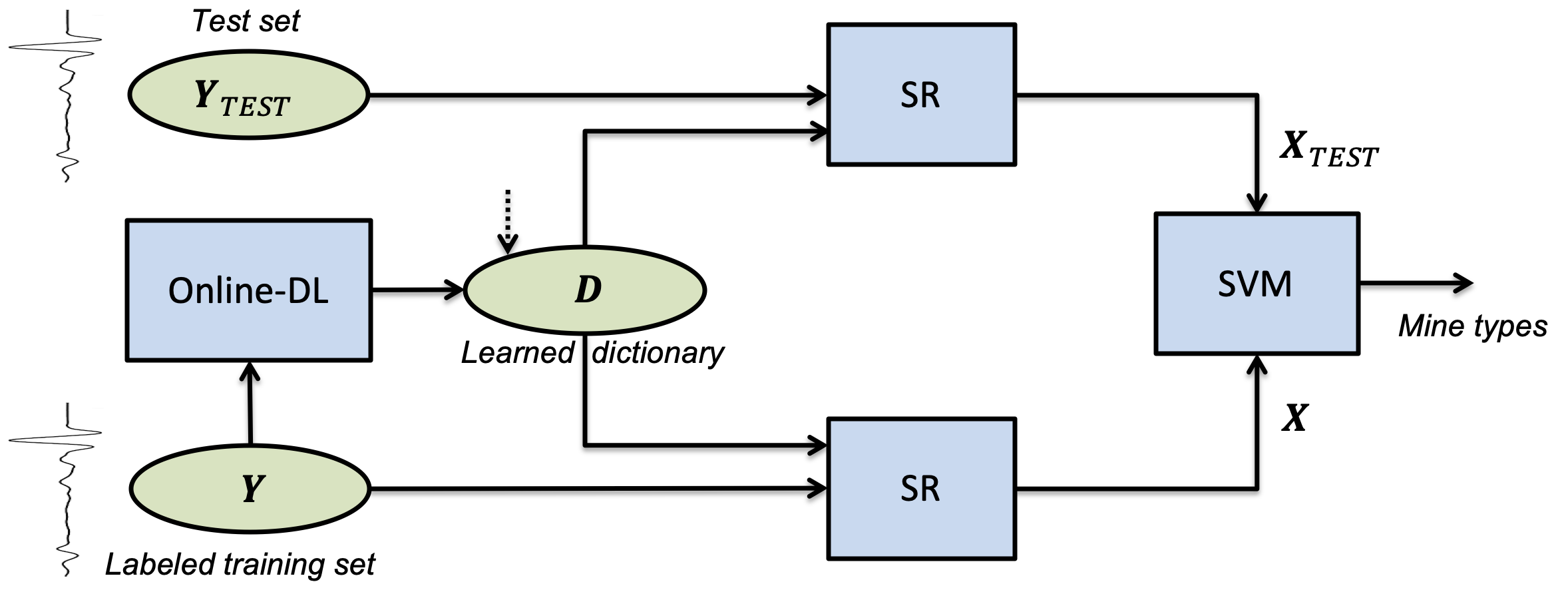}
  \caption{Flowchart for adaptive classification strategy.}
\label{fig:methodology}
\end{figure}

After selecting the input parameters of the proposed DL strategies (see table \ref{tbl:paramresume}), we use the resulting dictionaries for the sparse decomposition of both training and test sets. The resulting sets of sparse coefficients are the input to the SVM classifier. Figure~\ref{fig:methodology} summarizes the flowchart of methodology. Note that the labeled training set $\mathbf{Y}$ is used both for DL and classification.
The threshold $C$ and the kernel function parameter $\gamma$ for SVM have been selected through cross validation. Our key objective is to assess whether online DL algorithms (and in particular DOMINODL) lead to an improvement in the classification accuracy over batch learning strategies or not. As a comparison with a popular state-of-the-art classification method, we also show the classification results with a deep-learning approach based on CNN. Finally, we will show classification performances when the original samples of the range profiles are randomly reduced.
\begin{figure}[t]
\centering
  \includegraphics[width=0.75\textwidth]{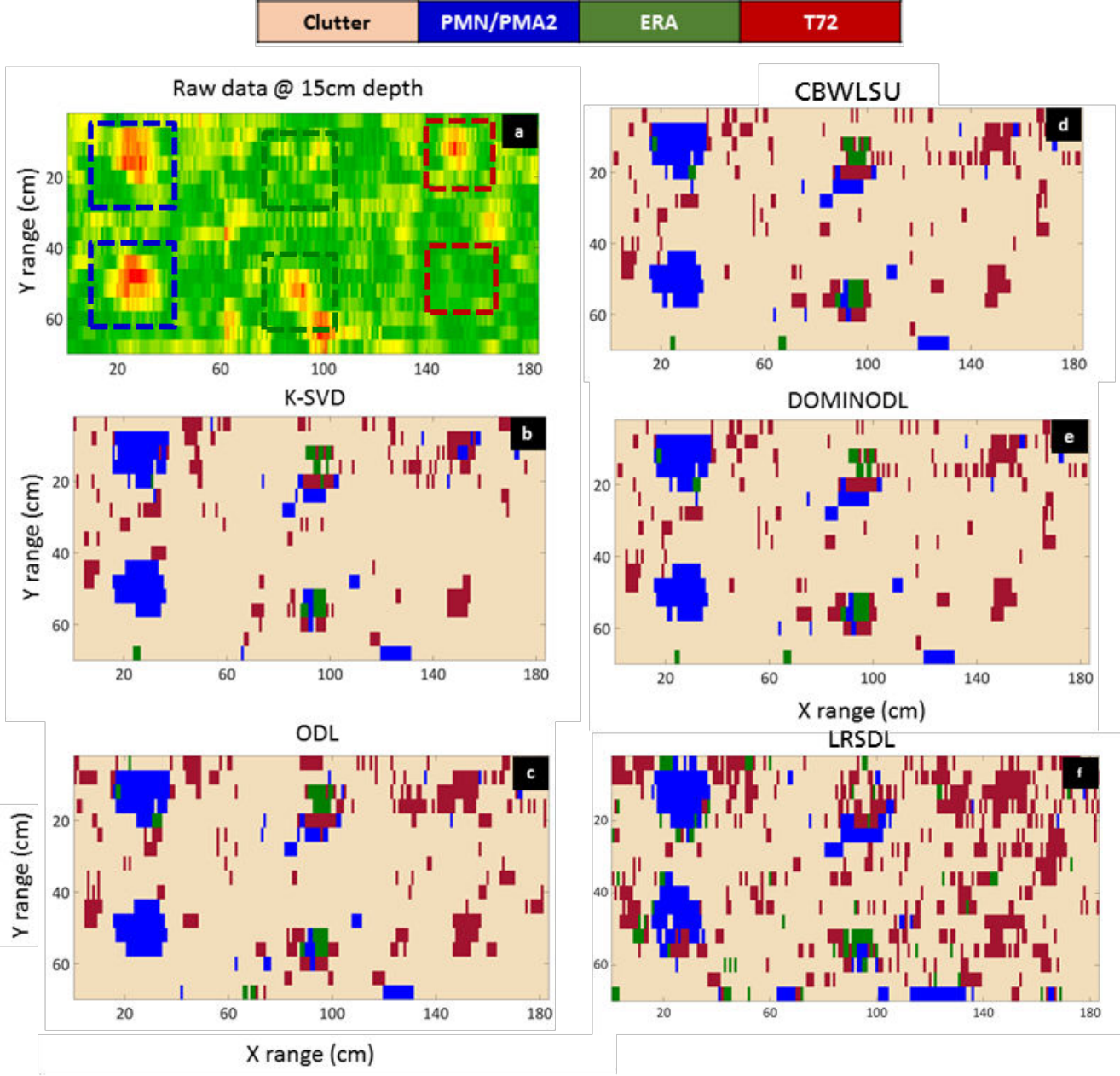}
  \caption{(a) Raw data at $15$ cm depth. The classification maps of the same area containing 6 buried landmines (enclosed by rectangles with dotted boundaries) using an SR-based approach with dictionary learned using (b) K-SVD, (c) ODL, (d) CBWLSU, (e) DOMINODL and (f) LRSDL algorithms with optimally selected input parameters.}
\label{fig:CMAP_full}
\end{figure}
\paragraph{Classification with Optimal Parameters}
\label{subsec:class_opt}
For a comprehensive analysis of the classification performance, we provide both classification maps and confusion matrices for the test set $\mathbf{Y}_{\text{TEST}}$ using the optimal DL input parameters that we selected following our parametric evaluation in Section~\ref{sec:parameval}. The classification maps depict the predicted class of each range profile of the survey under test. The pixel dimension of these maps is dictated by the sampling of the GPR in X and Y directions (see Table~\ref{tbl:techparams}). We stacked together 3 of the 6 surveys from the test set $\mathbf{Y}_{\text{TEST}}$ where each survey had 2 buried landmines of a specific class (PMN/PMA2, ERA and Type-72).

The support of the measurements on our learned dictionaries varies from 1 to 4. Due to coherency, these few non zero elements may appear in the same position for different vector class but with slightly different values. This proved to be enough for the classifier to correctly discriminate between different classes.

Figure~\ref{fig:CMAP_full} shows the classification maps for different DL methods along with the raw data at depth $15$ cm. 
The raw data in Fig.~\ref{fig:CMAP_full}(a) shows that only four of the six mines exhibit a strong reflectivity while the other two mines have echoes so weak that they are not clearly visible in the raw data. Figures~\ref{fig:CMAP_full}(b)-(d) show the results of the SR-based classification (SRC) approaches using DL. All methods clearly detect and correctly classify the large PMN/PMA2 mines. In case of the medium-size ERA, the echoes are certainly detected as non-clutter but some of its constituent pixels are incorrectly classified as another mine. It is remarkable that the left ERA mine is recognized by our method even though it cannot be discerned visually in the raw data. Most of the false alarms in the map belong to the smallest Type-72 mines. This is expected because their small sizes produce echoes very similar to the ground clutter. On the other hand, when T-72 is the ground truth, it is correctly identified.

Using accurate ground truth information, we defined \textit{target halos} as the boundaries of the buried landmines. The dimension of the target halos varied depending on the mine size. Let the number of pixels and the declared mine pixels  inside the target halo (for a certain class) be $n_t$ and $n_m$, respectively. Similarly, we denote the number of true and declared clutter pixels outside the target halo by $n_c$ and $n_d$, respectively. Then, the probabilities of correct classification ($P_{CC}$) for each target class and clutter are, respectively,
\begin{align}
P_{CC_{\text{mines}}} = \frac{n_m}{n_t},\:\:\textrm{and}\:\:P_{CC_{\text{clutter}}} = \frac{n_d}{n_c}.
\end{align}
The $P_{CC}$ being the output of a classifier should not be mistaken as the radar's probability of detection $P_d$ which is the result of a detector. A detector declares the presence of a mine when only a few pixels inside the halo have been declared as mine. For statistical detectors, we refer the reader to \cite{pambudi2019forward}. Here, we describe a more practical approach. $P_{CC}$ provides a fairer and more accurate evaluation of the classification result. This per-pixel information can be easily used to improve the final detection result. For instance, the operator could set a threshold for the minimum number of pixels to be detected in a cluster so that a circle with center at the cluster centroid could be used as the detected mine. However, such a circle may exclude some of the mine pixels leading to a potential field danger. The per-pixel classification is then employed to determine the guard area around the mine circle.

As an additional test, we computed the final detection result using two-third of pixels in the entire target halo to declare a successful detection. These values are based on only six mines in the field. It will be, obviously, more useful to repeat this procedure for a broader field with higher number of mines for conclusive $P_d$/$P_{fa}$ results. Table~\ref{tbl:detection} lists $P_d$ and probability of false alarm $P_{fa}$ for the case when optimal parameters were used for DL algorithms.\\

\begin{table}
\centering
	\begin{threeparttable}[b]
    	\caption{Probability of detection and false alarm for various DL methods when optimal parameters are used}
		\label{tbl:detection}
		\begin{tabular}{ l | l | l | l |l |l }
			\hline
         	\noalign{\vskip 1pt}
            	& $K-SVD$ & $ODL$ & $CBWLSU$ & $DOMINODL$ & $LRSDL$ \\[1pt]
            	
			\hline
        	\hline
        	\noalign{\vskip 1pt}
           		$P_d$ & $0.75$ & $0.75$ & \cellcolor{blue!25}$0.833$ & \cellcolor{blue!25}$0.833$ & $0.75$  \\[1pt]
			\hline
			
        	\noalign{\vskip 1pt}
           		$P_{fa}$ & $0.124$ & $0.128$ & $0.129$ & \cellcolor{blue!25}$0.108$ & $0.565$  \\[1pt]
			\hline

        	\noalign{\vskip 1pt}
		\end{tabular}
       	\begin{tablenotes}
     		\item[1] Blue denotes the best performance among all DL algorithms
		\end{tablenotes}
    \end{threeparttable}
\end{table}

A \textit{confusion matrix} is a quantitative representation of the classifier performance. The matrix lists the probability of classifying the ground truth as a particular class. The classes listed column-wise in the confusion matrix are the ground truths while the row-wise classes are their predicted labels. Therefore, the diagonal of the matrix is the $P_{CC}$ while off-diagonal elements are probabilities of misclassification. 
\begin{table}[t]
\centering
	\begin{threeparttable}[b]
 		\caption{Confusion matrix with optimal DL input parameter selection.}
		\label{tbl:CM_DL_full}
		\begin{tabular}{ c | l | l | l | l | l}
			\hline
    		\noalign{\vskip 1pt}
            \multicolumn{2}{c|}{} & Clutter & PMN/PMA2 & ERA & Type-72\\[1pt]
        	\hline
        	\hline
        	\noalign{\vskip 1pt}
             						& Clutter & \cellcolor{black!25}0.892& 0.044 & 0.25 & 0.37 \\[1pt]
             \multirow{2}{*}{K-SVD} & PMN/PMA2 & 0.022 & \cellcolor{black!25}0.938\tnote{1} & 0.166 & 0.074 \\[1pt]
             						& ERA & 0.021 & 0.017 & \cellcolor{black!25}0.472 & 0.018 \\[1pt]
             						& Type-72 & 0.064 & 0 & 0.111 & \cellcolor{black!25}0.537 \\[1pt]
			\hline
		    \noalign{\vskip 1pt}
            						& Clutter & \cellcolor{black!25}0.435 & 0.061 & 0.111 & 0.351 \\[1pt]
            \multirow{2}{*}{LRSDL (SRC)} & PMN/PMA2 & 0.155 & \cellcolor{black!25}0.289 & 0.319 & 0.259 \\[1pt]
            						& ERA & 0.172 & 0.372 & \cellcolor{black!25}0.361 & 0.278 \\[1pt]
            						& Type-72 & 0.237 & 0.272 & 0.208 & \cellcolor{black!25}0.111 \\[1pt]
 			\hline
        	\noalign{\vskip 1pt}
            					   	& Clutter & \cellcolor{black!25}0.871 & 0 & 0.194 & 0.333 \\[1pt]
            \multirow{2}{*}{ODL} 	& PMN/PMA2 & 0.022 & \cellcolor{black!25}0.973 & 0.139 & 0 \\[1pt]
            					   	& ERA & 0.018 & 0.026 & \cellcolor{black!25}0.583 & 0.018 \\[1pt]
            					   	& Type-72 & 0.088 & 0 & 0.083 & \cellcolor{black!25}0.648 \\[1pt]
        	\hline
        	\noalign{\vskip 1pt}
           							& Clutter & \cellcolor{black!25}0.872 & 0.017 & 0.181 & 0.314 \\[1pt]
            \multirow{2}{*}{CBWLSU} & PMN/PMA2 & 0.023 & \cellcolor{black!25}0.973 & 0.153 & 0 \\[1pt]
            						& ERA & 0.025 & 0.008 & \cellcolor{black!25}0.528 & 0\\[1pt]
            						& Type-72 & 0.08 & 0 & 0.138 & \cellcolor{black!25}0.685 \\[1pt]
			\hline
	    	\noalign{\vskip 1pt}
            						& Clutter & \cellcolor{black!25}0.876 & 0.017 & 0.167 & 0.315 \\[1pt]
            \multirow{2}{*}{DOMINODL} & PMN/PMA2 & 0.023 & \cellcolor{black!25}0.974 & 0.138 & 0 \\[1pt]
            						& ERA & 0.027 & 0.008 & \cellcolor{black!25}0.58 & 0 \\[1pt]
            						& Type-72 & 0.077 & 0 & 0.11 & \cellcolor{black!25}0.685 \\[1pt]
 			\hline
			\hline
        	\noalign{\vskip 1pt}
		\end{tabular}
       	\begin{tablenotes}
			\item[1] Gray denotes the $\mathbf{P_{CC}}$ value for a specified class and DL algorithm
		\end{tablenotes}
    \end{threeparttable}
\end{table}    

For the classification map of Fig.~\ref{fig:CMAP_full}, table~\ref{tbl:CM_DL_full} shows the corresponding confusion matrices for each DL-based classification approach. In general, we observe an excellent classification of PMN/PMA2 landmines ($\sim98$\%), implying that almost every range profile in the test set which belongs to this class is correctly labeled. The $P_{cc}$ for the clutter is also quite high ($\sim90$\%). This can also be concluded from the classification maps where the false alarms within the actual clutter regions are very sparse (i.e. they do not form a cluster) and, therefore, unlikely to be interpreted as an extended target. As noted previously, most of the clutter misclassification is associated with the Type-72 class. The ERA test targets show some difficulties with correct classification. However, most of the pixels within its target halo are declared at least as some type of mine (which is quite useful in terms of issuing safety warnings in the specific field area). This result can be explained by the fact that ERA test targets (being simulant landmines, i.e. SIMs) do not represent a specific mine but have general characteristics common to most landmines. The Type-72 mines exhibit a $P_{cc}$  which is slightly higher with respect to ERA targets. This is a remarkable result because Type-72 targets were expected to be the most challenging to classify due to their small size.

Conventionally, as mentioned in \cite{vu2017fast}, LRSDL is associated with a SRC (Sparse Representation Based Classification) technique. However, applying this approach to our problem resulted in very low accuracy (an average of ca. $20$\% across all classes as evident from Table~\ref{tbl:CM_DL_full}) and semi-random classification maps (Fig.~\ref{fig:CMAP_full}). This can be explained by the extreme similarity between the training set examples of different classes; mines and clutter are only slightly dissimilar in their responses and mine responses are generally hidden in the ground reflections. Each learned ``block'' $D_{c}$ differed only slightly from the other and, therefore, poor classification results are achieved with this dataset. LRSDL won't be used for further evaluations.

All DL algorithms used for our sparse classification approach show very similar results for the clutter and PMN/PMA2 classes. However, online DL methods show higher $P_{CC}$ for the ERA and Type-72 targets with respect to K-SVD. From Table~\ref{tbl:CM_DL_full}, the detection enhancement using the best of the online DL algorithms for PMN/PMA2 over K-SVD is $((0.974-0.938)\times100)/0.938 \approx4$\%. The improvements for ERA and T-72 are computed similarly as $23$\% and $28$\%, respectively.

We also evaluated the performance of the conventional classification approach, i.e. SVM without DL-SR. In this case, the confusion matrix (Table~\ref{tbl:CNN_SVMonly} shows that the clutter recognition does not deteriorate. However, the smaller mines (ERA and T-72) are incorrectly classified and the accuracy for PMN/PMA2 is poorer than DL-based approaches. It follows that it is critical to extract features prior to classification. This step is included in our methods, which combine SR with DL to generate discriminative features for the particular case of APM classification. 

\begin{table}[t]
\centering
    \caption{Confusion matrix for SVM-based classification}
	\label{tbl:CNN_SVMonly}
	\begin{tabular}{ l | l | l | l | l}
		\hline
         \noalign{\vskip 1pt}
            & Clutter & PMN/PMA2 & ERA & Type-72\\[1pt]
		\hline
        \hline
        \noalign{\vskip 1pt}
            Clutter & \cellcolor{black!25}0.960& 0.184 & 0.625 & 0.685 \\[1pt]
            PMN/PMA2 & 0.024 & \cellcolor{black!25}0.807 & 0.305 & 0.203 \\[1pt]
            ERA & 0.005 & 0.008 & \cellcolor{black!25}0.069 & 0.111 \\[1pt]
            Type-72 & 0 & 0 & 0 & \cellcolor{black!25}0 \\[1pt]
		\hline
		\hline
        \noalign{\vskip 1pt}
	\end{tabular}
\end{table}

\paragraph{Classification with Non-Optimal Parameters}
\label{subsec:class_nonopt}
In order to demonstrate how the quality of the learned dictionary affects the final classification, we now show the confusion matrices for a non-optimal selection of input parameters in different DL algorithms. Our goal is to emphasize the importance of learning a good dictionary by selecting the optimal parameters rather than specifying how each parameter affects the final classification result. We arbitrarily selected the number of trained atoms $K$ to be only $300$ for all DL approaches, reduce the number of iterations to $25$ for ODL and KSVD and, for DOMINODL, we use $N_r$=30, $N_b$=5 and $N_u$=2. Table~\ref{tbl:CM_DL_bad} shows the resulting confusion matrix. While the clutter classification accuracy is almost the same as in Table~\ref{tbl:CM_DL_full}, the $P_{cc}$ for PMN/PMA2 landmines decreased by $\sim10$\% for most of the algorithms except ODL where it remains unchanged. The classification accuracy for ERA and Type-72 mines is only slightly worse for online DL approaches. However, in the case of K-SVD, the $P_{CC}$ reduces by $\sim30$\% and $\sim10$\% for ERA and Type-72, respectively. Clearly, the reconstruction and correct classification of range profiles using batch algorithms such as K-SVD is strongly affected by a non-optimal choice of DL input parameters. As discussed earlier in Section~\ref{subsec:param_eval_dl}, this degradation is likely due to the influence of $K$ rather than $N_t$.
\begin{table}
\centering
 	\caption{Confusion matrix with non-optimal DL input parameter selection}
	\label{tbl:CM_DL_bad}
		\begin{tabular}{ c | l | l | l | l | l}
		\hline
    	\noalign{\vskip 1pt}
            \multicolumn{2}{c|}{} & Clutter & PMN/PMA2 & ERA & Type-72\\[1pt]
        \hline
        \hline
        \noalign{\vskip 1pt}
             						& Clutter & \cellcolor{black!25}0.853& 0.07 & 0.305 & 0.222 \\[1pt]
            \multirow{2}{*}{K-SVD} 	& PMN/PMA2 & 0.037 & \cellcolor{black!25}0.851 & 0.222 & 0.111 \\[1pt]
            						& ERA & 0.032 & 0 & \cellcolor{black!25}0.194 & 0.241 \\[1pt]
            						& Type-72 & 0.077 & 0.078 & 0.277 & \cellcolor{black!25}0.426 \\[1pt]
		\hline
        \noalign{\vskip 1pt}
            						& Clutter & \cellcolor{black!25}0.86 & 0.017 & 0.181 & 0.444\\[1pt]
            \multirow{2}{*}{ODL} 	& PMN/PMA2 & 0.016 & \cellcolor{blue!25}0.973 & 0.097 & 0 \\[1pt]
            						& ERA & 0.022 & 0.008 & \cellcolor{blue!25}0.638 & 0 \\[1pt]
            						& Type-72 & 0.1 & 0 & 0.083 & \cellcolor{black!25}0.555 \\[1pt]
        \hline
        \noalign{\vskip 1pt}
            						& Clutter & \cellcolor{black!25}0.887 & 0.078 & 0.319 & 0.352 \\[1pt]
            \multirow{2}{*}{CBWLSU} & PMN/PMA2 & 0.019 & \cellcolor{black!25}0.877 & 0.097 & 0 \\[1pt]
            						& ERA & 0.018 & 0.043 & \cellcolor{black!25}0.541 & 0\\[1pt]
            						& Type-72 & 0.074 & 0 & 0.042 & \cellcolor{blue!25}0.648 \\[1pt]
		\hline
         \noalign{\vskip 1pt}
            						& Clutter & \cellcolor{blue!25}0.888 & 0.078 & 0.319 & 0.352 \\[1pt]
            \multirow{2}{*}{DOMINODL} & PMN/PMA2 & 0.019 & \cellcolor{black!25}0.877 & 0.097 & 0 \\[1pt]
            						& ERA & 0.018 & 0.043 & \cellcolor{black!25}0.54 & 0 \\[1pt]
            						& Type-72 & 0.074 & 0 & 0.042 & \cellcolor{blue!25}0.648 \\[1pt]
 		\hline
		\hline
        \noalign{\vskip 1pt}
	\end{tabular}
\end{table}   
\paragraph{Comparison with Deep Learning Classification}
\label{subsec:cnn}
The core idea of SRC is largely based on the assumption that signals are linear combinations of a few atoms. In practice, this is often not the case. This has led to a few recent works that suggest employing deep learning for radar target classification. However, these techniques require significantly large datasets for training.

We compared classification results of our methods with a deep learning approach. In particular, we constructed a CNN because these networks are known to efficiently exploit structural or locational information in the data and yield comparable learning potential with far fewer parameters. We modeled our proposed CNN framework as a classification problem wherein each class denotes the type of mine or clutter. The training data set for our CNN structure is the matrix $\mathbf{Y}$ (see \ref{subsec:data_organization}).

Building up a synthetic database is usually an option for creating (or extending) a training set for deep learning applications. However, accurately modeling a GPR scenario is still an ongoing challenge in the GPR community because of the difficulties in accurately reproducing the soil inhomogeneities (and variabilities), the surface and underground clutter, the antenna coupling and ringing effects, etc. Even though some applications have been promising \cite{giannakis2016realistic}, this remains a cumbersome task. 
The input layer of our CNN took one-dimensional range profiles with $211$ samples. It was followed by two convolutional layers with $20$ and $5$ filters of size $20$ and $10$, respectively. The output layer consisted of four units wherein the network classifies the given input data as clutter or one of the three mines. There were rectified linear units (ReLU) after each convolutional layer; the ReLU function is given by $\text{ReLU}(x) = \text{max}(x,0)$.

The architecture of the CNN was selected through an arduous process of testing many combination of layers/filters and hyperparameters which would lead to better accuracy during training. A deeper network slightly increased the accuracy in the training phase but led to poorer performance when classifying new data (i.e. the test set $\mathbf{Y}_{test}$). Since our data are limited, adding more layers (i.e. more weights) only led to overfitting and made the network incapable to generalize on new datasets. A multi-dimensional CNN formed by clustering 2D and 3D data would have further reduced the training set. Augmenting the data was also envisioned but commonly used transformations such as scaling/rotations are not useful in our case because the mines were always in the same inclination and their dimension defines the class itself. We also attempted adding different levels of noise but this did not lead to better results considering the available data are already very noisy.

\begin{figure}[t]
\centering
  \includegraphics[width=0.75\textwidth]{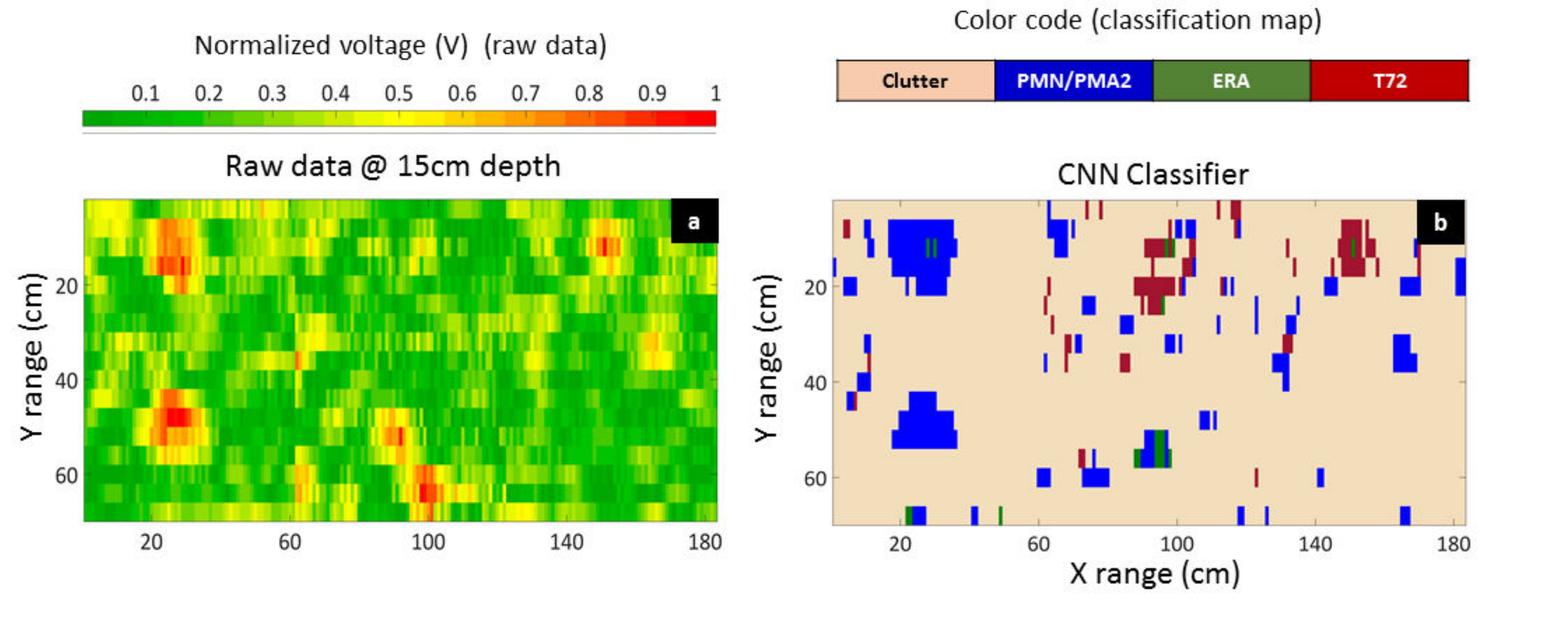}
  \caption{(a) Raw data at $15$ cm depth. (b) Classification maps of the same area containing 6 buried landmines using CNN-based classification.}
\label{fig:classmap_CNN}
\end{figure}

\begin{figure}[t]
\centering
  \includegraphics[width=0.75\textwidth]{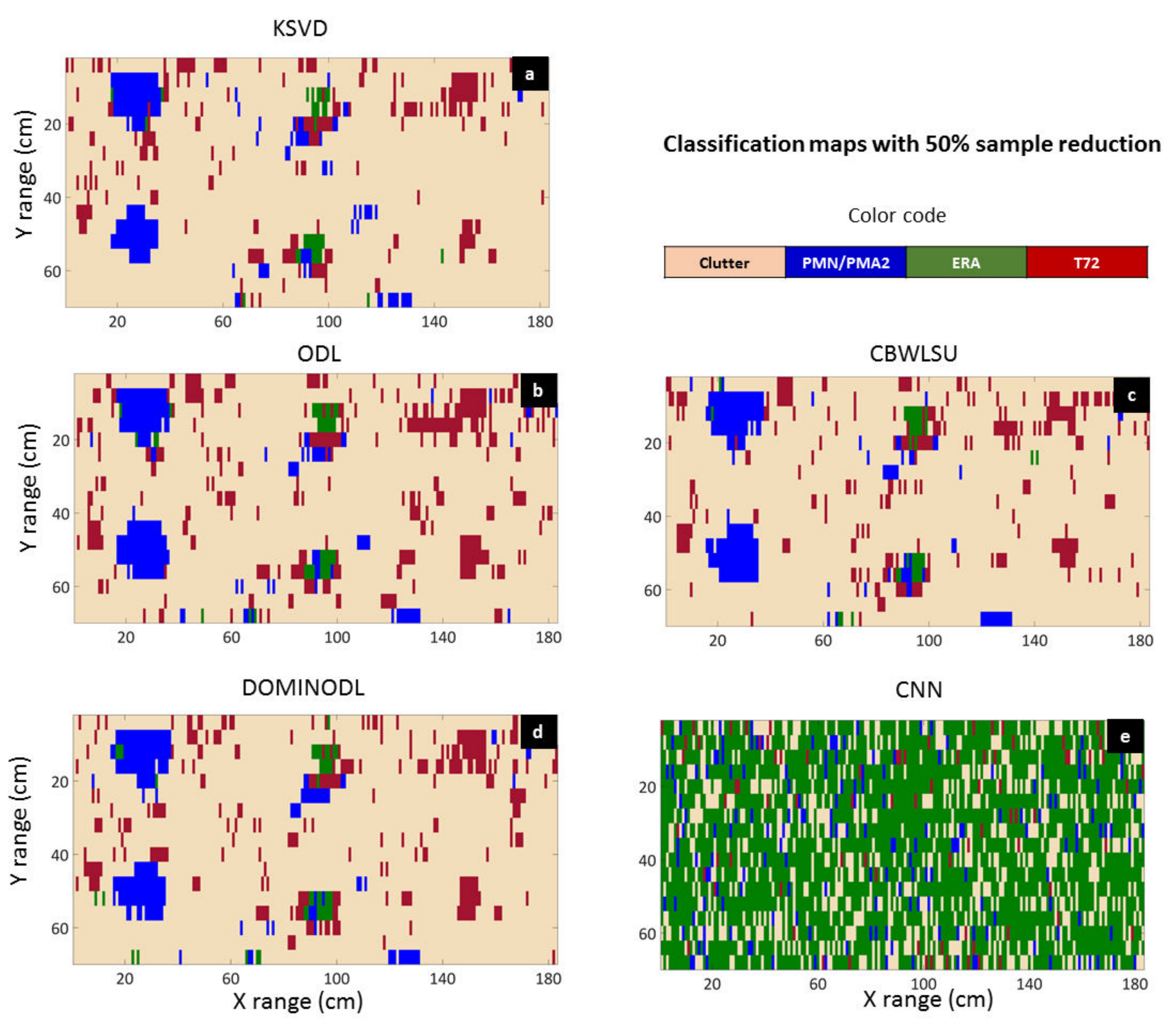}
  \caption{The classification maps of the same area containing 6 buried landmines using an SR-based approach with dictionary learned using (a) K-SVD, (b) ODL, (c) CBWLSU, and (d) DOMINODL algorithms. The input parameters were optimally selected and the number of samples were reduced by $50$\%. (e) The corresponding result with reduced samples for CNN-based classification.}
\label{fig:classmap_50}
\end{figure}

We trained the network with the labeled training set $\mathbf{Y}$, selecting $\sim20$\% of the training data for validation. Specifically, the validation set employed $100$, $25$, $25$, and $25$ range profiles for clutter, PMN/PMA2, ERA and Type-72, respectively. We used a stochastic gradient descent algorithm for updating the network parameters with the learning rate of $0.001$ and mini-batch size of $20$ samples for $2000$ epochs.
We realized the proposed network in TensorFlow on a Windows 7 PC with 8-core CPU. The network training took $3.88$ minutes. Figure~\ref{fig:classmap_CNN} shows the classification map obtained using CNN. The corresponding confusion matrix is listed in Table~\ref{tbl:CNN_CM}. We note that the CNN classifier shows worse $P_{CC}$ than our SR-based techniques, particularly for ERA and Type-72 target classes.  
\begin{table}[t]
\centering
    \caption{Confusion matrix for CNN-based classification}
	\label{tbl:CNN_CM}
	\begin{tabular}{ l | l | l | l | l}
		\hline
         \noalign{\vskip 1pt}
            & Clutter & PMN/PMA2 & ERA & Type-72\\[1pt]
		\hline
        \hline
        \noalign{\vskip 1pt}
            Clutter & \cellcolor{black!25}0.909& 0.14 & 0.38 & 0.574 \\[1pt]
            PMN/PMA2 & 0.036 & \cellcolor{black!25}0.807 & 0.181 & 0 \\[1pt]
            ERA & 0.022 & 0.053 & \cellcolor{black!25}0.319 & 0.056 \\[1pt]
            Type-72 & 0.033 & 0 & 0.111 & \cellcolor{black!25}0.370 \\[1pt]
		\hline
		\hline
        \noalign{\vskip 1pt}
	\end{tabular}
\end{table}   
\paragraph{Classification with Reduced Range Samples}
\label{subsec:sparse_sensing}
We now analyze the robustness of our DL-based adaptive classification method to the reduction of the number of samples in the raw data. Assuming the collected data $\mathbf{Y}_{\text{TEST}}$ is sparse in dictionary $\mathbf{D}$, we undersampled the original raw data $\mathbf{Y}_{\text{TEST}}$ in range to obtain its row-undersampled version $\widetilde{\mathbf{Y}}_{\text{TEST}}$ by randomly reducing the samples. We then applied the same random sampling pattern to the dictionary $\mathbf{D}$ for obtaining the sparse coefficients. We also analyzed the CNN classifier when the signals are randomly reduced in the same way. Figure~\ref{fig:classmap_50} illustrates the classification map for all DL approaches when the sampling is reduced by $50$\%. Table \ref{tbl:CM_DL_50} shows the corresponding confusionn matrix.
In comparison to the results in Table~\ref{tbl:CM_DL_full} which used all samples of the raw data, the DL approaches maintain similar classifier performance even when we reduce the samples by 50\% (i.e. just 105 samples in total). In contrast, the CNN classifier result which is fails completely for $50$\% sampling rate.

Reducing the number of signal samples when using a dictionary which minimizes the number of non-zero entries in the sparse representation, still assures an exact reconstruction of the signal itself and, consequently its correct classification. The features for classifying the traces are thus robust to the reduction of the original samples. Deep learning strategies use the signal samples directly as classification features. They also require enormous amount of data for training. Therefore, the degradation in their performance is expected. 
\begin{table}[t]
\centering
 	\caption{Confusion matrices for different DL algorithms and CNN with 50\% samples reduction}
	\label{tbl:CM_DL_50}
		\begin{tabular}{ c | l | l | l | l | l}
		\hline
    	\noalign{\vskip 1pt}
            \multicolumn{2}{c|}{} & Clutter & PMN/PMA2 & ERA & Type-72\\[1pt]
        \hline
        \hline
        \noalign{\vskip 1pt}
             						& Clutter & \cellcolor{blue!25}0.882& 0.026 & 0.291 & 0.37 \\[1pt]
            \multirow{2}{*}{K-SVD} 	& PMN/PMA2 & 0.018 & \cellcolor{black!25}0.947 & 0.153 & 0.037 \\[1pt]
            						& ERA & 0.021 & 0.026 & \cellcolor{black!25}0.5 & 0 \\[1pt]
            						& Type-72 & 0.078 & 0 & 0.055 & \cellcolor{black!25}0.592 \\[1pt]
		\hline
        \noalign{\vskip 1pt}
            						& Clutter & \cellcolor{black!25}0.868 & 0 & 0.208 & 0.333\\[1pt]
            \multirow{2}{*}{ODL} 	& PMN/PMA2 & 0.021 & \cellcolor{black!25}0.965 & 0.18 & 0.018 \\[1pt]
            						& ERA & 0.018 & 0.035 & \cellcolor{black!25}0.5 & 0 \\[1pt]
            						& Type-72 & 0.09 & 0 & 0.111 & \cellcolor{black!25}0.648 \\[1pt]
        \hline
        \noalign{\vskip 1pt}
            						& Clutter & \cellcolor{black!25}0.872 & 0.017 & 0.25 & 0.40 \\[1pt]
            \multirow{2}{*}{CBWLSU} & PMN/PMA2 & 0.023 & \cellcolor{blue!25}0.973 & 0.111 & 0 \\[1pt]
            						& ERA & 0.02 & 0.008 & \cellcolor{black!25}0.541 & 0\\[1pt]
            						& Type-72& 0.083 & 0 & 0.097 & \cellcolor{black!25}0.592 \\[1pt]
		\hline
         \noalign{\vskip 1pt}
            						& Clutter & \cellcolor{black!25}0.868 & 0.035 & 0.194 & 0.296  \\[1pt]
            \multirow{2}{*}{DOMINODL} & PMN/PMA2 & 0.023 & \cellcolor{black!25}0.929 & 0.138 & 0  \\[1pt]
            						& ERA & 0.024 & 0.035 & \cellcolor{black!25}0.527 & 0.018 \\[1pt]
            						& Type-72 & 0.083 & 0 & 0.138 & \cellcolor{blue!25}0.685 \\[1pt]
 		\hline
         \noalign{\vskip 1pt}
            						& Clutter & \cellcolor{black!25}0.265 & 0.166 & 0.181 & 0.148  \\[1pt]
            \multirow{2}{*}{CNN} & PMN/PMA2 & 0.062 & \cellcolor{black!25}0.096 & 0.069 & 0.018\\[1pt]
            						& ERA  & 0.645 & 0.728 & \cellcolor{blue!25}0.736 & 0.75 \\[1pt]
            						& Type-72 & 0.027 & 0.088 & 0.014 & \cellcolor{black!25}0.074  \\[1pt]
 		\hline 		
		\hline
        \noalign{\vskip 1pt}
	\end{tabular}
\end{table}   

\section{Summary}
\label{sec:summ}
In this chapter, we reviewed modern methods for GPR target recognition which includes sparse representation, data learning methods, adpaptive processing and advanced classifiers. Most applications combine these techniques for specific application. In particular, we presented effective online DL strategies for sparse decomposition of GPR traces. The online methods outperform K-SVD thereby making them a good candidate for SRC. Our algorithm DOMINODL is always the fastest providing near real-time performance and high clutter rejection while also maintaining a classifier performance that is comparable to other online DL algorithms. DOMINODL and CBWLSU generally classify smaller targets better than ODL and K-SVD. Unlike previous works that rely on RMSE, we used metrics based on statistical inference to tune the DL parameters for enhanced operation.

Most importantly, we illustrated application of each technique using real data from GPR for landmine detection. Our analyses show that adoption of SR-based DL classification is not only feasible in real applications but also fast enough to be implemented in real-time. We believe that such practical implementations pave the way towards the next step of cognition in modern GPR operation, wherein the system uses previous measurements to optimize the processing performance and is capable of sequential sampling adaptation based on the learned dictionary. 


\bibliographystyle{vancouver-modified}
\bibliography{main}

\end{document}